\documentclass[11pt,a4paper,oneside]{article}
\usepackage[top=3cm, bottom=3cm, left=2cm, right=2cm]{geometry}
\usepackage{float}
\usepackage{amsthm,amsmath,amssymb,mathrsfs,dsfont,mathtools}

\usepackage{caption}
\usepackage{bm}
\usepackage{enumitem}
\usepackage{graphicx}
\usepackage{multirow,makecell}

\usepackage{xcolor}
\usepackage[colorlinks=true, citecolor=blue,
linkcolor = black]{hyperref}
\usepackage{xurl}
\usepackage{booktabs,subcaption}
\usepackage[all]{nowidow}
\usepackage{setspace}
\usepackage[ruled,vlined]{algorithm2e}
\SetAlCapNameFnt{\small}
\SetAlCapFnt{\small}

\makeatletter
\renewcommand{\algocf@captiontext}[2]{#1\algocf@typo. \AlCapFnt{}#2} % text of caption
\def\@algocf@capt@plain{top}
\renewcommand{\algocf@makecaption}[2]{%
	\addtolength{\hsize}{\algomargin}%
	\sbox\@tempboxa{\algocf@captiontext{#1}{#2}}%
	\ifdim\wd\@tempboxa >\hsize%     % if caption is longer than a line
	\hskip .5\algomargin%
	\parbox[t]{\hsize}{\algocf@captiontext{#1}{#2}}% then caption is not centered
	\else%
	\global\@minipagefalse%
	\hbox to\hsize{\box\@tempboxa}% else caption is centered
	\fi%
	\addtolength{\hsize}{-\algomargin}%
}
\makeatother

\allowdisplaybreaks

\usepackage{tikz} % For the plot
\usepackage[sf]{titlesec}
\usepackage{sectsty}
\allsectionsfont{\centering \normalfont\scshape} % Make all sections centered, the default font and small caps
\usepackage{adjustbox}
\usepackage{natbib}
\usepackage[resetlabels]{multibib}
\newcites{Supp}{References}
\usepackage[noabbrev,capitalise]{cleveref}
\creflabelformat{equation}{#2\textup{(#1)}#3}

\usepackage{indentfirst}

\usepackage{authblk}
\title{Poisson process factorization for modeling mutational processes\\along cancer genomes}
\date{}
\author[1,2]{Alessandro Zito}
\author[1,2]{Giovanni Parmigiani}
\author[1]{Jeffrey W. Miller}

\affil[1]{\small{\emph{Department of Biostatistics, Harvard T.H. Chan School of Public Health, Boston, MA, U.S.A.}}}
\affil[2]{\small{\emph{Department of Data Sciences, Dana-Farber Cancer Institute, Boston, MA, U.S.A.}}}

\newtheorem{lemma}{Lemma}
\newtheorem{proposition}{Proposition}

\theoremstyle{definition}

\newcommand{\bx}{\mathbf{x}}

\newcommand{\bbeta}{\boldsymbol{\beta}}

\newcommand{\bkx}{\bbeta_k^\top\bx}

\newcommand{\dt}{\mathrm{d}t}

\newcommand{\bz}{\mathbf{z}}
\newcommand{\decr}{$\textcolor{blue}{\boldsymbol{\searrow}}$}
\newcommand{\incr}{$\textcolor{red}{\boldsymbol{\nearrow}}$}
\definecolor{sketch}{HTML}{A41034}

\newcommand{\sbs}[1]{SBS#1}

\begin{document}
\maketitle

\begin{abstract}
Cancer cells acquire DNA mutations through many processes, such as environmental exposures and dysregulated repair mechanisms, and each process consistently produces distinct mutation types at characteristic frequencies, referred to as its signature. The usual approach for inferring these signatures is to decompose the matrix of mutation counts from a sample of tumors via non-negative matrix factorization (NMF). However, existing methods do not model the heterogeneity of mutation rates along the genome, which is driven in part by observed genomic features. In this paper, we introduce Poisson process factorization (PPF), which addresses this limitation by employing an inhomogeneous Poisson point process model to infer mutational signatures and their activities as they vary across the genome. PPF generalizes the baseline NMF model by representing a patient's exposure to each signature as a locus-specific function that depends on genomic covariates and patient-specific copy numbers via a log-linear model. We apply PPF to a sample of 113 breast tumors with $707{,}104$ mutations, using covariates representing histone modifications, cell replication timing, nucleosome positioning, and DNA methylation. Our analysis quantifies the joint effects of these features on the mutational processes in breast cancer, and estimates patient-specific activities of each signature as a function of genomic position.
\end{abstract}

\newpage
\section{Introduction}

Cancer is associated with the accumulation of mutations in the genomes of somatic cells \citep{Stratton2009}. These mutations manifest in distinctive patterns called \emph{mutational signatures}, which are latent vectors encoding the relative frequencies of various mutation types \citep{Alexandrov_2013}. Several empirically derived signatures have been experimentally linked to biological processes such as DNA repair deficiencies, oxidative stress, or exposure to exogenous carcinogens like tobacco smoking or ultraviolet light. Detecting the activity of certain signatures in patients has led to advances in precision treatment \citep{Aguirre_2018, samur2020genome}, making mutational signature analysis an exciting emerging direction in cancer research \citep{Alexandrov_2020}.%

However, nearly all methods for mutational signature analysis operate at an aggregate scale, considering only the total number of each mutation type detected in the genome of each patient.
This implicitly assumes a constant mutation rate across the genome, which is not generally the case; for example, see \cref{fig:Mutations_all_breast}(a).
This inhomogeneity in mutation rates is well documented and relates to variation in GC content, DNA accessibility, replication timing, and epigenetic modifications, as well as copy number alterations in cancer.
For example, histone modification H3K9me3 can account for nearly 40\% of the variation in mutation rates at the megabase scale in many tumors \citep{schuster2012chromatin}; see \cref{fig:Mutations_all_breast}(b) for example.
Existing methods for handling this heterogeneity are either based on partitioning the genome and fitting a separate model on each region \citep{Fischer_2013}, stratifying mutations according to discrete epigenetic states \citep{TensorSignatures_2021}, or \emph{post hoc} association of genomic features with the estimated activity of pre-defined signatures
\citep{Otlu_2023,Blokzijl_2018,Timmons_2022,kups75586}.
Relatedly, approaches based on multinomial mixture models capture dependence between consecutive mutations \citep{StickySig} and the effect of discrete mutation-level covariates \citep[such as replication strand;][]{Kahane_2023}, but do not model the joint effect of continuous epigenetic features on mutational process activity.

In this article, we address this by introducing \emph{Poisson process factorization} (PPF), a statistical framework for jointly inferring mutational signatures, their activities, and the effect of genomic covariates on such activities.
Specifically, we model mutations from each signature as realizations from an inhomogeneous Poisson point process \citep{kingman-poisson-processes, Daley_verejones_2003} with a locally varying intensity function that captures the multiplicative effects of genomic covariates and copy number alterations via a log-linear model. In PPF, each signature is associated with a set of regression coefficients that quantify the enrichment or depletion of mutational burden associated with each covariate. In contrast to existing methods, these coefficients are estimated \emph{jointly}, accounting for the effects of all covariates as well as patient-specific copy number. Furthermore, the model enables the attribution of individual mutations to signatures by providing a patient- and locus-specific probability distribution over signatures for a given mutation type.

PPF provides automatic selection of the number of active signatures using compressive Bayesian hyperpriors \citep{zito2024compressive}, which avoid computationally intensive rank selection procedures by shrinking the activities of redundant signatures to negligible values. We develop an algorithm for \emph{maximum a posteriori} estimation using majorization-minimization \citep{Lee_Seung_2000} coupled with Newton's method, and present a Markov chain Monte Carlo sampler for fully Bayesian inference. Both algorithms are applicable under any genomic resolution.

The article is structured as follows.
\cref{sec:breastCancer} describes the scientific questions of interest and the data used.
\cref{sec:model} introduces the Poisson process factorization (PPF) model. \cref{sec:inference} provides estimation algorithms. \cref{sec:simulations} presents the simulation studies, and \cref{sec:application} applies PPF to breast cancer data. Future directions are in \cref{sec:discussion}.

\subsection{Scientific questions of interest and datasets}\label{sec:breastCancer}

Breast cancer is the most common cancer among women, with an estimated lifetime risk of 1 in 8 \citep{Giaquinto_2024}. Over the last 15 years, several mutational signatures representing distinct biological processes have been identified in breast tumors  \citep{Nik_zainal_2012, Nik-Zainal_2016, Davies_2017}. These primarily involve DNA methylation and aging (Signatures SBS1 and SBS5), homologous recombination deficiencies (SBS3), mismatch repair deficiencies (SBS6, SBS26, and SBS44), dysregulated APOBEC (SBS2 and SBS13), and oxidative damage (SBS18); see the \emph{Catalogue of Somatic Mutations in Cancer} \citep{Tate_2019}. Quantifying the activity of each signature in each patient helps identify candidate subgroups for targeted treatments \citep{Gulhan_2019}.

Our goal is to develop a modeling framework to address the following questions:
\begin{enumerate}
    \setlength{\itemsep}{0.1pt}
    \setlength{\parskip}{0.5pt}
    \item[Q1.] Which mutational signatures are active in a tumor, and to what degree?
    \item[Q2.] At specific locations, which genomic features are associated with the activity of processes underlying cancer signatures? And with what direction and magnitude?
    \item[Q3.] Does accounting for genomic features influence signature attribution for mutations?
    \item[Q4.] Can genomic-covariate-aware signature analysis identify patient subgroups with biologically and clinically distinct mutational processes?
\end{enumerate}
In Section~\ref{subsec:previous_work}, we discuss why existing methods are insufficient to answer these questions.

\begin{figure}
    \centering
    \includegraphics[width=\linewidth]{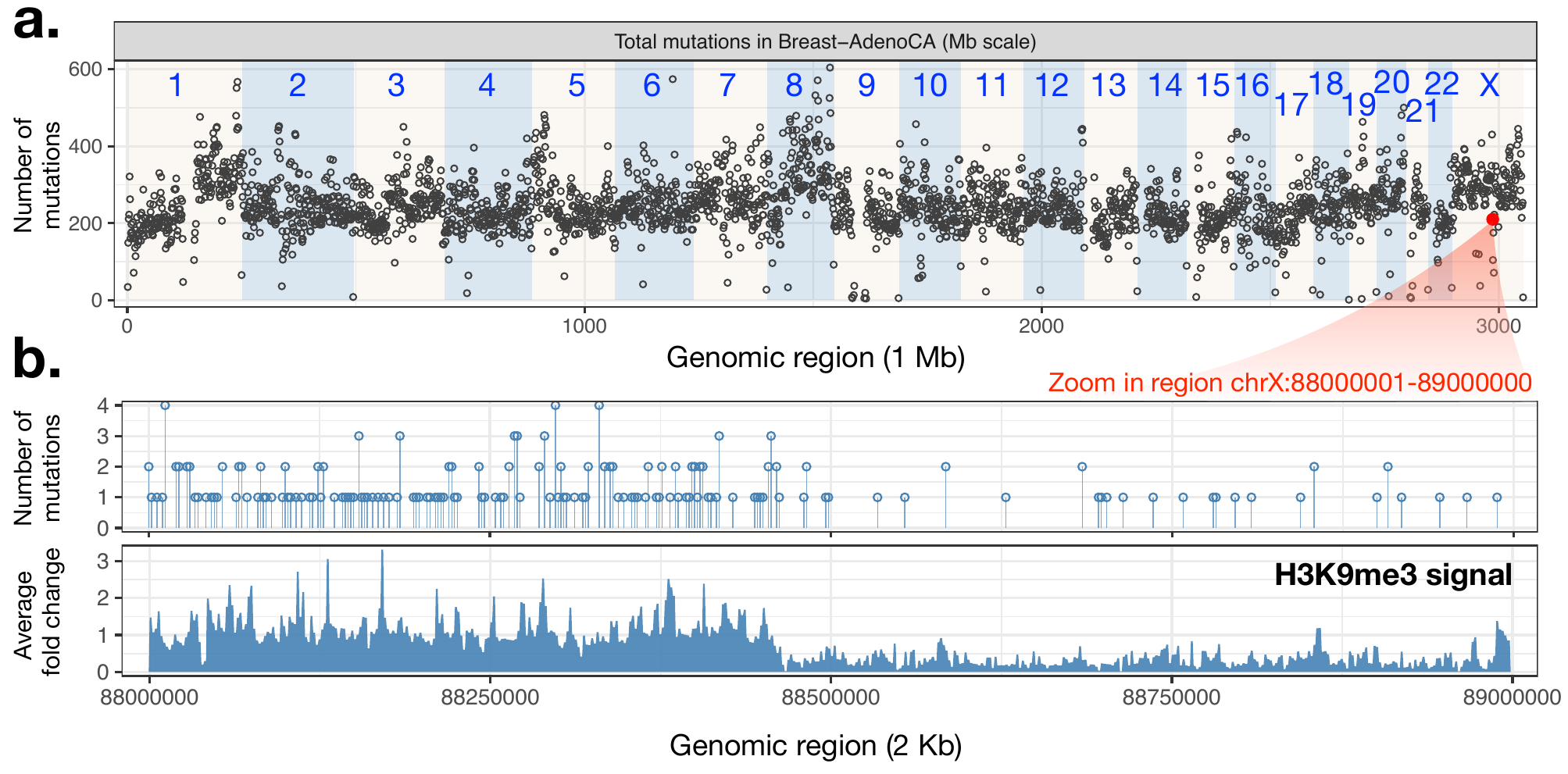}
    \caption{(a) Number of single-base substitutions in 1-megabase bins in 113 breast adenocarcinomas from the International Cancer Genome Consortium (ICGC). Alternating colored bands and numbers indicate chromosomes. (b) Top: Number of mutations in 2-kilobase bins for region chrX:88,000,001-89,000,000, highlighted by the red point. Bottom: Average signal (fold-change with respect to assay background) in the same bins for histone modification H3K9me3, in breast epithelium tissue; see Supplementary material.
    }
    \label{fig:Mutations_all_breast}
\end{figure}

To study these questions, we analyze whole-genome sequencing data from 113 women with breast cancer from the \texttt{Breast-AdenoCA} ICGC cohort of \citet{PCAWG_2020}. The data consist of $707{,}104$ mutations in total, spread across 23 chromosomes as depicted in \cref{fig:Mutations_all_breast}a, from two sub-collections: BRCA-UK (43 patients, triple-negative ductal) and BRCA-EU (70 patients, ER+ and HER2-- subtypes). For each mutation, we have knowledge of the genomic position (e.g., \texttt{chrX:77364730}) and the single-base substitution (SBS) with trinucleotide context (e.g., A[C$>$T]C). The median number of mutations per patient is $4{,}125$, and there are four hypermutated patients with more than $20{,}000$ mutations each, the largest of which has $64{,}464$ mutations (patient \textsc{DO1076}). To assess replicability, we also analyze 80 breast cancer tumors from \citet{Davies_2017}; see \cref{sec:breast80}.
Both collections are landmark datasets in cancer biology, as they first established the comprehensive catalog of mutational processes operating in breast cancer \citep{Nik-Zainal_2016} and showed that mutational signatures can identify BRCA1/2-deficient tumors \citep{Davies_2017}.%

In addition to SBS mutations, we have access to estimated copy numbers $c_j(t)$, genome-wide, for each patient in both cohorts. Importantly, variations in $c_j(t)$ are correlated with the mutation rate at the megabase scale ($\rho = 0.44$). For example, the average copy numbers in chromosome arms 1p and 1q are $2.42$ and $3.90$, respectively, in the ICGC data, with associated average mutation counts equal to $217$ and $304$, across all patients. This explains the jump in mutation rate in the middle of chromosome 1 in \cref{fig:Mutations_all_breast}.

We use
covariates for histone modifications, nucleosome occupancy, transcription factor bindings (CTCF), cell replication timing, DNA methylation, and GC content \citep{Morganella2016, Otlu_2023}. These genomic features have previously been found to influence mutation rates: \cref{tab:Epi_covariates} describes the 11 covariates considered, as well as their expected direction of effect on aggregate mutation rate based on the literature; see \cref{subsec:genomic_covariates} for more background. We focus our analyses on these 11 covariates since they represent the most important and well-studied components of the epigenome,
making them suitable for answering our questions. Data sources and preprocessing are described in \cref{sec:sources_prep} of the Supplementary material.

\begin{spacing}{1}
\begin{table}[t]
\caption{\footnotesize{Description of genomic and epigenetic covariates used in the application. The \textsc{Assay} column indicates the assay used to obtain the covariate.
\textsc{Mutations} describes previously reported associations between the covariate and mutation rate in cancer (not necessarily breast). The \incr \ and \decr \ symbols indicate positive and negative effects, respectively. \textsc{Reference} indicates the work that reports the association.}}
\label{tab:Epi_covariates}
\centering
\footnotesize
\begin{adjustbox}{max width=1.1\textwidth,center}
\begin{tabular}{p{0.1\textwidth}lp{0.5\textwidth}cp{0.18\textwidth}}
\toprule
\textsc{Covariate} & \textsc{Assay} & \textsc{Description and role} & \textsc{Mutations} &  \textsc{Reference} \\
\midrule
Nucleosome occupancy & MNase-seq & Degree to which DNA is wrapped around nucleosomes; higher values indicate packed chromatin. & \makecell{Periodic\\patterns\vspace{-10pt}} &\citet{PICH20181074}  \\
\midrule
H3K27ac & Histone ChIP-seq & Marker of active enhancers and promoters; associated with gene activation.  & \incr & \citet{schuster2012chromatin} \\
\addlinespace[1pt]
H3K4me1 & Histone ChIP-seq & Marks poised and active enhancers; enriched at regulatory elements. & \decr &\citet{Hodgkinson_2012}\\
\addlinespace[1pt]
H3K4me3 & Histone ChIP-seq & Marks active promoters; associated with transcription initiation. & \decr & \citet{Hodgkinson_2012} \\
\addlinespace[1pt]
H3K9me3 & Histone ChIP-seq & Marker of constitutive heterochromatin; gene silencing/repression. & \incr & \citet{schuster2012chromatin}\\
\addlinespace[1pt]
H3K27me3 & Histone ChIP-seq & Repressive mark, associated with Polycomb-mediated gene silencing. & \incr & \citet{schuster2012chromatin}  \\
\addlinespace[1pt]
H3K36me3 & Histone ChIP-seq & Marker of transcriptional elongation within gene bodies. & \decr & \citet{Li_2013_H3K39me3} \\
\addlinespace[1pt]
\midrule
CTCF & TF ChIP-seq & Insulator protein, key architectural TF regulating chromatin looping and gene expression. & \incr & \citet{Katainen2015}\\
\addlinespace[1pt]
\midrule
Replication timing & Repli-seq & Timing of DNA replication during S-phase of a cell; reflects chromatin state and genome organization. & \makecell{\incr \ (later)\\ \decr \, (early)\vspace{-10pt}} & \citet{Supek_Lehner_2015_nature} \\
\midrule
Methylation & WGBS & Level of DNA methylation; regulates gene expression and silencing. & \incr \ at CpGs & \citet{Bird_19780_Methylation} \\
\addlinespace[1pt]
GC ~~~~ content & --- & Proportion of G and
C nucleotides in a region; influences DNA stability and nucleosome positioning. & \decr & \citet{Makova2015} \\
\bottomrule
\end{tabular}
\end{adjustbox}
\end{table}
\end{spacing}

\section{Model}\label{sec:model}

In this section, we provide background on mutational signature analysis (\cref{subsec:BaselinePoisson}), introduce the PPF model (\cref{subsec:PoisPro}) and its theoretical properties (\cref{subsec:theory}), describe the log-linear parametrization for the effect of genomic covariates (\cref{subsec:loglinear}), and specify the prior distributions over model parameters (\cref{subsec:priors}).

\subsection{Background on mutational signature analysis}\label{subsec:BaselinePoisson}

Since DNA has a double-stranded structure where nucleotide A always binds with T, and C always binds with G, six distinct substitutions can happen at any genomic position: C$>$A, C$>$T, C$>$G, T$>$A, T$>$C, or T$>$G, where C$>$A denotes a mutation from the pair C:G into the pair A:T. Moreover, it has been observed that mutation rates depend strongly on the adjacent bases \citep{Nik_zainal_2012}. Hence, it is common practice to further categorize single-base substitutions (SBS) according to the trinucleotide context, leading to a total of $I = 4 \times 6 \times 4 = 96$ mutation types, sometimes called channels \citep{Alexandrov_2013}. For instance, T[C$>$A]G indicates a C$>$A mutation where T and G are the nucleotides flanking C on the 5' and 3' sides, respectively.

Let $N_{i j} \in \mathds{N} = \{0, 1, 2, \ldots, \}$ be the number of mutations of type $i = 1, \ldots, I$, with $I = 96$, for patient $j = 1, \ldots, J$, and let $N \in \mathds{N}^{I\times J}$ denote the matrix with entries $N_{ij}$. The standard approach to mutational signature analysis uses non-negative matrix factorization \citep[NMF;][]{Lee_Seung_2000} to approximate $N$ by the product $R\, \Theta$, where $R\in \mathds{R}^{I\times K}_+$ is the \emph{mutational signature} matrix, and $\Theta\in \mathds{R}^{K\times J}_+$ is the \emph{activity} matrix. Here, $K\geq 1$ is the rank of the factorization and represents the number of active mutational processes. In traditional NMF, $R$ and $\Theta$ are estimated by minimizing a loss function such as the squared Euclidean distance or the Kullback--Leibler divergence; see \citet{gillis2021nonnegative}.

From a probabilistic modeling perspective \citep[e.g.,][]{Zhou_Carin_2015}, it is common to assume
\begin{equation}\label{eq:PoissonBaseline}
N_{ij} \sim \mathrm{Poisson}\bigg(\sum_{k=1}^K r_{ik}\theta_{kj}\bigg)
\end{equation}
independently for each $i$ and $j$, where $\mathrm{Poisson}(\lambda)$ denotes the Poisson distribution with mean $\lambda$. Here, $r_{k} = (r_{1k}, \ldots, r_{Ik})$ and $\theta_{k} = (\theta_{k1},\ldots, \theta_{kJ})$ are the $k$th column of $R$ and the $k$th row of $\Theta$, respectively.  With the constraint $\sum_{i = 1}^{I} r_{ik} = 1$, the vector $r_{k}$ represents the relative frequencies with which each of the $96$ mutation types are generated by signature $k$, and $\theta_{kj}$ represents its activity in patient $j$.

\subsection{Poisson process factorization}\label{subsec:PoisPro}

A point process is a stochastic model for the locations of randomly placed points, which we interpret as mutations. Let $T\geq 0$ be the length of the reference genome and let $A\subseteq[0,T)$ represent a region of the genome. To simplify tractability, we assume that $[0,T)$ is a continuous approximation of the genome where any arbitrarily defined region $A$ is Borel measurable. Such approximation is reasonable, since in practice $T\approx 3\times 10^9$ in our application.
A \emph{Poisson point process} with intensity function $\lambda_{ij}: [0, T) \to \mathds{R}_+$ is a counting process $Z_{i j}$ such that (i) $Z_{ij}(A) \sim \mathrm{Poisson}\big(\int_A \lambda_{ij}(t) \mathrm{d}t\big)$ for all regions $A\subseteq[0,T)$, and (ii) $Z_{ij}(A_1),\ldots,Z_{ij}(A_m)$ are independent for all pairwise disjoint regions $A_1,\ldots,A_m\subseteq [0,T)$.
The intensity determines the expected number of points in $A$, since $\mathds{E}\big(Z_{ij}(A)\big) =\int_A \lambda_{ij}(t)\mathrm{d}t$. If $\lambda_{ij}(t)$ takes the same value for all $t$, then the process is called \emph{homogeneous}; otherwise, the process is \emph{inhomogeneous} \citep{kingman-poisson-processes}.

We model the mutations as following inhomogeneous Poisson point processes $Z_{i j}$, independently for all $i$ and $j$, where $Z_{i j}(A)$ denotes the number of mutations of type $i$ observed in region $A$ for patient $j$.
An attractive feature of Poisson point processes is that the likelihood is relatively simple.
Suppose that for a given $i$ and $j$, we observe $N_{ij}$ mutations at positions $t_{i j 1}$, \ldots, $t_{i j N_{ij}}$, where $t_{i j n} \in [0, T)$ for each $n = 1,\ldots, N_{ij}$.
It can be shown \citep[Chapter 2]{Daley_verejones_2003} that the likelihood of the intensity functions $\lambda = (\lambda_{i j})$ is
\begin{equation}\label{eq:lik_Lambda}
\mathscr{L}(\lambda) = \prod_{i=1}^I \prod_{j=1}^J\exp\bigg( -\int_{0}^{T} \lambda_{ij}(t)\dt \bigg) \prod_{n=1}^{N_{ij}} \lambda_{ij}(t_{i j n}).
\end{equation}

The flexibility of the specification arising from \cref{eq:lik_Lambda} lies in
the choice of the functional form of the intensity $\lambda_{ij}$. Mirroring \cref{eq:PoissonBaseline}, we let
\begin{equation}\label{eq:Intensity_function}
\lambda_{ij}(t) = \sum_{k = 1}^K r_{ik}\vartheta_{kj}(t), \qquad  t\in [0, T),
\end{equation}
for each mutation type $i =1, \ldots, I$ and patient $j = 1,\ldots, J$. As before, $K$ is the rank of the factorization and
$r_{k}=(r_{1 k},\ldots,r_{I k})$ is the $k$th mutational signature, but instead  now $\vartheta_{kj}: [0, T)\to \mathds{R}_+$ is a position-dependent \emph{activity function} that quantifies the rate of mutations due to signature $k$ at locus $t$ in patient $j$. We refer to the model specified by \cref{eq:lik_Lambda,eq:Intensity_function} as \emph{Poisson process factorization} (PPF).

\subsection{Theoretical properties}\label{subsec:theory}
We establish key theoretical properties of PPF. Our first result, in \cref{pro:baseline}, shows that if the activity functions are held constant along the genome, then the PPF model is equivalent to the standard Poisson NMF in \cref{eq:PoissonBaseline}. As a consequence, PPF is a flexible generalization of NMF where the activities depend on position $t$.

\begin{proposition}\label{pro:baseline}
Consider a collection of inhomogeneous Poisson point processes $Z_{ij}$ on $[0,T)$ with intensity functions $\lambda_{i j}$ as in \cref{eq:Intensity_function}.
 If $\vartheta_{kj}(t) = \theta_{kj}/T$ for all $t \in [0, T)$ and all $i,j,k$,
then inference for $r_{ik}$ and $\theta_{kj}$ is equivalent under \cref{eq:lik_Lambda} and \cref{eq:PoissonBaseline}.
\end{proposition}

The next two properties are a consequence of the superposition and the coloring theorems of Poisson processes \citep{kingman-poisson-processes}. \cref{pro:superposition} shows that the combined mutational process in PPF can be generated as a sum of $K$ independent Poisson processes, each representing the contribution of a distinct signature.  Conversely, \cref{pro:thinning} states that assigning the mutations generated by a PPF to distinct signatures yields $K$ mutually independent Poisson processes, each with its own intensity. Thus, signatures generate mutations along the genome independently of each other. This generalizes the additivity of Poisson NMF, for which counts arise from the sum of independent Poisson contributions.

\begin{proposition}\label{pro:superposition}
Let $Y_{ij1}, \ldots, Y_{ijK}$ be independent, inhomogenoeus Poisson point processes on $[0,T)$ such that $Y_{ijk}$ has intensity function $\lambda_{ijk}(t) = r_{ik}\vartheta_{kj}(t)$ for $t \in [0, T)$. Then $\sum_{k = 1}^K Y_{ijk}$ is a Poisson point process with the intensity function in \cref{eq:Intensity_function}.%
\end{proposition}

\begin{proposition}\label{pro:thinning}
Let $Z_{ij}$ be an inhomogeneous Poisson process on $[0,T)$ with intensity function $\lambda_{ij}(t) = \sum_{k=1}^K \lambda_{ijk}(t)$, where $\lambda_{ijk}(t) = r_{ik}\vartheta_{kj}(t)$, as in \cref{eq:Intensity_function}. Suppose we observe
points $t_{ij1},\ldots,t_{ijN_{ij}}$ from $Z_{ij}$ and, independently
for each $n = 1, \ldots, N_{ij}$, we draw
{\setlength{\abovedisplayskip}{6pt}
\setlength{\belowdisplayskip}{6pt}
\begin{equation*}
\big(W_{ij1}(t_{ijn}), \ldots, W_{ijK}(t_{ijn})\big) \sim \mathrm{Mult}\big(1;\, p_{ij1}(t_{ijn}),\ldots,
p_{ijK}(t_{ijn})\big),
\end{equation*}}
where $p_{ijk}(t) = \lambda_{ijk}(t)/\lambda_{ij}(t)$. Define $Y_{ijk} = \sum_{n=1}^{N_{i j}} \delta_{t_{i j n}} \mathds{1}(W_{ijk}(t_{ijn}) = 1)$.
Then
$Y_{ij1},\ldots,Y_{ijK}$ are realizations of independent Poisson processes with intensity functions $\lambda_{ij1}(t),\ldots,\lambda_{ijK}(t)$.
\end{proposition}
Here, $\delta_t$ denotes the point mass at $t$, so $Y_{i j k}$ has points at $(t_{i j n} : W_{ijk}(t_{ijn}) = 1)$.

\subsection{Log-linear model for effect of genomic covariates}
\label{subsec:loglinear}

In \cref{eq:Intensity_function}, the activities are allowed to vary with the genomic position $t$, whereas the signatures are constant in $t$.
Holding each $r_{k}$ constant allows us to interpret variations in the mutation rate as variations in the activity of specific signatures. This assumption is biologically justified: locally, the biochemical mechanisms producing mutations are largely invariant across the genome, while the opportunity for each process to act varies with genomic context. For example, SBS1 is due to deamination of 5-methylcytosine producing  C$>$T substitutions, but this only happens at methylated sites, so it depends on the amount of methylation in a region. Similarly,  SBS2 and SBS13 are due to APOBEC cytidine deaminases, producing C$>$T and C$>$G substitutions respectively, but these only act on single-stranded DNA, whose availability is governed by replication timing and transcriptional activity.
Some processes act more on one DNA strand than the other, such as transcription-coupled DNA repair; we discuss how to handle strand asymmetry in Section~\ref{sec:discussion}.

To parametrize the activity functions $\vartheta_{k j}(t)$, we propose a log-linear model that leverages genomic covariates as well as patient-specific factors.
Suppose there are $L$
known covariate functions $x_{\ell} : [0, T) \to  \mathds{R}$ for $\ell = 1, \ldots, L$, with $x_\ell(t)$ denoting the value of covariate $\ell$ at position $t$. Denote their vector representation by $\bx(t) = (x_1(t),\ldots,x_L(t)) \in \mathds{R}^L$.
For example, $x_\ell(t)$ may be the intensity of the signal indicating a histone modification such as H3K36me3, H3K27ac, or H3K9me3, as in \cref{fig:Mutations_all_breast}.
Additionally, let $c_j: [0, T)\to \mathds{R}_+$ be a non-negative function such that $c_j(t)$ is the copy number at position $t$ in patient $j$.
Both $\bx(t)$ and $c_j(t)$ are assumed to be fixed and known based on other data.
Note that $\bx(t)$ is shared across all patients, whereas $c_j(t)$ is patient-specific.

Our choice for the activity functions in
\cref{eq:Intensity_function} is
{\setlength{\abovedisplayskip}{6pt}
\setlength{\belowdisplayskip}{6pt}
\begin{equation}\label{eq:vartheta_t}
\vartheta_{kj}(t) = \frac{1}{2}\, \phi_{kj} \, c_j(t) \,e^{\boldsymbol{\beta}_{k}^\top \mathbf{x}(t)}, \qquad  t\in [0, T),
\end{equation}}
where $\phi_{kj} > 0$ for each $k, j$, and $\bbeta_{k} = (\beta_{k 1}, \ldots, \beta_{k L})\in\mathds{R}^L$ for each $k$.
Here, $\phi_{kj}$ is the baseline activity of signature $k$ in patient $j$, while $\bbeta_{k}$ is a vector of regression coefficients capturing the effect of $\bx(t)$ on the rate of mutations due to signature $k$ at position $t$. The copy number  $c_j(t) \geq 0$ affects the mutation rate at location $t$ in patient $j$, irrespective of the signature. In a normal genome, $c_j(t) = 2$ since there are two copies of each chromosome, except for X and Y in males.
Thus, the factor $c_j(t)/2$ is the ratio of observed to expected copy number.
The baseline function $\vartheta_{kj}(t) = \phi_{kj}$ is a special case occurring when $c_j(t) = 2$ and $\bbeta_k^\top \bx(t) = 0$ for all $t$. This is the case described by \cref{pro:baseline}, where $\phi_{kj} = \theta_{kj}/T$.%

The model in \cref{eq:vartheta_t} is motivated by scientific interpretability.
The genomic covariates tend to modulate the opportunities for each process to act, and thus have multiplicative effects on mutation rates.  Importantly, this multiplicative structure can capture effects that suppress mutational activity as well as those that amplify it, as opposed to a strictly additive structure. Specifically, a one-unit increase or decrease in $x_{\ell}(t)$ implies a $|e^{\beta_{k\ell}} - 1|$ percent increase or decrease, respectively, in the activity of signature $k$ at $t$. With an additive structure (such as an identity link), modeling activity suppression would be complicated by the non-negativity constraint on Poisson means, and the interpretation of coefficients would be less clear.
Nonparametric mutation rate models have also been proposed that offer greater flexibility \citep{Lawless01091987, Trucolo_2005, LLoyd_2016}, but they sacrifice interpretability, which is one of our central goals. In this respect, our model differs from previous approaches for point processes with a factorized intensity \citep{pmlr-v32-miller14, LLoyd_2016, Hosseini_2020}, which are primarily designed to capture spatial or temporal autoregressive trends rather than covariate-driven effects.

In our model,
the expected number of mutations of type $i$ in region $A$ for patient $j$ is
\begin{equation}\label{eq:expectation}
\mathds{E}\big(Z_{ij}(A) \,\big\vert\, R, \Phi, B\big) = \sum_{k = 1}^K r_{ik} \phi_{kj} \int_A \frac{1}{2} \,c_{j}(t) \,e^{\bbeta_k^\top\bx(t)}\mathrm{d}t,
\end{equation}
where $R = (r_{i k}) \in\mathds{R}^{I\times K}$, $\Phi = (\phi_{k j}) \in\mathds{R}^{K\times J}$ and $B = (\beta_{k \ell}) \in\mathds{R}^{K\times L}$ denote the matrices of signatures, activities, and regression coefficients, respectively. Given $R$, $\Phi$, and $B$, the integral in \cref{eq:expectation} can be calculated via binning, considering that metrics used to represent genomic features are typically stepwise constant; see \cref{sec:application}.

\subsection{Prior distributions}\label{subsec:priors}
Our prior construction extends the hierarchical model of \citet{zito2024compressive}, employing \emph{compressive hyperpriors} to provide automatic selection of the number of latent factors. Specifically, for each $k = 1, \ldots, K$, we take
{\setlength{\abovedisplayskip}{4pt}
\setlength{\belowdisplayskip}{4pt}
\begin{align}
    r_k = (r_{1k}, \ldots, r_{Ik}) &\sim \mathrm{Dir}(\alpha_{1k}, \ldots,\alpha_{Ik}), \label{eq:R_prior}\\
    \phi_{k1}, \ldots, \phi_{kJ}\mid \mu_k, \bbeta_k  &\sim \mathrm{Ga}\bigg(a,\; \frac{a}{\mu_k}\int_{0}^T\frac{1}{2}c_j(t) e^{\bbeta_k^\top\bx(t)}\mathrm{d}t\bigg),\label{eq:baseline_Priors}\\
    \mu_k &\sim \mathrm{InvGa}(a J + 1,\, \varepsilon a J),\label{eq:compressive}\\
    \boldsymbol{\beta}_k \mid \sigma^2_k &\sim \mathrm{N}(\mathbf{0}, \sigma^2_k I_L),\label{eq:prior_beta} \qquad \sigma^2_k \sim \mathrm{InvGa}(c_0, d_0),
\end{align}
}
where $a, \varepsilon, c_0, d_0 > 0$, $\mathrm{Dir}(\alpha_1, \ldots, \alpha_I)$ is a Dirichlet distribution, $\mathrm{Ga}(a, b)$ is a gamma distribution with mean $a/b$,  $\mathrm{InvGa}(a_0, b_0)$ is an inverse gamma with mean $b_0/(a_0-1)$ for $a_0>1$, $\mathrm{N}(\boldsymbol{m}, \boldsymbol{\Sigma})$ is the multivariate normal distribution with mean $\boldsymbol{m}\in \mathds{R}^L$ and positive definite covariance matrix $\boldsymbol{\Sigma}\in \mathds{R}^{L\times L}$, and $I_L$ is the $L\times L$ identity matrix.
As defaults, we set $a = 1.01$, $\varepsilon = 0.001$, $c_0 = 100$, $d_0 = 1$, and $\alpha_{i k} = 1.01$.

This choice of prior has several important features.
Using a Dirichlet prior in \cref{eq:R_prior} avoids the scaling ambiguities that arise in NMF, since it constrains $\sum_{i=1}^I r_{ik} = 1$.
The prior on baseline $\phi_{k j}$ in \cref{eq:baseline_Priors} is carefully constructed to control the total activity $\theta_{k j} = \int_{0}^T\vartheta_{k j}(t)\dt$ of signature $k$ over the whole genome, independently of $\bbeta_k$, because
{\setlength{\abovedisplayskip}{3pt}
\setlength{\belowdisplayskip}{3pt}
\begin{equation*}
\mathds{E}(\theta_{kj} \mid \mu_k, \bbeta_k) =  \int_0^T\mathds{E}(\phi_{kj} \mid \mu_k, \bbeta_k) \frac{1}{2}c_{j}(t)e^{\bbeta_k^\top \bx(t)}\mathrm{d}t =  \mu_k.
\end{equation*}}
Each hyperparameter $\mu_k$, referred to as the \emph{relevance weight} for signature $k$, is given the data-dependent \emph{compressive hyperprior} in \cref{eq:compressive}, which favors small values of $\mu_k$ since $\mathds{E}(\mu_k) = \varepsilon$ \emph{a priori}. Consequently, when signature $k$ is not needed, the posteriors on $\mu_k$ and $\theta_{k 1},\ldots,\theta_{k J}$ concentrate near $\varepsilon$, effectively removing it from the decomposition \citep{zito2024compressive}.
Finally, the coefficients $\bbeta_k$ and variance $\sigma^2_k$ follow a normal-inverse gamma prior, which eases posterior computation. The default settings of $c_0 = 100$ and $d_0=1$ encourage sparsity by shrinking $\bbeta_k$ near zero \emph{a priori}, so that $\beta_{k\ell}$ moves away from zero if covariate $\ell$ explains position-specific variations in mutation rate for signature $k$.

\section{Algorithms for Poisson process factorization}\label{sec:inference}

\subsection{Maximum a posteriori (MAP) estimation}\label{subsec:MAP}
We first derive a majorization-minimization algorithm to compute a \emph{maximum a posteriori} estimate of the PPF model parameters, inspired by the algorithm of \citet{Lee_Seung_2000} for NMF.
In simple terms, a majorizer for $f(x)$ is a function $g(y,x)$ of two arguments such that (i) $g(y,x) \geq f(y)$ for all $y$, and (ii) $g(x, x) = f(x)$ for all $x$. As a result, if $y = \arg \min_z g(z,x)$, then
$f(x) = g(x,x) \geq g(y,x) \geq f(y)$. Hence, minimizing $g$ implicitly minimizes $f$ as well, which is advantageous when the minima of the majorizer are analytically simple to compute, compared to minimizing the original function. In our model, the target $f$ is the negative log posterior density
{\setlength{\abovedisplayskip}{3pt}
\setlength{\belowdisplayskip}{3pt}
\begin{equation}\label{eq:logPosterior}
\begin{split}
-\log\pi(&R, \Phi, B,  \mu, \sigma^2 \mid \boldsymbol{t}) = \\
 & \sum_{jk} \phi_{kj} \int_0^T \frac{1}{2} \,c_{j}(t) \,e^{\bbeta_k^\top\bx(t)}\mathrm{d}t - \sum_{ij} \sum_{n = 1}^{N_{ij}} \log \Big(\sum_{k = 1}^K r_{ik}\phi_{kj}\frac{1}{2} \,c_{j}(t_{i j n}) \,e^{\boldsymbol{\beta}_{k}^\top \mathbf{x}(t_{i j n})}\Big)\\
 &\quad -\log \pi(R)\,\pi(\Phi\mid \mu, B)\,\pi(\mu)\,\pi(B\mid \sigma^2)\,\pi(\sigma^2) + \mathrm{const},
\end{split}
\end{equation}}
subject to the constraints that $r_{i k}$, $\phi_{k j}$, $\mu_k$, and $\sigma_k^2$ are positive and $\sum_{i=1}^I r_{ik} = 1$ for all $k$.
In \cref{eq:logPosterior}, the bottom line represents the priors from \cref{subsec:priors} and $\boldsymbol{t} = (t_{i j n})$ denotes the collection of observed mutation positions for all $i$, $j$, and $n = 1,\ldots,N_{i j}$.

\begin{algorithm}[t]
\caption{MAP updates for Poisson process factorization}
\label{algo:MAP_rules}
\footnotesize
\nl\textbf{Signatures:} \For{$k = 1, \ldots, K$ }{
\vspace{0.3em}
Update each probability $r_{ik}$ for $i = 1,\ldots, I$ in signature $k$ as
$$
r_{i k}' \gets \alpha_{ik} - 1 +  \sum_{j} \sum_{n = 1}^{N_{i j}} \frac{r_{i k}\phi_{k j} e^{\bkx(t_{i j n})}}{\sum_{s = 1}^K r_{i s}\phi_{s j} e^{\bbeta_s^\top \bx(t_{i j n})}}, \qquad r_{i k} \gets \frac{r_{i k}'}{\sum_{i} r_{i k}'},
$$
where $\phi_{k j} = \theta_{k j}/q_j(\bbeta_k)$ and $q_j(\bbeta_k) = \int_0^T \frac{1}{2}\, c_j(t)\,e^{\bbeta_k^\top \bx(t)}\dt$.
}
\nl \textbf{Activities}: \For{$j = 1, \ldots, J$ $\textbf{\textup{and}}$ $k =1,\ldots, K$}{
\vspace{0.3em}
Update the total activities $\theta_{k j}$ as
$$
\theta_{k j} \gets \frac{\mu_k}{a + \mu_k}\Big(a-1 + \bar{S}_{kj}\Big), \qquad \bar{S}_{kj} = \sum_{i} \sum_{n=1}^{N_{ij}} \frac{r_{i k}\phi_{k j}e^{\boldsymbol{\beta}_{k}^\top \mathbf{x}(t_{i j n})}}{\sum_{s=1}^K r_{i s}\phi_{s j}e^{\boldsymbol{\beta}_{s}^\top \mathbf{x}(t_{i j n})}}.
$$
}
\nl \textbf{Regression coefficients:} \For{$k =1,\ldots, K$}{
\vspace{0.3em}
Update the regression coefficients with Newton's method as $\bbeta_k \gets \bbeta_k - \mathbf{H}_k^{-1}\mathbf{g}_k$, where
\begin{align*}
\mathbf{H}_k ={}& \sum_j \bar S_{kj}\bigg(\frac{Q_j(\bbeta_k)}{q_j(\bbeta_k)} - \frac{\bm{m}_j(\bbeta_k)\bm{m}_j(\bbeta_k)^\top}{q_j(\bbeta_k)^2}\bigg) + \frac{1}{\sigma^2_k}I_L,\\
\mathbf{g}_k ={}& \sum_j \bar S_{kj}\frac{\bm{m}_j(\bbeta_k)}{q_j(\bbeta_k)} - \sum_{ij}\sum_{n = 1}^{N_{i j}} \frac{r_{i k} \phi_{kj} e^{\bkx(t_{i j n})}}{\sum_{s = 1}^K r_{i s}\phi_{s j} e^{\bbeta_s^\top \bx(t_{i j n})}} \bx(t_{i j n}) + \frac{1}{\sigma^2_k}\bbeta_k,
\end{align*}
where $\bm{m}_j(\bbeta_k) = \int_0^T \frac{1}{2}c_j(t)e^{\bbeta_k^\top\bx(t)}\bx(t)\dt$ and $Q_j(\bbeta_k) = \int_0^T \frac{1}{2}c_j(t) e^{\bbeta_k^\top\bx(t)}\bx(t)\bx(t)^\top\dt$.
}
\nl \textbf{Relevance weights and variance parameters:} \For{$k =1,\ldots, K$}{
Update each parameter as
$$
\mu_k \gets \frac{\varepsilon a J + a \sum_{j}\theta_{k j}}{2aJ + 2}, \qquad \sigma^2_k \gets \frac{d_0 + \bbeta_k^\top \bbeta_k/2}{c_0 + L/2 + 1}.
$$
}
\hspace{-1.25em}\textbf{Output}: Updated point estimates for $R, \Theta, B, \mu, \sigma^2$.
\end{algorithm}

Algorithm \ref{algo:MAP_rules} presents the steps for one iteration of our method for minimizing \cref{eq:logPosterior}. A complete derivation is in \cref{sec:proofs_derivs}. Steps 1 and 2 satisfy the positivity requirements, provided that $a > 1$ and $\alpha_{ik} > 1$.
Step 3 updates $\bbeta_k$ by majorizing \cref{eq:logPosterior} with respect to $\bbeta_k$ and then applying Newton's method using the gradient  $\mathbf{g}_k$ and the Hessian  $\mathbf{H}_k$ of the majorizer with respect to $\bbeta_k$.
Step 4 updates $\mu_k$ and $\sigma^2_k$ by minimizing their full conditional densities, which have a closed-form solution. We cycle through Steps 1--4 of Algorithm~\ref{algo:MAP_rules} until the relative difference between two successive evaluations of \cref{eq:logPosterior} (excluding the constant) is lower than a tolerance, set at $10^{-7}$ by default.

\subsection{Posterior inference with Markov chain Monte Carlo}\label{subsec:MCMC}
\begin{algorithm}[t]
\caption{MCMC sampler updates for Poisson process factorization}
\label{algo:Gibbs_rules}
\footnotesize
\nl \textbf{Latent attributions:} \For{$i = 1,\ldots, I$ $\textbf{\textup{and}}$ $j = 1, \ldots, J$ $\textbf{\textup{and}}$ $n = 1, \ldots, N_{ij}$}{
\vspace{0.3em}
Update the latent attribution for the mutation at position $t_{i j n}$ by drawing
$$W_{ij}(t_{i j n}) \sim \mathrm{Mult}\big(1;\, p_{ij1}(t_{i j n}), \ldots, p_{ijK}(t_{i j n})\big), \qquad p_{ijk}(t_{i j n}) =  \frac{r_{i k} \phi_{kj} e^{\bkx(t_{i j n})}}{\sum_{s = 1}^K r_{i s}\phi_{s j} e^{\bbeta_s^\top \bx(t_{i j n})}}, $$
where $\phi_{k j} = \theta_{k j}/q_j(\bbeta_k)$ and $q_j(\bbeta_k) = \int_0^T \frac{1}{2}\, c_j(t)\,e^{\bbeta_k^\top \bx(t)}\dt$.
}
\nl \textbf{Signatures:} \For{$k = 1, \ldots, K$ }{
\vspace{0.3em}
Define $M_{i k} = \sum_j\sum_{n= 1}^{N_{ij}} W_{ijk}(t_{i j n})$ and update $r_k$ by drawing
$$
r_{k} \sim  \mathrm{Dir}\big(\alpha_{1k} + M_{1 k}, \ldots, \alpha_{Ik} + M_{I k}\big).
$$
}
\nl \textbf{Total activities:} \For{$j = 1, \ldots, J$ $\textbf{\textup{and}}$ $k =1,\ldots, K$}{
\vspace{0.3em}
Define $S_{k j} = \sum_i\sum_{n= 1}^{N_{ij}} W_{ijk}(t_{i j n})$ and update $\theta_{k j}$ by drawing
$$
\theta_{k j} \sim  \mathrm{Ga}\big(a + S_{k j},\; 1 + a/\mu_k\big).
$$
}
\nl \textbf{Regression coefficients:} \For{$k =1,\ldots, K$}{
\vspace{0.3em}
Update $\bbeta_k$ using adaptive generalized elliptical slice sampling with target density
$$
\begin{aligned}
f(\bbeta_k) \propto \exp \Big(-\sum_j S_{kj} \log q_j(\bbeta_k) + \sum_{ij} \sum_{n = 1}^{N_{ij}} W_{ijk}(t_{i j n}) \bbeta_k^\top \bx(t_{i j n})\Big) \mathrm{N}(\bbeta_k; 0, \sigma_k^2 I_L).
\end{aligned}
$$
}
\nl \textbf{Relevance weights and variances:} \For{$k =1,\ldots, K$}{
Update $\mu_k$ and $\sigma_k^2$ by drawing
\begin{align*}
\mu_k &\sim \mathrm{InvGa}\big(2a J + 1,\; \varepsilon a J + a \textstyle\sum_{j} \theta_{k j}\big),\qquad
\sigma^2_k \sim \mathrm{InvGa}\big(c_0 + L/2,\; d_0 + \bbeta_k^\top\bbeta_k/2\big).
\end{align*}
}
\hspace{-1.25em}\small{\textbf{Output}: Updated sample from the joint posterior of $R, \Theta, B$, $\mu$, and $\sigma^2$.}
\end{algorithm}
The optimization approach in \cref{subsec:MAP} is a comparatively fast way of obtaining point estimates for PPF; see \cref{tab:sim_time_results,tab:mcmc_results} for runtimes relative to baseline NMF.
However, it is often important to quantify uncertainty in the parameters, especially when assessing the effects of the covariates.
In Algorithm \ref{algo:Gibbs_rules}, we provide a Markov chain Monte Carlo (MCMC) sampler for the PPF posterior.  Relying on \cref{pro:thinning}, we augment the data by introducing one multinomial random variable for each observed mutation: for the $n$th mutation of type $i$ in patient $j$, we let $W_{ij}(t_{i j n}) = \big(W_{ij1}(t_{i j n}), \ldots, W_{ijK}(t_{i j n})\big)  \sim\mathrm{Mult}\big(1;\, p_{ij1}(t_{i j n}), \ldots, p_{ijK}(t_{i j n})\big)$ given $R$, $\Phi$, $B$, and $\boldsymbol{t}$, where
\begin{equation}\label{eq:AssignentProbs}
p_{ijk}(t_{i j n}) = \mathds{P}(W_{ijk}(t_{i j n}) = 1\mid R, \Phi, B, \boldsymbol{t}) =  \frac{r_{i k} \phi_{kj} e^{\bkx(t_{i j n})}}{\sum_{s = 1}^K r_{i s}\phi_{s j} e^{\bbeta_s^\top \bx(t_{i j n})}}.
\end{equation}
Then, conditional on these multinomial random variables, denoted $W$ collectively, we obtain conjugate updates for the signatures, the activities, and the relevance weights (Steps 2, 3, and 5, respectively, in Algorithm \ref{algo:Gibbs_rules}).
The full conditional distribution for the regression coefficients does not take a form that can be sampled from directly; however, we can use elliptical slice sampling \citep{pmlr-v9-murray10a} thanks to the Gaussian prior over $\bbeta_k$. Although evaluating the target is expensive due to the large genome dimension, this method is practical because it requires no tuning. To improve mixing, we use the adaptive generalized elliptical slice sampler of \citet{marco2026adaptivegeneralizedellipticalslice}, with their default settings.

In \cref{eq:AssignentProbs}, $p_{ijk}(t_{i j n})$ represents the probability that the mutation of type $i$ in patient $j$ at position $t_{i j n}$ is due to signature $k$. Unlike in Bayesian Poisson NMF \citep{Zhou_Carin_2015}, these probabilities are not constant but vary along the genome depending on covariate values.
This can provide valuable insights, since mutations of type $i$ may be attributed to one signature at some locations, and another signature elsewhere.

\section{Simulations}\label{sec:simulations}

\cref{secSupp:simulations} in the supplementary material presents extensive simulation analyses to test our model against baseline NMF models \citep{zito2024compressive, kim2016somatic}. We demonstrate that PPF accurately recovers the true signatures, exposures, regression coefficients, and patient mutations along the genome (\cref{subsec:Simulation_setup}). We also benchmark PPF in several scenarios with increasing model misspecification, including patient-specific covariates that differ from the observed shared covariates (i.e., $\bx_j(t) \neq \bx(t)$), clusters of mutations, noisy copy numbers, and misspecified trinucleotide opportunities (\cref{subsec:Simulation_misspec}). In all cases, we show that PPF can still correctly retrieve the true number of signatures and attribute individual mutations to signatures with well-calibrated probabilities.

\section{Mutational signatures in breast adenocarcinoma}\label{sec:application}

\subsection{Overview}\label{subsec:overview}
We use PPF to address the scientific questions from \cref{sec:breastCancer} in the context of breast adenocarcinoma. In \cref{subsec:breast_denovo}, we infer signatures \emph{de novo}, and in \cref{subsec:breast_semi_sup} we use fixed COSMIC signatures. Additional results are in the Supplementary material: \cref{sec:sensitivity_analysis} presents sensitivity analyses, varying the choice of model settings $K$, $c_0$, and $d_0$ and evaluating robustness under covariate ablation; \cref{sec:breast80} assesses replicability of our findings using an independent breast cancer dataset from \citet{Davies_2017, Davies_MMRd}; and \cref{sec:tensorsignatures} compares PPF and \texttt{TensorSignatures} \citep{TensorSignatures_2021} using chromatin states.

In all applications, we use the same data preprocessing detailed in \cref{subsec:data_preprocessing}. This includes binning each file at 2kb and 10kb resolution, filtering blacklisted regions, and substituting copy numbers with $c_j(t) = 2$ (i.e., the normal condition) in regions where missing (on average, for $<1.5\%$ of the genome). All covariates in \cref{tab:Epi_covariates} are centered and scaled to have a mean of zero and a variance of one along the genome.

\subsection{De novo analysis}\label{subsec:breast_denovo}

We first perform a \emph{de novo} analysis, that is, estimating all of the signatures without using reference signatures from COSMIC to inform priors or parameter values.

\subsubsection{Mutation rate reconstruction and inferred parameters}\label{subsec:sig_clusters_denovo}
We use the PPF model with the default settings described in \cref{subsec:priors}, setting $K = 12$ as an upper bound.  We run Algorithm \ref{algo:MAP_rules} from three different starting points generated at random from the prior, and stop the updates when the relative difference in log-posterior falls below $10^{-7}$ (details in \cref{subsec:details_denovo}). We selected the solution with the highest log-posterior density as the initial value for Algorithm \ref{algo:Gibbs_rules}, and ran it for $10{,}000$ iterations, using the last $5{,}000$ to compute posterior means and 95\% credible intervals. We exclude signatures for which $\hat{\mu} < 5\varepsilon$, which leads to a final solution having $\hat{K} = 10$ signatures; see \cref{sec:sensitivity_analysis} for analysis of sensitivity to the choice of $K, c_0$, and $d_0$.

\begin{figure}[t]
    \centering
    \includegraphics[width=\linewidth]{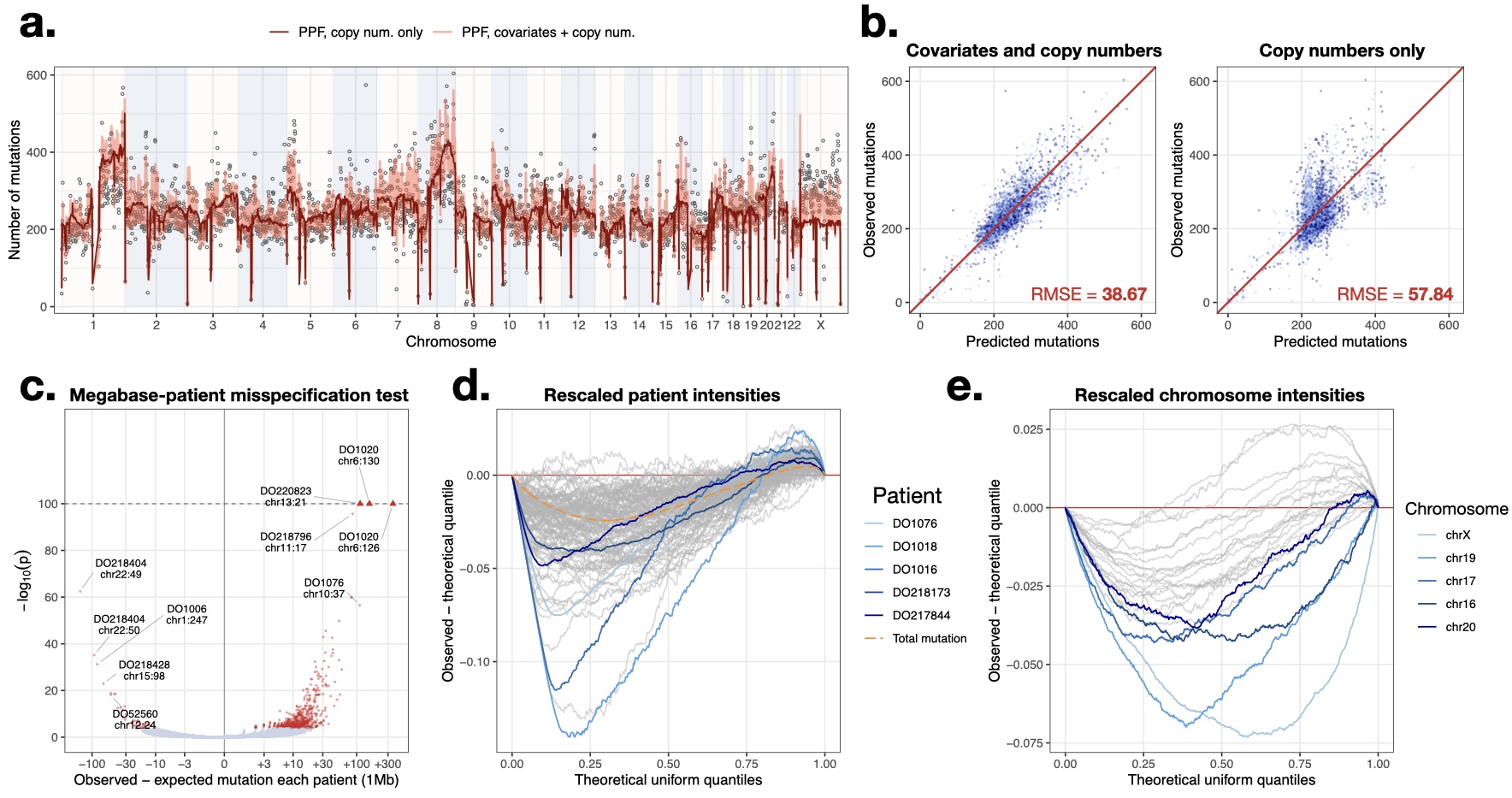}
    \caption{\textbf{a.} True and predicted total number of mutations from the PPF model at the megabase scale under PPF, both when using covariates and copy numbers (light red line) and only copy numbers (dark red line). Chromosomes are indicated by alternating white/blue vertical bands. \textbf{b.} Scatterplot of true and predicted values in both versions of PPF. \textbf{c.} Volcano plot for goodness-of-fit tests in each patient-megabase pair, where chr6:130 denotes the 130th megabase in chromosome 6. Red points indicate pairs where \texttt{poisson.test} rejects the null hypothesis, after multiple testing adjustment. $\log_{10}$ p-values are capped at 100 to ease visualization. \textbf{d.} Difference between observed and theoretical quantiles for the time-rescaled intensity function of each patient. Coloured curves are the five patients whose Kolmogorov--Smirnov distance is largest relative to the critical value. The dashed orange curve denotes the full cohort. \textbf{e.} The same plot by chromosome, highlighting the five chromosomes with the largest Kolmogorov--Smirnov distance.
}
    \label{fig:Results_unsupervised_track}
\end{figure}

We start by assessing how well PPF reconstructs the aggregate mutation rates along the genome. \cref{fig:Results_unsupervised_track}a shows that using covariates substantially improves the reconstruction error of the total number of mutations compared to using copy-number only; this is evident also from the scatterplots in \cref{fig:Results_unsupervised_track}b, which show that adding covariates reduces the error by 33\%. Furthermore, PPF reveals which regions are not captured by the covariates. For every patient $j$ and megabase $b$ pair, we test whether the mean of the observed number of mutations $N_{bj}$ is equal to the PPF intensity (denoted as $\lambda_{bj}$) using an exact two-sided Poisson test.
Since this requires running $323{,}971$ tests, we adjust for multiple testing using the Benjamini--Hochberg procedure to control the false discovery rate at $5\%$.
We detect significant excesses in $709$ pairs $(j,b)$; see \cref{fig:Results_unsupervised_track}c (right arm). These pairs feature heavily rearranged loci, such as chr6:126--132 in patient DO1020, and are mostly due to localized hypermutations not captured by any covariate. Moreover, we see $44$ significant depletions (\cref{fig:Results_unsupervised_track}c, left arm), and of these, $64\%$ have copy numbers $\geq 20$.
While these suggest possible model improvements, only only 0.23\% of patient–megabase cells show significant departures (753 out of $323{,}971$).

To further assess model misspecification, we employ the time-rescaling diagnostics of \citet{Brown_2002}: if a Poisson process model is correctly specified, then the difference $\Lambda(t_{n}) - \Lambda(t_{n-1})$ between the cumulative intensity function at every two consecutive points is such that $1 - \exp\big(\Lambda(t_{n-1}) - \Lambda(t_{n})\big)\sim \mathrm{Uniform}(0,1)$ independently. Hence, comparing the quantiles of the rescaled fitted mutation rate against the quantiles of the uniform distribution provides a visual check for misspecification. We run this procedure for each patient (\cref{fig:Results_unsupervised_track}d) and chromosome  (\cref{fig:Results_unsupervised_track}e). We see a systematic excess of short inter-mutation distances for most patients and chromosomes, especially chromosome X, which shows large unexplained deviations from the fitted values, and patient DO1076, which is heavily hypermutated. Additional model checking results are presented in \cref{subsec:checking}.

\begin{figure}[t]
    \centering
    \includegraphics[width=\linewidth]{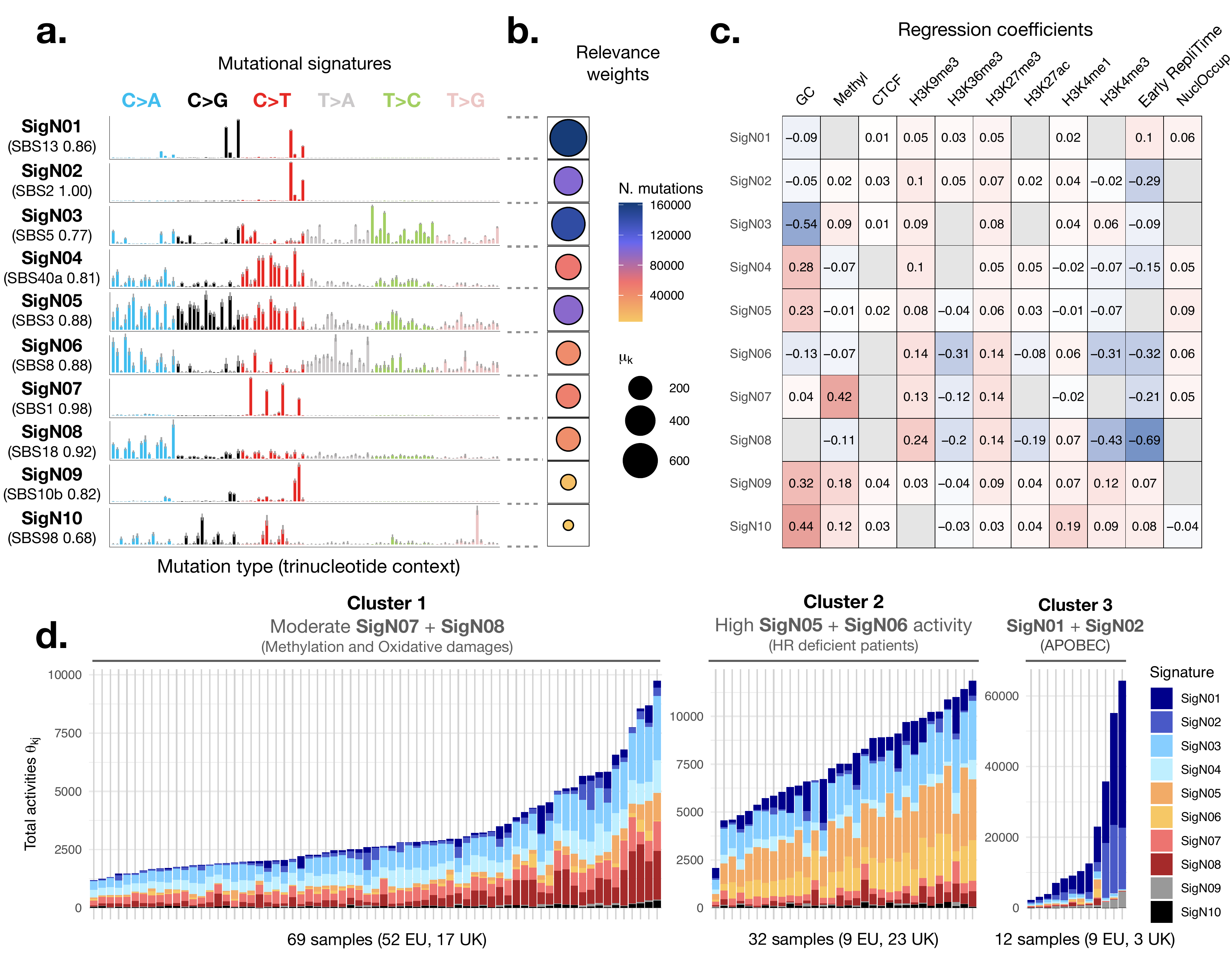}
    \caption{\textbf{a.} Posterior mean for the \emph{de novo} mutational signatures, with gray bars indicating 95\% credible intervals. Numbers in parentheses on the left indicate the highest cosine similarity in the COSMICv3.4 catalog. \textbf{b.} Posterior mean of the relevance weights $\hat{\mu}_k$ associated with each signature, with larger point size indicating larger values. Color intensity denotes the number of mutations in the data assigned to the signature via \cref{eq:AssignentProbs}. \textbf{c.} Posterior mean of the regression coefficients, $\hat{\bbeta}_k$. Gray cells with no number indicate estimates for which the 95\% credible interval contains zero. \textbf{d.} Total activities. Columns correspond to patients, split by the clusters inferred from the normalized activities.}
    \label{fig:Results_unsupervised}
\end{figure}

To address question Q1 from \cref{sec:breastCancer} (which signatures are present), \cref{fig:Results_unsupervised}a reports the posterior mean and credible intervals for the signatures. Values in parentheses on the left indicate the closest COSMICv3.4 signature in terms of cosine similarity. We see that our model recovers both APOBEC-related signatures: SigN02 is a perfect match to \sbs{2}, while SigN01 matches \sbs{13} in COSMICv3.4 with a score of 0.86, which however becomes 0.99 in COSMICv2, an earlier version of the database. Overall, the similarities are moderate-to-high, except for SigN10, whose similarity to \sbs{98} is too low for a confident match. SigN03 and SigN05 also have large contributions to the mutation rate, as demonstrated by the size of the relevance weights $\hat{\mu}_k$ in \cref{fig:Results_unsupervised}b. SigN03 matches with \sbs{5}, which is associated with aging, while SigN05 is close to \sbs{3} (homologous recombination deficiency, HRD). Moreover, we find moderate evidence of \sbs{8} (SigN06), \sbs{18} (SigN08), and \sbs{10b} (SigN09): these are important signatures in breast cancer, related to transcription-coupled nucleotide excision repair, reactive oxygen species (ROS),
and DNA replication proofreading, respectively. Finally, SigN04 resembles SBS40a, which is ubiquitous in cancer \citep{Alexandrov_2020}, and we recover \sbs{1} with high accuracy.

To answer question Q4 (patient subgroups), we perform a cluster analysis on the total activity of each signature in each patient, computed as $\theta_{kj} = \int_{0}^T \vartheta_{kj}(t)\dt$. \cref{fig:Results_unsupervised}d shows the cluster profiles obtained from the normalized activities, where the optimal number of clusters was obtained by maximizing the silhouette score. We find three distinct groups: a large cluster characterized by relatively high rates of methylation-driven mutations and ROS damage (Cluster 1: SigN07 and SigN08),  a medium-sized cluster with a high activity of \sbs{3} and SBS8  (Cluster 2: SigN05 and SigN06), and a small APOBEC-related cluster (Cluster 3: SigN01 and SigN02; hypermutated patients). Interestingly, these clusters separate triple-negative tumors by HRD status, because Cluster 2 identifies putative HRD tumors that primarily belong to BRCA-UK. Using copy number data, we also find that all 32 patients in Cluster 2 show loss of heterozygosity (LOH) at genes BRCA1 or BRCA2, compared to 33 in Cluster 1 (48\%) and 8 in Cluster 3 (67\%). For HRD tumors, PARP inhibitors have been shown to be effective in personalized therapies \citep{Gulhan_2019}.

\subsubsection{Covariate effects in the \emph{de novo} solution}
A central feature of the PPF model is its ability to associate location covariates with signatures.  This is illustrated by  \cref{fig:Results_unsupervised}c, which shows the posterior means of the regression coefficients $\hat{\bbeta}_k$. Since we standardized the covariates, all values are on the same scale.
Gray cells without numbers indicate coefficients that are not significantly different from zero, in the sense that their 95\% posterior credible intervals contain zero. We highlight the following key findings, which address question Q2 from \cref{sec:breastCancer}.

\textbf{1. Methylation enables clean recovery of SBS1 and lowers the estimated mutation rates for C$>$A-enriched signatures}. Signature SigN07 has a cosine similarity of 0.98 with \sbs{1}, an almost perfect match. In turn, its coefficient for methylation is large and positive (0.42), which is consistent with the proposed biological mechanism, namely C$>$T mutations arising from methylated CpG sites.
Meanwhile, CompNMF and SignatureAnalyzer recover contaminated versions of SBS1: CompNMF finds one signature that matches SBS1 with 0.82 cosine similarity, while SignatureAnalyzer finds two SBS1-like signatures with 0.83 and 0.80 similarity; see \cref{fig:SigComparison}. Contamination of \sbs{1} with other signatures, especially \sbs{5}, is a common issue in many tools since their activities tend to be correlated \citep{Wu2022}. However, using methylation as a covariate appreciably mitigates this issue in our model. Interestingly, SigN04, SigN05, SigN06, and SigN08, which all feature a high probability of C$>$A substitutions, are inversely associated with methylation. Since we are using 2kb binning with piecewise constant covariate values, this suggests that being in areas near methylated sites may lead to lower activities from other processes, particularly oxidative stress and HR deficiencies. More generally, GC content has a strong effect on several signatures, as reflected in the size of all coefficients.

\textbf{2. Late replication timing leads to large increases in mutational burden, but some signatures are more active in early replicating regions}.
Replication timing is defined such that higher values indicate earlier replication, which means that a negative $\beta_{k\ell}$ indicates enrichment in later replicating regions. Our results indicate strong associations with several signatures, with SigN08, SigN06, SigN02 being the most affected.
Our estimates indicate that a one standard deviation increase in late replication timing translates to a 99\% increase in the expected mutations due to oxidative damage (SigN08, $e^{0.69} -1 \approx 0.99$), a 23\% increase in \sbs{1} (SigN07, $e^{0.21} - 1 \approx 0.23$), and a 34\% increase in the faulty APOBEC signature \sbs{2} (SigN02, $e^{0.29} - 1 \approx 0.34$), holding all other covariates constant. In contrast, the other APOBEC-related signature, SigN01, shows decreased activity with later replication timing, with a 10\% decrease per standard deviation.

Existing approaches like \texttt{SigProfilerTopography} \citep{Otlu_2025} assess the impact of replication timing by tiling the total genomic signal into deciles and counting the number of mutations from a signature belonging to each bin. Then, these approaches test for a difference between this partition and the one obtained by placing mutations at random genomic locations. Mutations are typically assigned to a signature only if their attribution probability exceeds a certain threshold (e.g., 90\%), leading to a loss of information.
Our model-based approach jointly quantifies the covariate effects, adjusting for other covariates and copy numbers, without loss of information.
Consequently, our results may differ from previously reported results in some cases: for instance, COSMIC reports that \sbs{10b} is mildly enriched at later replicating areas in colorectal and endometrial cancer. In breast cancer, when controlling for several other covariates we find that our SigN09 (which matches to \sbs{10b}) is mildly but positively associated with early replication.

\textbf{3. All activities increase in heterochromatic domains and vary at active enhancers and promoters.}
H3K9me3 and H3K27me3, which are markers for heterochromatin and gene suppression, respectively, have nonnegative coefficients for every signature.
H3K36me3 exhibits a mild positive association with APOBEC signatures SigN01 and SigN02 (exhibiting 3\% and 5\% increases per standard deviation, respectively).
Meanwhile, we find that H3K36me3 has a strong inhibitory effect on SigN05, SigN06, SigN07 and SigN08, which is consistent with its known role in recruiting mismatch repair factors and thus limiting the accumulation of replication‑associated mismatches \citep{Li_2013_H3K39me3}.
The effects of H3K27ac, H3K4me1, and H3K4me3, which are markers for enhancers and promoters and favor gene activity, differ across signatures. In particular, SigN06 and SigN08 are the most affected by H3K4me3. Finally, several signatures are slightly positively associated with CTCF sites. The small coefficients might be due to the 2kb aggregation, since individual TF's are much smaller. For further results on covariate associations with signature activity, we refer to the analysis in \cref{sec:tensorsignatures} using chromatin states.

\subsection{Analysis using pre-specified signatures}\label{subsec:breast_semi_sup}

In this section, we take the signatures to be fixed and known based on a set of reference COSMIC signatures, and fit the rest of the parameters.
This is commonly referred to as \emph{signature refitting} analysis, and is often done using non-negative least squares. Since many COSMIC signatures have been linked to known biological processes, this can improve interpretability. By fixing the signatures, our PPF model can be used to implement a refitting approach that infers position-specific signature activities based on genomic covariates.

\subsubsection{Activities and covariate effects for COSMIC signatures}

\begin{figure}[t]
    \centering
    \includegraphics[width=0.95\linewidth]{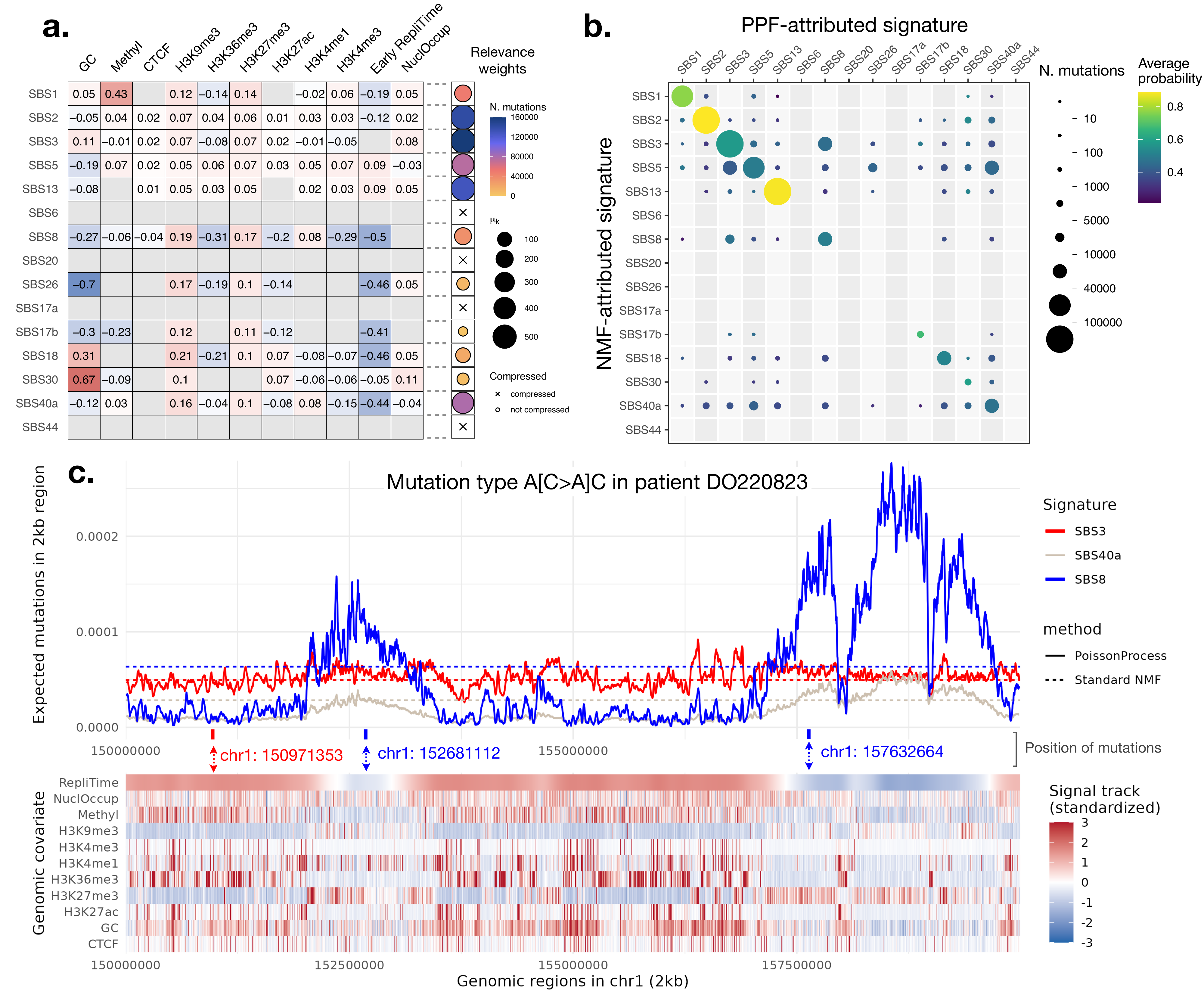}
    \caption{\textbf{a.} Regression coefficients and relevance weights for the signatures in the analysis. Gray boxes indicate entries where the 95\% posterior credible interval contains zero. The $\times$ marks indicate cases where $\hat{\mu}_k \approx \varepsilon$. \textbf{b.} Number of mutations attributed to a signature in the standard NMF case (y-axis) and in the PPF model (x-axis). \textbf{c.} Top panel: mutation intensity for \sbs{3}, \sbs{40a}, and \sbs{8} on substitution type A[C$>$A]C in patient DO220823 in genomic region chr1:150000000-160000000. The three ticks indicate the exact mutation position. Solid and dashed lines indicate the intensity from the PPF and the CompNMF model with fixed signatures, respectively. Bottom: standardized values for the covariates in the region. }
    \label{fig:supervised_results}
\end{figure}
Consider a PPF with \emph{$K = 15$} and with the same covariates as in \cref{tab:Epi_covariates}, but with $r_1, \ldots, r_K$ set to the following COSMICv3.4 signatures: two aging signatures (\sbs{1} and \sbs{5}), two APOBEC-related signatures (\sbs{2} and \sbs{13}), four associated with mismatch repair deficiencies (MMRd; \sbs{6}, \sbs{20}, \sbs{26}, and \sbs{44}), three related to oxidative damage (\sbs{18}, \sbs{17a}, and \sbs{17b}), two HRD-related (\sbs{3} and \sbs{8}), one related to base excision repair deficiency (\sbs{30}), and \sbs{40a}, which appears in almost all tumors \citep{Alexandrov_2020}.
We run Algorithm \ref{algo:Gibbs_rules}---but skip the updates to the signatures $r_k$ (Step 2)---for $10{,}000$ iterations starting from the MAP, discarding the first \emph{$5{,}000$} iterations as burn-in, using default hyperparameters as in \cref{subsec:priors}.

\cref{fig:supervised_results}a shows the posterior means of the regression coefficients $\hat{\bbeta}_k$ and the relevance weights $\hat{\mu}_k$. The ``$\times$'' mark shown for \sbs{6}, \sbs{20}, \sbs{17}a and \sbs{44} indicates that $\hat{\mu}_k \approx\varepsilon$. The associated regression coefficients are also shrunk near zero, making the contribution of these signatures negligible.
The values of $\hat{\bbeta}_k$ for \sbs{1} closely resemble the estimates in the \emph{de novo} results in \cref{fig:Results_unsupervised}c, where SigN07 corresponds to \sbs{1}. We find increased activity of \sbs{13} with early replication timing, which differs from previous analyses \citep{Morganella2016}. For all signatures, we still detect increased mutation rates in association with heterochromatic markers H3K9me3 and H3K27me3, while the effect of H3K36me3 varies but is mostly negative. As before, the largest contributors are \sbs{2}, \sbs{3}, \sbs{5}, and \sbs{13}, and GC content exhibits relatively large effects. Interestingly, \sbs{8} and \sbs{17}b have fairly strongly correlated coefficients. The signs are usually concordant. This suggests that \sbs{8} and \sbs{17}b tend to be active in similar regions. Our model also detects the presence of the MMRd signature \sbs{26}, and finds that it is especially influenced by replication timing and GC content. The baseline NMF model does not detect this signature, which suggests that accounting for covariates also improves signature detection. To replicate these findings, in \cref{sec:breast80} we run PPF on the 80 additional breast tumors from \citet{Davies_2017, Davies_MMRd}, using a coarser 10kb resolution for the same covariates. We find that most coefficients are consistent across datasets, except for \sbs{44}, which is not detected in the ICGC data but is found in the Breast80 data. Indeed, this second dataset features a high proportion of MMRd tumors \citep{Davies_MMRd}.

\subsubsection{PPF changes attribution of mutations to signatures}

An important aspect of the PPF model is that we can calculate mutation-specific signature attribution probabilities, as described in \cref{subsec:MCMC}, and infer which signatures are most likely to have generated any given mutation. By comparison, in the baseline NMF model (\cref{eq:PoissonBaseline}), the attribution probabilities in  \cref{eq:AssignentProbs} are the same at every location $t_{i j n}$ (for a given $i, j, k$). To quantify these differences and address question Q3 of \cref{sec:breastCancer}, we fit the baseline NMF model using CompNMF  (\cref{algo:compressive_nmf_map}) with the same set of fixed signatures, and compute the attribution probabilities for each mutation.

\cref{fig:supervised_results}b shows a cross-tabulation of the mutation attribution to each signature, by model, and the average assignment probabilities for these mutations. The majority of attributions are consistent between the two models, as demonstrated by the large values on the main diagonal. However, $130{,}044$ mutations (18.4\% of the total) are estimated to have a different attribution, though the classification probabilities are lower on average (0.43 against 0.72).  In \cref{subsec:sens_covariate_ablation}, we find that replication timing is the single most important covariate responsible for the re-assignment of mutations (\cref{fig:Stability_betas}b).

Specific examples include differences in attributions to \sbs{3}, \sbs{8}, and, to a lesser extent, \sbs{5}. The first two signatures show a similar pattern of mutations at C$>$A and T$>$A substitutions, which naturally leads to a higher \emph{a priori} uncertainty in the attribution probability. \sbs{8} has been posited to originate from homologous recombination deficiencies \citep{Singh2020}, but, unlike \sbs{3}, experimental validation for this conjecture has not been achieved yet. In \cref{fig:supervised_results}c, which displays the posterior for the intensity function in region \texttt{chr1:150000000-160000000} for patient DO220823 and mutation type $i = \mathrm{A[C>A]C}$, we see that such unclear biological identification might be related to a higher propensity of \sbs{8} to appear in heterochromatic regions characterized by later replication timing.  The three ticks between the facets indicate the genomic position where mutations were detected in the region. In the plot, we smooth the intensity using a rolling mean with a 10-region window to aid visualization. PPF attributes the first mutation to \sbs{3}, since it falls in a region characterized by early replication timing, with strong peaks of H3K36me3 and low values of H3K9me3 and H3K27me3. Meanwhile, the other two mutations are detected in regions of predominantly late replication timing, leading PPF to attribute them to \sbs{8} due to the higher predicted activity of that signature.  Interestingly, this activity drops dramatically when H3K36me3 is high, which explains the two dips in the predicted \sbs{8} activity following the third mutation. Hence, our analysis suggests that \sbs{8} is more likely to appear in late-replicating heterochromatic regions \citep{Singh2020}, demonstrating the advantages of the PPF framework in uncovering empirical relationships that are candidates for biological validation through experiments.

\vspace{-1em}
\section{Discussion}\label{sec:discussion}

Mutational signature analysis is now routinely performed in studies of tumors with whole-genome or whole-exome sequencing data, and remains an area of active methodological development. The method presented in this article further advances the state-of-the-art by introducing a model that enables joint inference on mutational signatures and their dependence on genomic features such as histone modifications, transcription factors, GC content, and DNA methylation, as well as patient-specific copy number alterations. Our analyses illustrate how adding covariates to the factorization improves interpretability, while also improving the detection of signatures. Furthermore, the model naturally provides position-specific activities for each signature, which allows one to infer which signature generated each observed mutation, based on genomic context and covariates. This helps in generating mechanistic hypotheses in a novel way compared to earlier methods.

In our model, the signatures do not vary with genomic position and only differentiate the 96 single-base substitution types.
However, some  COSMIC signatures display asymmetries with respect to replication and transcription strands \citep{Otlu_2023}. For example, signature SBS8 is associated with the transcription-coupled nucleotide excision repair (NER) pathway in addition to HRD, thus, its activity may vary markedly between transcribed and untranscribed strand status; see \citet{Singh2020}. Simultaneously accounting for replication timing and transcription strand is possible using PPF. For instance, we can expand the mutation types from 96 to 192, as in \citet{Otlu_2025}, or further model each strand separately by allowing signatures to depend on genomic coordinates, paired with hierarchical Dirichlet priors,
or topic-modeling priors as in \citet{Kahane_2023}.

Another limitation of our present implementation is that we use the same genomic covariate values for all patients, taken from a reference epigenome. Moreover, most of these covariates come from healthy breast tissue, and the nucleosome-occupancy track comes from a K562 leukemia cell line; these choices are motivated by data availability and comparability with previous analyses \citep{Morganella2016, Otlu_2023}.
However, epigenomes do vary across individuals, tissues, and cell-specific chromatin domains \citep{Gopi_2021}.
Some mutational processes may lead to hypermutations in some genomes on top of their baseline rate; for example, the formation of single-stranded DNA, necessary for APOBEC activity, is often connected to hypermutation and \emph{kataegis} \citep{Sakofsky_2014, Seplyarskiy_2016}. While our simulations show that the model remains robust to hypermutations when recovering signatures and regression coefficients, it may improve performance to adapt PPF to incorporate patient-specific genomic covariates such as chromatin accessibility \citep{Corces_2018} and methylation, patient-specific signatures, and strand-specific signature effects.
Estimating the model parameters would follow similar steps to those detailed in \cref{sec:inference}, albeit with increased computational costs.

Another direction for future work would be to use a nonparametric approach to flexibly model spatial dependencies in signature activity.  In addition to dependency between nearby genomic positions, recent analyses suggest that the three-dimensional structure of DNA may affect the mutational landscape of cancer genomes \citep{Akdemir_2020}.
Spatial dependency in signature activity could be modeled using a Gaussian process with a biologically informed covariance function based on estimated three-dimensional maps of intrachromosomal distance between genomic regions \citep{Lieberman-Aiden_2009}.

All extensions mentioned above have the potential to offer further insight into the mutational processes in cancer, with the ultimate goal of improving precision therapies.

%%%%%%%%%%%%%%%%%%%%%%%%%%%%%%%%%%%%%%%%%%%%%%%
% Acknowledgements
%%%%%%%%%%%%%%%%%%%%%%%%%%%%%%%%%%%%%%%%%%%%%%%
\section*{Acknowledgments}
We are deeply grateful to Mehmet Samur, Jenna Landy, and Catherine Xue for many helpful discussions. J.W.M. and A.Z. were supported in part by the National Cancer Institute of the NIH under award number R01CA240299.
G.P. and A.Z. were supported by grant number 5R01CA262710-05.
The content is solely the responsibility of the authors and does not necessarily represent the official views of the National Institutes of Health. A.Z. also acknowledges support from the Paula and Rodger Riney Foundation.
The authors acknowledge using Opus 5 in Claude Code for coding support, including figures, improvements to the \texttt{C++} implementation of the method, and organization of the R package. All AI-generated code was carefully reviewed and validated by the authors.

%%%%%%%%%%%%%%%%%%%%%%%%%%%%%%%%%%%%%%%%%%%%%%%
% Code availability
%%%%%%%%%%%%%%%%%%%%%%%%%%%%%%%%%%%%%%%%%%%%%%%
\section*{Code availability}
Code to reproduce the analyses is accessible at \url{https://github.com/alessandrozito/SignaturePPF-paper}. The R package to run the model is accessible at \url{https://github.com/alessandrozito/SignaturePPF}, with a tutorial at \url{https://alessandrozito.github.io/SignaturePPF/vignette.html}.

\bibliographystyle{chicago}
\bibliography{references}

\begin{thebibliography}{}

\bibitem[\protect\citeauthoryear{Abascal, Acosta, Addleman, et~al.}{Abascal et~al.}{2020}]{ENCODE_2020}
Abascal, F., R.~Acosta, N.~J. Addleman, et~al. (2020).
\newblock Expanded encyclopaedias of {DNA} elements in the human and mouse genomes.
\newblock {\em Nature\/}~{\em 583\/}(7818), 699--710.

\bibitem[\protect\citeauthoryear{Alexandrov, Jones, Wedge, Sale, Campbell, Nik-Zainal, and Stratton}{Alexandrov et~al.}{2015}]{Alexandrov_clock_2015}
Alexandrov, L.~B., P.~H. Jones, D.~C. Wedge, J.~E. Sale, P.~J. Campbell, S.~Nik-Zainal, and M.~R. Stratton (2015).
\newblock Clock-like mutational processes in human somatic cells.
\newblock {\em Nature Genetics\/}~{\em 47\/}(12), 1402--1407.

\bibitem[\protect\citeauthoryear{Amemiya, Kundaje, and Boyle}{Amemiya et~al.}{2019}]{blacklist}
Amemiya, H.~M., A.~Kundaje, and A.~P. Boyle (2019).
\newblock The {ENCODE} blacklist: Identification of problematic regions of the genome.
\newblock {\em Scientific Reports\/}~{\em 9\/}(1), 9354.

\bibitem[\protect\citeauthoryear{Bird}{Bird}{1980}]{Bird_19780_Methylation}
Bird, A.~P. (1980, 04).
\newblock {DNA methylation and the frequency of CpG in animal DNA}.
\newblock {\em Nucleic Acids Research\/}~{\em 8\/}(7), 1499--1504.

\bibitem[\protect\citeauthoryear{Blokzijl, Janssen, van Boxtel, and Cuppen}{Blokzijl et~al.}{2018}]{Blokzijl_2018}
Blokzijl, F., R.~Janssen, R.~van Boxtel, and E.~Cuppen (2018).
\newblock Mutational{P}atterns: comprehensive genome-wide analysis of mutational processes.
\newblock {\em Genome Medicine\/}~{\em 10\/}(1), 33.

\bibitem[\protect\citeauthoryear{Davies, Glodzik, Morganella, et~al.}{Davies et~al.}{2017}]{Davies_2017}
Davies, H., D.~Glodzik, S.~Morganella, et~al. (2017).
\newblock {HRDetect is a predictor of BRCA1 and BRCA2 deficiency based on mutational signatures}.
\newblock {\em Nature Medicine\/}~{\em 23\/}(4), 517--525.

\bibitem[\protect\citeauthoryear{Davies, Morganella, Purdie, et~al.}{Davies et~al.}{2017}]{Davies_MMRd}
Davies, H., S.~Morganella, C.~A. Purdie, et~al. (2017).
\newblock Whole-genome sequencing reveals breast cancers with mismatch repair deficiency.
\newblock {\em Cancer Research\/}~{\em 77\/}(18), 4755--4762.

\bibitem[\protect\citeauthoryear{Dunson and Herring}{Dunson and Herring}{2005}]{Dunson_Herring_2005}
Dunson, D.~B. and A.~H. Herring (2005).
\newblock {Bayesian latent variable models for mixed discrete outcomes}.
\newblock {\em Biostatistics\/}~{\em 6\/}(1), 11--25.

\bibitem[\protect\citeauthoryear{Ernst and Kellis}{Ernst and Kellis}{2017}]{Ernst2017_ChromHMM}
Ernst, J. and M.~Kellis (2017).
\newblock Chromatin-state discovery and genome annotation with {ChromHMM}.
\newblock {\em Nature Protocols\/}~{\em 12\/}(12), 2478--2492.

\bibitem[\protect\citeauthoryear{Fischer, Illingworth, Campbell, and Mustonen}{Fischer et~al.}{2013}]{Fischer_2013}
Fischer, A., C.~J. Illingworth, P.~J. Campbell, and V.~Mustonen (2013).
\newblock {EMu}: probabilistic inference of mutational processes and their localization in the cancer genome.
\newblock {\em Genome Biology\/}~{\em 14\/}(4), R39.

\bibitem[\protect\citeauthoryear{Hansen, Thomas, Sandstrom, Canfield, Thurman, Weaver, Dorschner, Gartler, and Stamatoyannopoulos}{Hansen et~al.}{2010}]{Hansen_repliseq}
Hansen, R.~S., S.~Thomas, R.~Sandstrom, T.~K. Canfield, R.~E. Thurman, M.~Weaver, M.~O. Dorschner, S.~M. Gartler, and J.~A. Stamatoyannopoulos (2010).
\newblock Sequencing newly replicated {DNA} reveals widespread plasticity in human replication timing.
\newblock {\em Proceedings of the National Academy of Sciences\/}~{\em 107\/}(1), 139--144.

\bibitem[\protect\citeauthoryear{Islam, D{\'\i}az-Gay, Wu, et~al.}{Islam et~al.}{2022}]{Islam_2022}
Islam, S.~A., M.~D{\'\i}az-Gay, Y.~Wu, et~al. (2022).
\newblock Uncovering novel mutational signatures by de novo extraction with {SigProfilerExtractor}.
\newblock {\em Cell Genomics\/}~{\em 2\/}(11), 100179.

\bibitem[\protect\citeauthoryear{Jaenisch and Bird}{Jaenisch and Bird}{2003}]{Jaenisch2003}
Jaenisch, R. and A.~Bird (2003).
\newblock Epigenetic regulation of gene expression: how the genome integrates intrinsic and environmental signals.
\newblock {\em Nature Genetics\/}~{\em 33\/}(3), 245--254.

\bibitem[\protect\citeauthoryear{Joe}{Joe}{2006}]{JOE20062177}
Joe, H. (2006).
\newblock Generating random correlation matrices based on partial correlations.
\newblock {\em Journal of Multivariate Analysis\/}~{\em 97\/}(10), 2177--2189.

\bibitem[\protect\citeauthoryear{Kahane, Leiserson, and Sharan}{Kahane et~al.}{2023}]{Kahane_2023}
Kahane, I., M.~D.~M. Leiserson, and R.~Sharan (2023).
\newblock A mutation-level covariate model for mutational signatures.
\newblock {\em PLOS Computational Biology\/}~{\em 19\/}(6), 1--10.

\bibitem[\protect\citeauthoryear{Katainen, Dave, Pitk{\"a}nen, et~al.}{Katainen et~al.}{2015}]{Katainen2015}
Katainen, R., K.~Dave, E.~Pitk{\"a}nen, et~al. (2015).
\newblock {CTCF}/cohesin-binding sites are frequently mutated in cancer.
\newblock {\em Nature Genetics\/}~{\em 47\/}(7), 818--821.

\bibitem[\protect\citeauthoryear{Kim, Mouw, Polak, et~al.}{Kim et~al.}{2016}]{kim2016somatic}
Kim, J., K.~W. Mouw, P.~Polak, et~al. (2016).
\newblock {Somatic ERCC2 mutations are associated with a distinct genomic signature in urothelial tumors}.
\newblock {\em Nature Genetics\/}~{\em 48\/}(6), 600--606.

\bibitem[\protect\citeauthoryear{Kingman}{Kingman}{1993}]{kingman-poisson-processes}
Kingman, J. F.~C. (1993).
\newblock {\em Poisson Processes}, Volume~3 of {\em Oxford Studies in Probability}.

\bibitem[\protect\citeauthoryear{Last and Penrose}{Last and Penrose}{2017}]{Last_Penrose_2017}
Last, G. and M.~Penrose (2017).
\newblock {\em Lectures on the Poisson Process}.
\newblock Institute of Mathematical Statistics Textbooks. Cambridge University Press.

\bibitem[\protect\citeauthoryear{Lawrence, Gentleman, and Carey}{Lawrence et~al.}{2009}]{rtracklayer}
Lawrence, M., R.~Gentleman, and V.~Carey (2009, 05).
\newblock rtracklayer: an {R} package for interfacing with genome browsers.
\newblock {\em Bioinformatics\/}~{\em 25\/}(14), 1841--1842.

\bibitem[\protect\citeauthoryear{Lawrence, Huber, Pagès, Aboyoun, Carlson, Gentleman, Morgan, and Carey}{Lawrence et~al.}{2013}]{GenomicRanges}
Lawrence, M., W.~Huber, H.~Pagès, P.~Aboyoun, M.~Carlson, R.~Gentleman, M.~T. Morgan, and V.~J. Carey (2013, 08).
\newblock Software for computing and annotating genomic ranges.
\newblock {\em PLOS Computational Biology\/}~{\em 9\/}(8), 1--10.

\bibitem[\protect\citeauthoryear{Lee, Lee, Lee, Park, Kwon, Park, Chun, Ju, and Hong}{Lee et~al.}{2018}]{Lee_2018_mutalisk}
Lee, J., A.~Lee, J.-K. Lee, J.~Park, Y.~Kwon, S.~Park, H.~Chun, Y.~S. Ju, and D.~Hong (2018, 05).
\newblock Mutalisk: a web-based somatic mutation analyis toolkit for genomic, transcriptional and epigenomic signatures.
\newblock {\em Nucleic Acids Research\/}~{\em 46\/}(W1), W102--W108.

\bibitem[\protect\citeauthoryear{Lee, Wang, Cho, Xi, and Kim}{Lee et~al.}{2022}]{Lee_regional_2022}
Lee, S.-Y., H.~Wang, H.~J. Cho, R.~Xi, and T.-M. Kim (2022).
\newblock The shaping of cancer genomes with the regional impact of mutation processes.
\newblock {\em Experimental \& Molecular Medicine\/}~{\em 54\/}(7), 1049--1060.

\bibitem[\protect\citeauthoryear{Li, Mao, Tong, et~al.}{Li et~al.}{2013}]{Li_2013_H3K39me3}
Li, F., G.~Mao, D.~Tong, et~al. (2013).
\newblock The histone mark {H3K36me3} regulates human {DNA} mismatch repair through its interaction with {MutSalpha}.
\newblock {\em Cell\/}~{\em 153\/}(3), 590--600.

\bibitem[\protect\citeauthoryear{Loyfer, Magenheim, Peretz, et~al.}{Loyfer et~al.}{2023}]{Loyfer2023}
Loyfer, N., J.~Magenheim, A.~Peretz, et~al. (2023).
\newblock A {DNA} methylation atlas of normal human cell types.
\newblock {\em Nature\/}~{\em 613\/}(7943), 355--364.

\bibitem[\protect\citeauthoryear{Makova and Hardison}{Makova and Hardison}{2015}]{Makova2015}
Makova, K.~D. and R.~C. Hardison (2015).
\newblock The effects of chromatin organization on variation in mutation rates in the genome.
\newblock {\em Nature Reviews Genetics\/}~{\em 16\/}(4), 213--223.

\bibitem[\protect\citeauthoryear{Marco and Tokdar}{Marco and Tokdar}{2026}]{marco2026adaptivegeneralizedellipticalslice}
Marco, N. and S.~T. Tokdar (2026).
\newblock Adaptive generalized elliptical slice sampling.
\newblock arXiv:2605.21659.

\bibitem[\protect\citeauthoryear{Miller and Carter}{Miller and Carter}{2020}]{miller2020inference}
Miller, J.~W. and S.~L. Carter (2020).
\newblock Inference in generalized bilinear models.
\newblock arXiv:2010.04896.

\bibitem[\protect\citeauthoryear{Morganella, Alexandrov, Glodzik, et~al.}{Morganella et~al.}{2016}]{Morganella2016}
Morganella, S., L.~B. Alexandrov, D.~Glodzik, et~al. (2016).
\newblock The topography of mutational processes in breast cancer genomes.
\newblock {\em Nature Communications\/}~{\em 7\/}(1), 11383.

\bibitem[\protect\citeauthoryear{Murray, Adams, and MacKay}{Murray et~al.}{2010}]{pmlr-v9-murray10a}
Murray, I., R.~Adams, and D.~MacKay (2010).
\newblock Elliptical slice sampling.
\newblock In {\em Proceedings of the 13th International Conference on Artificial Intelligence and Statistics}, Volume~9, pp.\  541--548.

\bibitem[\protect\citeauthoryear{Nik-Zainal, Alexandrov, Wedge, et~al.}{Nik-Zainal et~al.}{2012}]{Nik_zainal_2012}
Nik-Zainal, S., L.~B. Alexandrov, D.~C. Wedge, et~al. (2012).
\newblock Mutational processes molding the genomes of 21 breast cancers.
\newblock {\em Cell\/}~{\em 149\/}(5), 979--993.

\bibitem[\protect\citeauthoryear{Nishiyama and Nakanishi}{Nishiyama and Nakanishi}{2021}]{Nishiyama_2021}
Nishiyama, A. and M.~Nakanishi (2021).
\newblock Navigating the {DNA} methylation landscape of cancer.
\newblock {\em Trends in Genetics\/}~{\em 37\/}(11), 1012--1027.

\bibitem[\protect\citeauthoryear{Otlu and Alexandrov}{Otlu and Alexandrov}{2025}]{Otlu_2025}
Otlu, B. and L.~B. Alexandrov (2025).
\newblock Evaluating topography of mutational signatures with {SigProfilerTopography}.
\newblock {\em Genome Biology\/}~{\em 26\/}(1), 134.

\bibitem[\protect\citeauthoryear{Otlu, Díaz-Gay, Vermes, Bergstrom, Zhivagui, Barnes, and Alexandrov}{Otlu et~al.}{2023}]{Otlu_2023}
Otlu, B., M.~Díaz-Gay, I.~Vermes, E.~N. Bergstrom, M.~Zhivagui, M.~Barnes, and L.~B. Alexandrov (2023).
\newblock Topography of mutational signatures in human cancer.
\newblock {\em Cell Reports\/}~{\em 42\/}(8), 112930.

\bibitem[\protect\citeauthoryear{Pagès, Aboyoun, Gentleman, and DebRoy}{Pagès et~al.}{2023}]{biostrings}
Pagès, H., P.~Aboyoun, R.~Gentleman, and S.~DebRoy (2023).
\newblock {\em Biostrings: Efficient manipulation of biological strings}.
\newblock R package version 2.68.1.

\bibitem[\protect\citeauthoryear{Polak, Karli{\'c}, Koren, et~al.}{Polak et~al.}{2015}]{Polak2015}
Polak, P., R.~Karli{\'c}, A.~Koren, et~al. (2015).
\newblock Cell-of-origin chromatin organization shapes the mutational landscape of cancer.
\newblock {\em Nature\/}~{\em 518\/}(7539), 360--364.

\bibitem[\protect\citeauthoryear{Robinson, Sharan, and Leiserson}{Robinson et~al.}{2019}]{Robinson_2019}
Robinson, W., R.~Sharan, and M.~D.~M. Leiserson (2019, 07).
\newblock Modeling clinical and molecular covariates of mutational process activity in cancer.
\newblock {\em Bioinformatics\/}~{\em 35\/}(14), i492--i500.

\bibitem[\protect\citeauthoryear{Sason, Wojtowicz, Robinson, Leiserson, Przytycka, and Sharan}{Sason et~al.}{2020}]{StickySig}
Sason, I., D.~Wojtowicz, W.~Robinson, M.~D.~M. Leiserson, T.~M. Przytycka, and R.~Sharan (2020).
\newblock A sticky multinomial mixture model of strand-coordinated mutational processes in cancer.
\newblock {\em iScience\/}~{\em 23\/}(3).

\bibitem[\protect\citeauthoryear{Schmalohr}{Schmalohr}{2025}]{kups75586}
Schmalohr, C.~L. (2025).
\newblock {\em Modeling the tissue-specific somatic mutation rate along the genome based on genomic features}.
\newblock Ph.\ D. thesis, Universit{\"a}t zu K{\"o}ln.

\bibitem[\protect\citeauthoryear{Schuster-B{\"o}ckler and Lehner}{Schuster-B{\"o}ckler and Lehner}{2012}]{schuster2012chromatin}
Schuster-B{\"o}ckler, B. and B.~Lehner (2012).
\newblock Chromatin organization is a major influence on regional mutation rates in human cancer cells.
\newblock {\em Nature\/}~{\em 488\/}(7412), 504--507.

\bibitem[\protect\citeauthoryear{Stamatoyannopoulos, Adzhubei, Thurman, Kryukov, Mirkin, and Sunyaev}{Stamatoyannopoulos et~al.}{2009}]{Stamatoyannopoulos2009}
Stamatoyannopoulos, J.~A., I.~Adzhubei, R.~E. Thurman, G.~V. Kryukov, S.~M. Mirkin, and S.~R. Sunyaev (2009).
\newblock Human mutation rate associated with {DNA} replication timing.
\newblock {\em Nature Genetics\/}~{\em 41\/}(4), 393--395.

\bibitem[\protect\citeauthoryear{Supek and Lehner}{Supek and Lehner}{2019}]{Supek_lehner_2019}
Supek, F. and B.~Lehner (2019).
\newblock Scales and mechanisms of somatic mutation rate variation across the human genome.
\newblock {\em DNA Repair\/}~{\em 81}, 102647.

\bibitem[\protect\citeauthoryear{Thurman, Day, Noble, and Stamatoyannopoulos}{Thurman et~al.}{2007}]{Thurman01062007}
Thurman, R.~E., N.~Day, W.~S. Noble, and J.~A. Stamatoyannopoulos (2007).
\newblock Identification of higher-order functional domains in the human {ENCODE} regions.
\newblock {\em Genome Research\/}~{\em 17\/}(6), 917--927.

\bibitem[\protect\citeauthoryear{Timmons, Morris, and Harrigan}{Timmons et~al.}{2022}]{Timmons_2022}
Timmons, C., Q.~Morris, and C.~F. Harrigan (2022, 12).
\newblock Regional mutational signature activities in cancer genomes.
\newblock {\em PLOS Computational Biology\/}~{\em 18\/}(12), 1--23.

\bibitem[\protect\citeauthoryear{Tolstorukov, Volfovsky, Stephens, and Park}{Tolstorukov et~al.}{2011}]{Tolstorukov2011}
Tolstorukov, M.~Y., N.~Volfovsky, R.~M. Stephens, and P.~J. Park (2011).
\newblock Impact of chromatin structure on sequence variability in the human genome.
\newblock {\em Nature Structural \& Molecular Biology\/}~{\em 18\/}(4), 510--515.

\bibitem[\protect\citeauthoryear{V{\"o}hringer, Hoeck, Cuppen, and Gerstung}{V{\"o}hringer et~al.}{2021}]{TensorSignatures_2021}
V{\"o}hringer, H., A.~V. Hoeck, E.~Cuppen, and M.~Gerstung (2021).
\newblock Learning mutational signatures and their multidimensional genomic properties with {TensorSignatures}.
\newblock {\em Nature Communications\/}~{\em 12\/}(1), 3628.

\bibitem[\protect\citeauthoryear{Wang, Tao, Wu, and Liu}{Wang et~al.}{2020}]{sigminer_2020}
Wang, S., Z.~Tao, T.~Wu, and X.-S. Liu (2020).
\newblock {Sigflow: an automated and comprehensive pipeline for cancer genome mutational signature analysis}.
\newblock {\em Bioinformatics\/}~{\em 37\/}(11), 1590--1592.

\bibitem[\protect\citeauthoryear{Wu, Chua, Ng, Boot, and Rozen}{Wu et~al.}{2022}]{Wu2022}
Wu, Y., E.~H.~Z. Chua, A.~W.~T. Ng, A.~Boot, and S.~G. Rozen (2022).
\newblock Accuracy of mutational signature software on correlated signatures.
\newblock {\em Scientific Reports\/}~{\em 12\/}(1), 390.

\bibitem[\protect\citeauthoryear{Zito and Miller}{Zito and Miller}{2024}]{zito2024compressive}
Zito, A. and J.~W. Miller (2024).
\newblock Compressive {B}ayesian non-negative matrix factorization for mutational signatures analysis.
\newblock arXiv:2404.10974.

\end{thebibliography}

%%%%%%%%%%%%%%%%%%%%%%%%%%%%%%%%%%%%%%%%%%%%%%%%%%%%%%%%%%
% Supplementary material
%%%%%%%%%%%%%%%%%%%%%%%%%%%%%%%%%%%%%%%%%%%%%%%%%%%%%%%%%%

\newpage
\setcounter{page}{1}
\setcounter{section}{0}
\setcounter{table}{0}
\setcounter{figure}{0}
\renewcommand{\thealgocf}{S\arabic{algocf}}
\setcounter{algocf}{0}
\setcounter{equation}{0}
\renewcommand{\theHsection}{SIsection.\arabic{section}}
\renewcommand{\theHtable}{SItable.\arabic{table}}
\renewcommand{\theHfigure}{SIfigure.\arabic{figure}}
\renewcommand{\thepage}{S\arabic{page}}
\renewcommand{\thesection}{S\arabic{section}}
\renewcommand{\theequation}{S\arabic{equation}}
\renewcommand{\thetable}{S\arabic{table}}
\renewcommand{\thefigure}{S\arabic{figure}}

\begin{center}
{\LARGE SUPPLEMENTARY MATERIAL \\``Poisson process factorization for modeling mutational processes along cancer genomes''}
\end{center}

\cref{subsec:genomic_covariates} provides details on the genomic covariates. \cref{subsec:previous_work}  discusses recent work on the relationships between mutational signatures and epigenetic covariates. \cref{sec:proofs_derivs} provides proofs of the theoretical statements and contains a detailed derivation of the estimation and inference algorithms. \cref{secSupp:simulations} presents two simulation studies. \cref{sec:add_breast} provides additional details and results for the breast adenocarcinoma application. \cref{sec:sources_prep} lists data sources and describes the preprocessing steps.

\section{Additional details on genomic covariates}\label{subsec:genomic_covariates}

Cellular activity is regulated by the interaction between the genome and the epigenome, largely through the action of molecular modifications that control DNA functions such as chromatin conformation, regulation of transcriptional activities of genes, and repair mechanisms \citepSupp{Jaenisch2003}. To measure these activities at various genomic locations, an abundance of data has been collected from various assays and experiments over the last decade, with the Encyclopedia of DNA Elements \citepSupp[\textsc{encode};][]{ENCODE_2020}
being the largest publicly accessible collection in the field. We now describe the covariates used in this analysis and their relationship to mutation rates.

\textbf{Nucleosome occupancy}.
The core building block of the spatial conformation of the genome is the nucleosome, a segment of 147 DNA base pairs wrapped around four pairs of histone proteins called H2A, H2B, H3, and H4. Nucleosomes are separated by 60-80 base pairs, forming ``linker DNA''. Their relative position determines the type of chromatin in a given region and, subsequently, its accessibility. When nucleosomes are closely packed together---forming the heterochromatin--- transcription is usually suppressed and DNA damage is harder to
repair. In such regions, a higher frequency of somatic mutations has been observed in many cancers \citepSupp{schuster2012chromatin}. On the contrary, looser regions---comprising the euchromatin---exhibit fewer mutations \citepSupp{Polak2015}. Current analyses suggest that mutational activity  exhibits periodic patterns relating to nucleosome occupancy, with behaviors varying depending on the mutational signatures involved \citepSupp{Otlu_2023}.
For background on the impact of chromatin conformation on mutations at the various scales, see \citetSupp{Makova2015} and \citetSupp{Supek_lehner_2019}.%

\textbf{Histone modifications and CTCF}.
The relative position and density of nucleosomes are influenced by histone modifications, which are chemical groups attached to certain proteins making up the histones. For instance, H3K36me3 indicates addition of three methyl groups (me3) on the lysine at position 36 (K36) in histone H3, causing elongations and favoring accessibility \citepSupp{Li_2013_H3K39me3}.
On the contrary, H3K9ac is short for addition of an acetyl group (ac) on the lysine at position 9 in H3, and is associated with gene activation. Depending on the modification at play, genomic regions can have higher or lower mutational burden; see  \cref{tab:Epi_covariates}.
Genomic architecture is also shaped by sequence-specific DNA-binding proteins such as the transcription factor CTCF, which acts as an insulator and helps organize chromatin loops and domains. Previous analyses of cancer genomes showed increased mutation rates in CTCF binding sites \citepSupp{Katainen2015}. Both CTCF and histone modifications are measured via chromatin immunoprecipitation (ChIP-seq assay).

\textbf{Replication timing}.
Another important factor is the timing of DNA replication during the S-phase of the cell cycle \citepSupp{Thurman01062007}. Early replicating regions are usually populated with genes and are highly accessible, while late replicating regions tend to be heterochromatic \citepSupp{Stamatoyannopoulos2009}. Replication timing is estimated via Repli-seq \citepSupp{Hansen_repliseq}, which labels and sequences DNA synthesized at different stages of S-phase; timing values are computed by smoothing a weighted average of the proportion of early and late replicating cells in 1kb windows. By construction, low values of replication timing indicate that the region tends to replicate later, while high values denote regions that replicate early. Late-replicating regions tend to be more prone to mutation \citepSupp{Otlu_2023}.

\textbf{Methylation and GC content}.
DNA methylation refers to the binding of a methyl group to a CpG locus (that is, C followed by G), forming 5-methylcytosine.  This configuration is prone to spontaneous deamination, causing a C$>$T substitution \citepSupp{Nishiyama_2021}.
This mechanism is a common cause of DNA mutations \citepSupp{Bird_19780_Methylation}.
This mutational process is represented by the signature SBS1 in COSMIC, which is active in essentially all subjects and correlates with age \citepSupp{Alexandrov_clock_2015}.  Finally, the proportion of G and C nucleotides in a given genomic region---referred to as GC content---is associated with several important genomic and epigenomic features, including gene expression levels, chromatin accessibility, and replication timing. The total GC content in the \texttt{hg19} reference genome is around 40\%, but regions may peak at 80\% at the kilobase scale.

\section{Previous work relating signature activity to genomic features}\label{subsec:previous_work}

There are a few methods that investigate the regional variations in mutational signature activity and their relationship with genomic features. In this section, we briefly describe this previous work and discuss how it differs from our proposed method, particularly with respect to the questions of interest Q1--Q4 posed in the main manuscript.

An early contribution in this direction is \texttt{EMu} \citepSupp{Fischer_2013}, which infers region-specific signatures and activities by dividing the genome into megabase regions and fitting a separate NMF for each bin. More recently, \texttt{GenomeTrackSig} \citepSupp{Timmons_2022} detects regional variation in signature activities by partitioning the genome into bins, and then performing a changepoint detection analysis. Each bin is recommended to have at least 100 mutations.
However, neither method incorporates genomic covariates into the model. Hence, these methods are suitable for inference on signatures and their exposures along the genome (Q1, and potentially Q3) but not their relationship with epigenetic covariates (Q2, Q4).

A step in this direction is taken by \texttt{TensorSignatures} \citepSupp{TensorSignatures_2021}, which applies Tucker-like factorizations to a tensor of mutations categorized into mutually exclusive chromatin states according to their position  \citepSupp{Ernst2017_ChromHMM}. While the model can differentiate mutations based on their replication and transcription strand, it cannot leverage continuous covariates in the factorization. As such, \texttt{TensorSignatures} can answer Q1, Q3, and Q4, but not Q2 for any type of feature. Moreover, benchmarking analyses indicated a need for improved performance \citepSupp{Islam_2022}. In Section~\ref{sec:tensorsignatures}, we compare PPF and \texttt{TensorSignatures} using covariates on chromatin states, finding that PPF effectively generalizes it while also exhibiting greater stability.

Several methods in the literature model mutational signatures and their relationship to covariates using topic-modeling approaches. For example, \citetSupp{Robinson_2019} introduces a latent Dirichlet allocation framework where patient-level (but not mutation-level) covariates shape signature activities via Dirichlet-multinomial regression priors. \citetSupp{StickySig}, instead, use multinomial mixtures to   model consecutive mutations and their co-occurrence along DNA strands. \citetSupp{Kahane_2023} further extends the multinomial mixture approach to explicitly model the effect of (categorical) mutation-level features on signature exposures, especially replication strand and genomic regions. Hence, PPF differs from these methods in terms of i) the generative model, namely the inhomogeneous Poisson process with low-rank intensity functions rather than a multinomial mixture over mutations; ii) the treatment of covariates, which enter jointly and multiplicatively as continuous marks rather than as discrete categories; and iii) the use of patient-specific copy number in the intensity function to account for variations in the mutation rate. Finally, none of these approaches explicitly answers Q2.

Recently, a compendium of the topographical features related to each signature has been assembled by \citetSupp{Otlu_2023} and subsequently by \citetSupp{Otlu_2025} using the \texttt{SigProfilerTopography} software. Their goal is to test for enrichment or depletion of mutations in the proximity of given topographic marks, by comparing the observed mutation data against simulated mutations placed uniformly at random along the genome. This is, however, a \emph{post hoc} analysis tool: each mutation in a patient is assigned to a specific signature without accounting for covariates, and then covariate values are averaged by position using a 2kb window centered around the mutations. Hence, the method is useful for testing the aggregate effects of individual topographic marks, but not their joint effect on the mutation rate. Moreover, the method considers only the subsets of mutations that are attributed to a signature with high confidence, specifically, attribution probability $>0.9$. Hence, it is not designed to answer Q1--Q4.

Additional \emph{post hoc} analyses on the regional variations in mutational signatures and their association with genomic covariates are described by \citetSupp{Blokzijl_2018}, \citetSupp{Lee_2018_mutalisk,Lee_regional_2022}, and \citetSupp{kups75586}. In all cases, these existing approaches consider a pre-estimated set of signatures and activities when inferring these associations, which does not provide direct answer to Q1--Q4 jointly.

\section{Proofs and derivations}\label{sec:proofs_derivs}
\subsection{Proofs of the statements}

\begin{proof}[\textbf{Proof of \cref{pro:baseline}}]
By assumption, $\vartheta_{kj}(t) = \theta_{kj}/T$, so \cref{eq:Intensity_function} becomes $\lambda_{i j}(t) = \sum_{k=1}^K r_{i k} \theta_{k j}/T$ for all $t$.
It follows that $\int_{0}^{T} \lambda_{i j}(t) \dt = \sum_{k=1}^K r_{i k} \theta_{k j}$ under this assumption.
Therefore, the likelihood function in \cref{eq:lik_Lambda} becomes
\begin{align*}
\mathscr{L}(\lambda) &= \prod_{i=1}^I \prod_{j=1}^J\bigg[\exp\bigg( -\int_{0}^{T} \lambda_{ij}(t)\dt \bigg) \prod_{n=1}^{N_{ij}} \lambda_{ij}(t_{i j n}) \bigg]\\
&= \prod_{i=1}^I \prod_{j=1}^J \bigg[\exp\bigg( - \sum_{k=1}^K r_{i k} \theta_{k j} \bigg) \bigg(\sum_{k=1}^K r_{i k} \theta_{k j}/T\bigg)^{N_{i j}}\bigg] \\
&\propto \exp\bigg( -\sum_{j=1}^J\sum_{k = 1}^K \theta_{kj} \bigg)  \prod_{i=1}^I \prod_{j=1}^J \bigg(\sum_{k=1}^K r_{ik}\theta_{kj}\bigg)^{N_{ij}}
\end{align*}
since $\sum_{i=1}^I r_{i k} = 1$, where the proportionality is as a function of $r_{ik}$ and $\theta_{kj}$.
This, in turn, is proportional to the likelihood function for the Poisson NMF model in \cref{eq:PoissonBaseline}.
\end{proof}

\begin{proof}[\textbf{Proof of \cref{pro:superposition}}]
The result follows from the superposition theorem of Poisson processes. See Theorem 3.3 in \citetSupp{Last_Penrose_2017}.
\end{proof}

\begin{proof}[\textbf{Proof of \cref{pro:thinning}}]
The result follows from the coloring theorem of Poisson processes. See \citetSupp{kingman-poisson-processes}, Chapter 5.1.
\end{proof}

\subsection{Derivation of MAP estimation algorithm for PPF}
We now present a detailed derivation of the majorization-minimization strategy to develop the steps used in Algorithm \ref{algo:MAP_rules}.

Define $q_{j}(\bbeta_k) = \int_0^T\frac{1}{2}c_j(t)e^{\bbeta_k^\top\bx(t)}\dt$. Then, \cref{eq:logPosterior} can be written as
\begin{align}\label{eq:logPosterior_all}
-&\,\log\pi(R, \Phi, B, \mu, \sigma^2\mid \bm{t}) = \\
 & \sum_{jk} \phi_{kj} q_{j}(\bbeta_k) - \sum_{ij} \sum_{n = 1}^{N_{ij}} \log \Big(\sum_{k = 1}^K r_{ik}\phi_{kj}\frac{1}{2} \,c_{j}(t_{i j n}) \,e^{\boldsymbol{\beta}_{k}^\top \mathbf{x}(t_{i j n})}\Big) &&\ \textrm{(Log-likelihood)} \notag\\
 &\quad -\sum_{ik}(\alpha_{ik} - 1)\log r_{ik} &&\ \textrm{(Signatures)} \notag\\
 &\quad - \sum_{j k} \bigg(- a\log\mu_k + a \log q_{j}(\bbeta_k)+ (a-1)\log\phi_{kj} -  \phi_{kj}\frac{a}{\mu_k} q_{j}(\bbeta_k)\bigg) &&\ \textrm{(Baseline activities)} \notag\\
 &\quad - \sum_{k} \bigg(-\frac{L}{2}\log\sigma_k^2 - \frac{1}{2\sigma^2_k}\bbeta_k^\top\bbeta_k\bigg) &&\ \textrm{(Regression coefficients)}\notag\\
 &\quad - \sum_k \bigg(- (a_0 + 1) \log\mu_k - \frac{b_0}{\mu_k}\bigg) &&\ \textrm{(Relevance weights)}\notag\\
 &\quad - \sum_k \bigg(- (c_0 + 1) \log\sigma_k^2 - \frac{d_0}{\sigma_k^2}\bigg) + \mathrm{const}\notag &&\ \textrm{(Variances)}
\end{align}
subject to $\sum_{i=1}^I r_{ik} = 1$ for each $k = 1, \ldots, K$. For generality, we consider the case of a generic hyperprior  $\mu_k\sim \mathrm{InvGa}(a_0, b_0)$ on the relevance weights; our compressive hyperprior in \cref{eq:baseline_Priors} is recovered by letting $a_0 = a J + 1$ and $b_0 = \varepsilon a J$.

We begin by providing two results which will be useful in our derivation. The first step in our approach consists in reparametrizing the model in terms of the total activities $\theta_{kj} = \int_0^T \vartheta_{kj}(t)\dt= \phi_{k j}q_{j}(\bbeta_k)$. This simplifies the computational tractability of the model and the optimization algorithm. The following proposition presents the form of the log-posterior in terms of $\theta_{kj}$.

\begin{lemma}\label{pro:model_reparam}
Let $\theta_{k j} = \phi_{k j}q_{j}(\bbeta_k)$, with $q_{j}(\bbeta_k) = \int_0^T\frac{1}{2}c_j(t)e^{\bbeta_k^\top\bx(t)}\dt$, and let $\Theta = (\theta_{k j})\in \mathds{R}_+^{K\times J}$ the matrix of total activities. Then, the log-posterior in Equation~\eqref{eq:logPosterior_all} becomes

\begin{align}\label{eq:logPosterior_all_new}
-&\,\log\pi(R, \Theta, B, \mu, \sigma^2\mid \bm{t}) = \\
 & \sum_{jk} \theta_{k j} - \sum_{ij} \sum_{n = 1}^{N_{ij}} \log \Big(\sum_{k = 1}^K r_{ik}\frac{\theta_{k j}}{q_j(\bbeta_k)}\, \frac{1}{2} \,c_{j}(t_{i j n}) \,e^{\boldsymbol{\beta}_{k}^\top \mathbf{x}(t_{i j n})}\Big) &&\ \textrm{(Log-likelihood)} \notag\\
 &\quad -\sum_{ik}(\alpha_{ik} - 1)\log r_{ik} &&\ \textrm{(Signatures)} \notag\\
 &\quad - \sum_{j k} \bigg(- a\log\mu_k + (a-1)\log\theta_{kj} -  \theta_{k j}\frac{a}{\mu_k} \bigg) &&\ \textrm{(Total activities)} \notag\\
 &\quad - \sum_{k} \bigg(-\frac{L}{2}\log\sigma_k^2 - \frac{1}{2\sigma^2_k}\bbeta_k^\top\bbeta_k\bigg) &&\ \textrm{(Regression coefficients)}\notag\\
 &\quad - \sum_k \bigg(- (a_0 + 1) \log\mu_k - \frac{b_0}{\mu_k}\bigg) &&\ \textrm{(Relevance weights)}\notag\\
 &\quad - \sum_k \bigg(- (c_0 + 1) \log\sigma_k^2 - \frac{d_0}{\sigma_k^2}\bigg)  + \mathrm{const}&&\ \textrm{(Variances)}\notag
\end{align}

\end{lemma}
\begin{proof}
We first calculate the Jacobian of the transformation $(\phi_{k 1}, \ldots, \phi_{k J}, \bbeta_k) \to (\theta_{k 1}, \ldots, \theta_{k J}, \bbeta_k)$, where $\theta_{k j} = \phi_{k j} q_j(\bbeta_k)$, for any signature $k$. The derivatives are

\begin{align*}
&\frac{\partial \theta_{k j}}{\partial \phi_{k j}} = q_{j}(\bbeta_k), &&\frac{\partial \theta_{k j}}{\partial \bbeta_{k}} = \int_{0}^T \frac{1}{2}c_{j}(t) e^{\bbeta_k^\top \bx(t)}\bx(t)\dt = \bm{m}_{j}(\bbeta_k),\\
& \frac{\partial \beta_{k\ell}}{\partial \phi_{k j}} = 0, &&\frac{\partial \bbeta_{k}}{\partial \bbeta_{k}} = I_L
\end{align*}

Let $\mathrm{C}_k = \mathrm{diag}\{q_1(\bbeta_k),\ldots, q_{J}(\bbeta_k)\}\in\mathds{R}^{J \times J}$ be the diagonal matrix containing the derivatives in the first block, and $\mathrm{M}_{k} = (\bm{m}_j(\bbeta_k))\in \mathds{R}^{L\times J}$ be the matrix with columns defined by stacking each $\bm{m}_j(\bbeta_k)$. Then, the determinant of the Jacobian matrix $\mathrm{J}_k \in \mathds{R}^{(J + L)\times(J + L)}$ is
\begin{equation}\label{eq:jacobian}
\det \mathrm{J}_k = \det
\begin{pmatrix}
\mathrm{C}_k & \mathrm{M}_{k}^\top\\
0_{L\times J} & I_L
\end{pmatrix} = \det(\mathrm{C}_k)\det(I_L) = \prod_{j} q_{j}(\bbeta_k).
\end{equation}
Subtracting the log-determinant in Equation~\eqref{eq:jacobian} and substituting $\phi_{k j} = \theta_{k j}/q_{j}(\bbeta_k)$ in Equation~\eqref{eq:logPosterior_all} leads to \cref{eq:logPosterior_all_new}.
\end{proof}

Next, the following result will be useful to majorize Equation~\eqref{eq:logPosterior_all_new}.

\begin{lemma}\label{lem:maj}
Suppose $f:\mathcal{X}\to\mathds{R}$ such that
$$ f(x) = -\log \sum_{k=1}^K h_k(x) $$
where $h_k(x) > 0$ is a strictly positive function for each $k = 1,\ldots,K$.
For $x,\tilde{x}\in \mathcal{X}$, define
$$ g(x,\tilde{x}) = -\sum_{k=1}^K w_k(\tilde{x}) \log \frac{h_k(x)}{w_k(\tilde{x})} $$
where $w_k(x) = h_k(x) / \sum_{s=1}^K h_s(x)$.
Then $f(x) \leq g(x,\tilde{x})$ and $f(x) = g(x,x)$ for all $x,\tilde{x}\in\mathcal{X}$.
\end{lemma}
In other words, the conclusion of \cref{lem:maj} is that $g(x,\tilde{x})$ is a majorizer of $f(x)$.  This provides a useful general strategy for obtaining majorizers of objective functions that involve a sum inside a logarithm.

\begin{proof}
The inequality $f(x) \leq g(x,\tilde{x})$ is a direct consequence of Jensen's inequality:
$$ f(x) = -\log\sum_k w_k(\tilde{x}) \frac{h_k(x)}{w_k(\tilde{x})}\leq \sum_k w_k(\tilde{x})\bigg(-\log\frac{h_k(x)}{w_k(\tilde{x})}\bigg) = g(x,\tilde{x}) $$
since $-\log(x)$ is convex.  The equality $f(x) = g(x,x)$ holds since, by the definition of $w_k(x)$,
$$ g(x,x) = -\sum_k w_k(x) \log \frac{h_k(x)}{w_k(x)} = -\sum_k w_k(x) \log \sum_s h_s(x) = -\log\sum_s h_s(x) = f(x). $$
\end{proof}

\subsubsection{Derivation of Step 1 - Signatures}

The update to the signature matrix $R$ is obtained via a majorization-minimization step on \cref{eq:logPosterior_all} with respect to $R$, subject to the constraint that $\sum_i r_{i k} = 1$ for all $k$.  Keeping only terms that depend on $R$, the objective function in \cref{eq:logPosterior_all_new} with $\phi_{k j} = \theta_{k j}/q_j(\bbeta_k)$ becomes
$$
f(R) = - \sum_{i j} \sum_{n = 1}^{N_{ij}} \log \Big(\sum_{k = 1}^K r_{i k}\phi_{k j}\, e^{\boldsymbol{\beta}_{k}^\top \mathbf{x}(t_{i j n})}\Big) - \sum_{i k} (\alpha_{ik} - 1)\log r_{ik},
$$
noting that $\frac{1}{2} \,c_{j}(t_{i j n})$ comes out in a term that does not depend on $R$.
To form a majorizer $g(R, \tilde{R})$ as a function of signature matrices $R$ and $\tilde{R}$, we apply \cref{lem:maj} for each $i$, $j$, and $n$, with $h_{i j n k}(R) = r_{i k}\phi_{k j}e^{\boldsymbol{\beta}_{k}^\top \mathbf{x}(t_{i j n})}$ and $w_{i j n k}(R) = h_{i j n k}(R) / \sum_s h_{i j n s}(R)$. Specifically, defining
$$
g(R, \tilde{R}) =
     -\sum_{i j k} \sum_{n = 1}^{N_{ij}} w_{i j n k}(\tilde{R})
    \log\Big(\frac{1}{w_{i j n k}(\tilde{R})} r_{i k}\phi_{k j} \,e^{\boldsymbol{\beta}_{k}^\top \mathbf{x}(t_{i j n})}\Big) - \sum_{i k} (\alpha_{ik} - 1)\log r_{ik},
$$
it follows that $g(R,\tilde{R})$ is a majorizer of $f(R)$ by \cref{lem:maj}.
To minimize $g(R,\tilde{R})$ with respect to each column $r_k$, we use the method of Lagrange multipliers. Define
$$\mathcal{L}(R) = g(R,\tilde{R}) -  \tau \Big(1 - \sum_i r_{i k}\Big),$$
where $\tau$ is a Lagrange multiplier for the constraint $\sum_{i = 1}^I r_{ik} = 1$. Differentiating $\mathcal{L}(R)$ with respect to $r_{i k}$ and setting the derivative to 0 yields
$$ 0 = \frac{\partial\mathcal{L}(R)}{\partial r_{i k}} = -\sum_j \sum_{n=1}^{N_{i j}} \frac{w_{i j n k}(\tilde{R})}{r_{i k}} - \frac{\alpha_{i k} - 1}{r_{i k}} + \tau $$
for $i = 1,\ldots,I$, or equivalently,
$$ r_{i k} = \frac{1}{\tau}\Big(\alpha_{i k} - 1 + \sum_j \sum_{n=1}^{N_{i j}} w_{i j n k}(\tilde{R}) \Big) = \frac{1}{\tau}\bigg(\alpha_{i k} - 1 + \sum_j \sum_{n=1}^{N_{i j}} \frac{\tilde{r}_{i k}\phi_{k j}e^{\boldsymbol{\beta}_{k}^\top \mathbf{x}(t_{i j n})}}{\sum_s \tilde{r}_{i s}\phi_{s j}e^{\boldsymbol{\beta}_{s}^\top \mathbf{x}(t_{i j n})}} \bigg). $$
Taking the derivative with respect to the Lagrange multiplier leads to $\sum_{i = 1}^I r_{ik} =1$. Plugging in this constraint and solving for $\tau$ yields the update in Algorithm \ref{algo:MAP_rules}, viewing $\tilde{r}_k$ as the current value and $r_k$ as the updated value.

\subsubsection{Derivation of Step 2 - Total activity}

The update to the total activity matrix $\Theta$ is obtained via a majorization-minimization step on \cref{eq:logPosterior_all_new} with respect to $\Theta$.  Keeping only terms that depend on $\Theta$, the objective function in \cref{eq:logPosterior_all_new} becomes
\begin{align*}
f(\Theta) &= \sum_{j k} \theta_{k j}
   - \sum_{i j} \sum_{n=1}^{N_{ij}} \log \Big( \sum_{k=1}^K r_{ik}\frac{\theta_{k j}}{q_{j}(\bbeta_k)} e^{\bbeta_{k}^\top \bx(t_{i j n})} \Big) \\
   &\quad - \sum_{j k} \bigg((a-1)\log\theta_{k j} - \frac{a}{\mu_k}\theta_{k j}\bigg).
\end{align*}
We use the same majorization trick as in Step 1, with $h_{i j n k}(\Theta)$ and $w_{i j n k}(\Theta)$ defined the same way but viewed as functions of $\Theta$.  This yields the majorizer
\begin{align*}
g(\Theta,\tilde{\Theta}) &= \sum_{j k} \theta_{k j}
   - \sum_{i j k} \sum_{n=1}^{N_{ij}} w_{i j n k}(\tilde{\Theta})\log \Big( \frac{1}{ w_{i j n k}(\tilde{\Theta})} r_{ik}\frac{\theta_{k j}}{q_{j}(\bbeta_k)}e^{\bbeta_{k}^\top \bx(t_{i j n})} \Big) \\
   &\quad - \sum_{j k} \bigg((a-1)\log\theta_{k j} -  \frac{a}{\mu_k}\theta_{k j}\bigg).
\end{align*}
To minimize $g(\Theta,\tilde{\Theta})$ with respect to $\theta_{k j}$, we take the derivative and set it to zero:
\begin{align*}
0 = \frac{\partial g(\Theta,\tilde{\Theta})}{\partial \theta_{k j}} = 1
   - \sum_{i} \sum_{n=1}^{N_{ij}} \frac{w_{i j n k}(\tilde{\Theta})}{\theta_{k j}}
   - \bigg(\frac{a-1}{\theta_{kj}} - \frac{a}{\mu_k}\bigg).
\end{align*}
Solving for $\theta_{k j}$ and plugging in the definition of $w_{i j n k}(\tilde{\Theta})$, we have
$$
\theta_{k j} = \frac{\mu_k}{a + \mu_k}\Big(a-1 + \bar{S}_{kj}\Big), \qquad \bar{S}_{kj} = \sum_{i} \sum_{n=1}^{N_{ij}} w_{i j n k}(\tilde{\Theta})
$$
which yields the update in Algorithm \ref{algo:MAP_rules}, treating $\tilde{\Theta}$ as the current value and $\Theta$ as the updated value.
In this case, there is no need to use Lagrange multipliers, since the only constraint is that $\theta_{k j} > 0$, which is guaranteed as long as $a > 1$.

\subsubsection{Derivation of Step 3 - Regression coefficients}

Following the same approach as in Step 2, but for $B$, the objective function in \cref{eq:logPosterior_all_new} becomes
\begin{align*}
f(B) =
   - \sum_{i j} \sum_{n=1}^{N_{ij}} \log \Big( \sum_{k=1}^K r_{ik}\frac{\theta_{k j}}{q_j(\bbeta_k)}e^{\bbeta_{k}^\top \bx(t_{i j n})} \Big) + \sum_k \frac{1}{2\sigma_k^2}\bbeta_k^\top \bbeta_k
\end{align*}
and the majorizer is
\begin{align*}
g(B,\tilde{B}) &=
   - \sum_{i j k} \sum_{n=1}^{N_{ij}} w_{i j n k}(\tilde{B})\log\!\bigg( \frac{r_{ik}\theta_{kj}e^{\bbeta_{k}^\top \bx(t_{i j n})}/q_j(\bbeta_k)}{ w_{i j n k}(\tilde{B})}  \bigg)
 + \sum_k \frac{1}{2\sigma_k^2}\bbeta_k^\top \bbeta_k,
\end{align*}
where
$$ w_{i j n k}(B) = \frac{r_{i k}\theta_{k j}e^{\bbeta_{k}^\top \mathbf{x}(t_{i j n})}/q_j(\bbeta_k)}{\sum_s r_{i s}\theta_{s j}e^{\bbeta_{s}^\top \mathbf{x}(t_{i j n})}/q_j(\bbeta_s)}.$$
Thus, the majorizer can be simplified as
\begin{align*}
g(B,\tilde{B}) =
   - \sum_{i j k} \sum_{n=1}^{N_{ij}} w_{i j n k}(\tilde{B})\bigg(\bbeta_k^\top\bx(t) - \log{q_j(\bbeta_k)} \bigg)
 + \sum_k \frac{1}{2\sigma_k^2}\bbeta_k^\top \bbeta_k + \mathrm{const},
\end{align*}

To minimize $g(B,\tilde{B})$ over $\bbeta_k$, we use a Newton's method step as follows. We introduce the following derivatives:
\begin{align*}
\bm{m}_j(\bbeta_k) &= \frac{\partial q_j(\bbeta_k)}{\partial\bbeta_k} = \int_0^T \frac{1}{2}c_j(t)e^{\bbeta_k^\top\bx(t)}\bx(t)\dt, \\
Q_j(\bbeta_k) &= \frac{\partial^2 q_j(\bbeta_k)}{\partial \bbeta_k\bbeta_k^\top} = \int_0^T \frac{1}{2}c_j(t) e^{\bbeta_k^\top\bx(t)}\bx(t)\bx(t)^\top\dt
\end{align*}
Call $\bar{S}_{kj} = \sum_{i}\sum_{n= 1}^{N_{ij}} w_{ijnk}(\tilde{B})$. The gradient and Hessian of $g(B,\tilde{B})$ with respect to $\bbeta_k$ are
\begin{align*}
\mathbf{g}_k &= \sum_{j} \bar{S}_{kj}\frac{\bm{m}_j(\bbeta_k)}{q_j(\bbeta_k)}
   - \sum_{i j} \sum_{n=1}^{N_{ij}} w_{i j n k}(\tilde{B})\bx(t_{i j n})
 + \frac{1}{\sigma_k^2}\bbeta_k \\
\mathbf{H}_k &= \sum_{j} \bar{S}_{kj} \bigg(\frac{Q_j(\bbeta_k)}{q_j(\bbeta_k)} - \frac{\bm{m}_j(\bbeta_k)\bm{m}_j(\bbeta_k)^\top}{q_j(\bbeta_k)^2}\bigg) + \frac{1}{\sigma_k^2}I_L.
\end{align*}

Treating $\tilde{B}$ as the current value, a Newton step would be to update $\bbeta_k$ to $\tilde{\bbeta}_k - \mathbf{H}_k^{-1} \mathbf{g}_k$, evaluating $\mathbf{g}_k$ and $\mathbf{H}_k$ at $\bbeta_k = \tilde{\bbeta}_k$.
This leads to the update in Algorithm~\ref{algo:MAP_rules}. To improve on numerical stability, we repeat this step two times each cycle of the algorithm, and cap the root-mean-square of the update at a fixed value $\rho > 0$ by using $\bbeta_k \gets \bbeta_k + \boldsymbol{\xi}_k\min\big\{1,\, \rho \sqrt{L}/\|\boldsymbol{\xi}_k\|\big\}$ where $\boldsymbol{\xi}_k = -\mathbf{H}_{k}^{-1}\mathbf{g}_{k}$ and $\|\cdot\|$ is the Euclidean norm, following \citetSupp{miller2020inference}; using $\rho = 0.5$ works well in our experiments.

\subsubsection{Derivation of Step 4 - Relevance weights and variance parameters}

The updates for
$\mu_k$ and $\sigma_k^2$ are derived by analytically maximizing the log posterior density.
Keeping only the terms that depend on $\mu_k$, the objective function in \cref{eq:logPosterior_all_new} becomes
\begin{align*}
    f(\mu_k) &= -\sum_{j} \Big(- a\log\mu_k - \theta_{kj}\frac{a}{\mu_k} \Big)
 - \Big(-(a_0 + 1) \log\mu_k - \frac{b_0}{\mu_k}\Big) \\
 &=  (a_0 + a J + 1)\log\mu_k + \Big(b_0 + a \sum_j  \theta_{kj}\Big) \frac{1}{\mu_k}.
\end{align*}
Since $\exp(-f(\mu_k)) \propto \mathrm{InvGa}\big(\mu_k \,\big\vert\, a_0 + a J,\, b_0 + a \sum_j  \theta_{k j}\big)$, the minimizer of $f(\mu_k)$ is the mode of this inverse gamma distribution, namely,
$$
\mu_k =
\frac{b_0 + a \sum_j \theta_{kj} }{a_0 + a J + 1}.
$$
Our choices of $a_0 = a J + 1$ and $b_0 = \varepsilon a J$ yield the update in Algorithm~\ref{algo:MAP_rules}.

Likewise, keeping only the terms that depend on $\sigma^2_k$, the objective function becomes
\begin{align*}
    f(\sigma^2_k) &= -\bigg(-\frac{L}{2}\log\sigma_k^2 - \frac{1}{2\sigma^2_k}\bbeta_k^\top\bbeta_k\bigg) - \bigg(- (c_0 + 1) \log\sigma_k^2 - \frac{d_0}{\sigma_k^2}\bigg) \\
    &= \Big(c_0 + \frac{L}{2} + 1\Big)\log\sigma_k^2 + \Big(d_0 + \frac{1}{2}\bbeta_k^\top\bbeta_k\Big) \frac{1}{\sigma_k^2}.
\end{align*}
Since $\exp(-f(\sigma^2_k)) \propto \mathrm{InvGa}\big(\sigma^2_k \,\big\vert\, c_0 + L/2,\, d_0 + \bbeta_k^\top \bbeta_k/2\big)$, the minimizer of $f(\sigma^2_k)$ is
$$
\sigma_k^{2} =
\frac{d_0 +  \bbeta_k^\top \bbeta_k/2}{c_0 + L/2 + 1},
$$
which is the update in Algorithm~\ref{algo:MAP_rules}.

\subsection{Derivation of MCMC algorithm for PPF}

Adopting the change of variable in \cref{pro:model_reparam}, the likelihood function is
$$
\mathscr{L} = \exp\bigg(-\sum_{jk} \theta_{k j}\bigg) \prod_{ij} \prod_{n=1}^{N_{ij}} \bigg(\sum_{k = 1}^K  \frac{1}{2}c_{j}(t_{i j n})r_{ik}\frac{\theta_{kj}}{q_j(\bbeta_k)}e^{\boldsymbol{\beta}_{k}^\top \mathbf{x}(t_{i j n})}\bigg).
$$
We adopt a data augmentation approach typical of Poisson factorization \citepSupp{Dunson_Herring_2005}. Specifically, we introduce a multinomial random variable for each of the observed mutations such that $$W_{ij}(t_{i j n}) = \big(W_{ij1}(t_{i j n}), \ldots, W_{ijK}(t_{i j n})\big) \sim \mathrm{Mult}\big(1;\, p_{ij1}(t_{i j n}), \ldots, p_{ijK}(t_{i j n})\big),$$
where
$$
p_{ijk}(t_{i j n}) = \mathds{P}(W_{ijk}(t_{i j n}) = 1\mid R, \Theta, B, \bm{t}) =  \frac{r_{i k} \theta_{kj} e^{\bkx(t_{i j n})}/q_j(\bbeta_k)}{\sum_{s = 1}^K r_{i s}\theta_{sj} e^{\bbeta_s^\top \bx(t_{i j n})}/q_j(\bbeta_s)}.
$$
Then one can write the augmented posterior as
\begin{align}\label{eq:Posterior}
\pi(&R, \Theta, B, \mu, \sigma^2, W\mid \bm{t}) \propto \\ & \exp\Big(-\sum_{jk} \theta_{k j}\Big) \prod_{ij} \prod_{n=1}^{N_{ij}} \prod_k \Big( \frac{1}{2}c_{j}(t_{i j n})r_{ik}\frac{\theta_{kj}}{q_j(\bbeta_k)}e^{\boldsymbol{\beta}_{k}^\top \mathbf{x}(t_{i j n})}\Big)^{W_{ijk}(t_{i j n})} &&\ \textrm{(Likelihood)}\notag\\
 &\quad\times \prod_k\prod_{i} r_{ik}^{\alpha_{ik} - 1} &&\ \textrm{(Signatures)} \notag\\
 &\quad\times \prod_k\prod_{j} \mu_k^{-a} \theta_{kj}^{a-1} \exp\bigg(- \theta_{kj}\frac{a}{\mu_k}\bigg) &&\ \textrm{(Activities)}\notag\\
 &\quad\times \prod_k (\sigma_k^2)^{-L/2}\exp\!\Big(-\frac{1}{2\sigma^2_k}\bbeta_k^\top\bbeta_k\Big) &&\ \textrm{(Coefficients)}\notag\\
 &\quad\times \prod_k \mu_k^{-a_0 - 1}\exp(-b_0/\mu_k) &&\ \textrm{(Weights)}\notag\\
 &\quad\times \prod_k (\sigma_k^2)^{-c_0 - 1}\exp(-d_0/\sigma_k^2). &&\ \textrm{(Variances)}\notag
\end{align}

\noindent All the full conditionals in Algorithm \ref{algo:Gibbs_rules} are standard conjugate updates based on \cref{eq:Posterior}.
The $\bbeta_k$ update is done using elliptical slice sampling \citepSupp{pmlr-v9-murray10a}, with the adaptive variant of \citetSupp{marco2026adaptivegeneralizedellipticalslice}, using their default hyperparameters.

\subsection{MAP for Poisson NMF with compressive hyperpriors}\label{subsec:map_compnNMF}

The CompNMF model of \citetSupp{zito2024compressive} is a special case of the PPF model where $\bbeta_k = \bm{0}$, $c_j(t) = 2$, and $\phi_{k j} = \theta_{k j}/T$ for all $j,k,t$, since then $\vartheta_{kj}(t) = \theta_{kj}/T$ and \cref{pro:baseline} applies. Hence, the \emph{maximum a posteriori} (MAP) estimate can be obtained using the same steps as Algorithm \ref{algo:MAP_rules}. Notice that these steps are valid if $a\geq 1$ and $\alpha_{ik}\geq 1$.

The following algorithm reduces Algorithm~\ref{algo:MAP_rules} to this special case, which yields a MAP algorithm for the baseline Poisson NMF model in \cref{eq:PoissonBaseline} with automatic selection of the number of factors.  Unlike the PPF algorithm, these updates can be written using matrix multiplication operations, which are particularly fast. We stop the algorithm when two successive iterations of the log posterior density have a relative difference less than $10^{-7}$.
The compressive hyperprior is obtained by using $a_0 = a J + 1$ and $b_0 = \varepsilon a J$.

\begin{algorithm}[h]
\caption{MAP updates for CompNMF}\label{algo:compressive_nmf_map}
\small

\hspace{-1em}\textbf{Input:} Count matrix $N$, where $N_{i j}$ = number of mutations of type $i$ in patient $j$.

\vspace{1em}
\nl\textbf{Signatures:}
Update the signature matrix $R$ as
$$
R \gets \mathrm{normalize}\bigg( (A-1) + R \circ \Big( \frac{N}{R\Theta} \Theta^\top \Big) \bigg)
$$
where $A = (\alpha_{i k}) \in \mathds{R}^{I\times K}$ is the matrix with $(i,j)$ entry  $\alpha_{i k}$. Here, $\circ$ denotes element-wise multiplication (Hadamard product), $\frac{X}{R\Theta}$ is element-wise division, and $\mathrm{normalize}(\cdot)$ makes each column of $R$ sum to one.

\nl\textbf{Activities:}
Let $\mu = (\mu_1, \ldots, \mu_K)$ be the vector of relevance weights, and let $M$ be the $K \times J$ matrix with every column equal to $\mu$. Update the activity matrix $\Theta$ as
$$
\Theta \gets \bigg((a-1) + \Theta \circ \Big(R^\top \frac{N}{R\Theta}\Big) \bigg) \circ \Big(\frac{M}{a + M}\Big).
$$

\nl\textbf{Relevance weights:}
Update each relevance weight $\mu_k$ as
\[
\mu_k \gets \frac{b_0 + a \sum_{j=1}^{J} \theta_{kj}}{a_0 + a J + 1}.
\]

\hspace{-1em}\textbf{Output:} Updated point estimates for $R$, $\Theta$, $\mu$.
\end{algorithm}

\clearpage
\section{Simulations}\label{secSupp:simulations}

This section benchmarks the performance of PPF against several synthetic mutation datasets, both when the model is correctly specified (\cref{subsec:Simulation_setup}) and under various sources of misspecification (\cref{subsec:Simulation_misspec}). Our goal is to determine whether using covariates improves mutation recovery compared to the baseline NMF model in each setting.

\subsection{Simulation I - Comparison of PPF and baseline NMF}\label{subsec:Simulation_setup}

\subsubsection{Data generation strategy}\label{subsec:sim_data}
The simulated covariates, copy numbers, true model parameters, and mutation data were generated as follows for each dataset.

\textbf{Covariates.} We set $T = 2{,}000{,}000$ and divide $[0, T)$ into equally sized bins $[\tau_m, \tau_{m+1})$ of length $\tau_{m+1} - \tau_{m} = 100$, so that $[0, T) = \bigcup_{m = 1}^{M-1} [\tau_m, \tau_{m+1})$ with $M = 20{,}000$, $\tau_1 = 0$, and $\tau_M = T$. We simulate autocorrelated signals as $\bz_m \sim \mathrm{N}(D\bz_{m-1}, \boldsymbol{\Sigma}_0)$ where $\bz_1 \sim N(0, \boldsymbol{\Sigma}_0)$, $D= \mathrm{diag}(0.99, \ldots, 0.99)\in\mathds{R}^{L \times L}$, and $\boldsymbol{\Sigma}_0 \in \mathds{R}^{L\times L}$ is a covariance matrix that varies by scenario.  In Scenario A, we simulate genomic covariates as independent by setting $\boldsymbol{\Sigma}_0 = I_{L}$; in Scenario B, we make them correlated by sampling $\boldsymbol{\Sigma}_0$ using the function \texttt{genPositiveDefMat} with \texttt{covMethod = "onion"} from the \textsc{R} package \texttt{clusterGeneration}; refer to \citetSupp{JOE20062177}.
We then set $\bx(t) = (\bz_m - \overline{\bz})/\mathrm{sd}(\bz)$ for $t \in [\tau_m, \tau_{m+1})$, where $\overline{\bz} = \frac{1}{M}\sum_{m = 1}^M \bz_m$,  the scaling $\mathrm{sd}(\bz) = \big(\frac{1}{M-1}\sum_{m = 1}^M (\bz_m - \overline{\bz})^2\big)^{1/2}$ is computed coordinate-wise, and ``$/$'' denotes coordinate-wise division.
Hence, our synthetic genome comprises $M = 20{,}000$ regions and has piecewise constant covariates with mean zero and variance one.

\textbf{Copy numbers.} For each patient $j$, we generate the copy numbers $c_j(t)$ as follows. Draw $S_j \sim \mathrm{Poisson}(20)$ and split $[0, T)$ into $S_{j}$ contiguous segments by randomly selecting segment boundaries uniformly without replacement from $\{\tau_1, \ldots, \tau_M\}$.
For each segment $s = 1,\ldots,S_j$, we sample $V_{j s} \sim \mathrm{NegBin}(1, 10)$ independently, where $\mathrm{NegBin}(m, v)$ denotes the negative binomial with mean $m$ and variance $m(1 + m/v)$. We then set $c_j(t) = V_{j s} + 2$ for all positions $t$ in segment $s$.

\textbf{Model parameters}. We define the true signature matrix $R_0 \in \mathds{R}^{I \times K_0}$ with $I = 96$ by setting the first four columns equal to the COSMIC signatures frequently appearing in breast cancer, namely \sbs{1}, \sbs{2}, \sbs{13}, and \sbs{3}; see \cref{sec:application} for their interpretation. The remaining $K_0 - 4$ columns are sampled as $r^0_{k} \sim \mathrm{Dir}(0.1, \ldots, 0.1)$ independently.  We generate the true baseline activities $\Phi_0 \in \mathrm{R}^{K\times J}$ as $\phi^0_{kj} = \mu_k^0 \nu^0_{kj}/\int_0^T \frac{1}{2}c_j(t)\dt$, with $\mu_k^0 \sim \mathrm{Ga}(100, 1)$ and $\nu^0_{kj} \sim \mathrm{Ga}(0.5, 0.5)$ independently. We simulate the true coefficients by sampling $\bbeta^0_k \sim \mathrm{N}(\mathbf{0}, \sigma^2_0 I_L)$, with $\sigma^2_0 = 1/2$, and we define $B_0 \in \mathds{R}^{K\times L}$ as the matrix with columns $\bbeta_k^0$.

\textbf{Mutations}. For each $i = 1, \ldots, I$ and $J = 1,\ldots, J$, we simulate mutations along $[0, T)$ by sampling from a Poisson process using the inverse cumulative intensity function. First, we sample the total number of mutations as $N_{ij} \sim \mathrm{Poisson}(\Lambda_{ij})$, where $\Lambda_{ij} = \int_0^T\lambda_{ij}(t)\dt = \sum_{m = 1}^M (\tau_{m+1} - \tau_{m}) \lambda_{ij, m}$, and $\lambda_{ij, m} = \sum_{k =1}^K r_{ik}^0 \phi_{kj}^0 \frac{1}{2}c_{j m}e^{{\bbeta^0_k}^\top \bx_m}$ with $c_{j m}$ and $\bx_m$ denoting the values of the copy number and the covariates in bin $m$.
Then, we sample $U_n \sim \mathrm{U}(0, \Lambda_{ij})$ for $n=1, \ldots, N_{ij}$ independently, and for each $n$, we find the bin index $m_n$ for which $\sum_{m = 1}^{m_n} \lambda_{ij, m} \leq U_{n} < \sum_{m = 1}^{m_n + 1}\lambda_{ij, m}$. Finally, $t_{i j n} \sim \mathrm{U}(\tau_{m_n}, \tau_{m_n + 1})$.

\textbf{Data generation}. We simulate data for $J = 40$ patients, with $L = 10$ covariates and $K_0 = 8$ signatures (4 from COSMIC---SBS1, SBS2, SBS13 and SBS3--- and 4 random). We use only the first $L_{\mathrm{true}} = 5$ covariates to generate the mutations, setting $\beta_{k\ell} = 0$ for $\ell = 6, \ldots, 10$, so that the last 5 covariates do not affect the mutation rate. We consider two cases: in Scenario A, the covariates are independent, while in Scenario B, they reflect the strong correlations exhibited by certain genomic features (see \cref{fig:correlation}).
Under each scenario, we generate 20 datasets---each with its own covariates, copy numbers, true parameters, and mutations---resulting in an average total number of mutations across replicates equal to $94{,}577$ and $92{,}860$ in Scenarios A and B, respectively.

\subsubsection{Methods compared}
On each simulated dataset, we compare results from the following methods:
\begin{enumerate}[itemsep=0pt, start=0, parsep=0pt, label=M\arabic*), ref=M\arabic*, leftmargin=*, align=left]
    \item \underline{MAP, true $\bx$}: the MAP with only the $L = 5$ covariates used to generate the data, and with $c_{j}(t)$ set to the true values.
    \label{simmod:add_MAPTrue}
    \item \underline{MAP,  CompNMF}: the  MAP of the Poisson NMF model in \cref{eq:PoissonBaseline} with prior as in \citetSupp{zito2024compressive}. (No covariates, no copy numbers.)
    \label{simmod:add_compNMF}
    \item \underline{MAP, no $\bx$}: the MAP with no covariates, and with $c_{j}(t)$ set to the true copy numbers.
    \label{simmod:add_noCovs}
    \item \underline{MAP, all $\bx$}: the MAP with all $L = 10$ covariates, and  $c_{j}(t)$ set to the true copy numbers.
    \label{simmod:add_MAP}
    \item \underline{MCMC, all $\bx$}: the posterior mean for the same model as \ref{simmod:add_MAP}, calculated by starting Algorithm \ref{algo:Gibbs_rules} from the MAP.
    \label{simmod:add_MCMC}
    \item \underline{MCMC, all $\bx$, $\Delta_b = 200$}: same as \ref{simmod:add_MCMC}, but with covariates aggregated by further averaging their values over groups of two consecutive of bins.
    \label{simmod:add_MCMC200}
    \item \underline{MCMC, all $\bx$, $\Delta_b = 500$}: same as \ref{simmod:add_MCMC200}, but averaging over groups of 5 consecutive bins. \label{simmod:add_MCMC500}
    \item \underline{SignatureAnalyzer}: the NMF method of \citetSupp{kim2016somatic}. The number of signatures is selected in one single model via automatic relevance determination. \label{simmod:sig_auto}
\end{enumerate}
All the models considered are variants of PPF except SignatureAnalyzer and CompNMF, although the latter can also be viewed as a special case of PPF. We use the default PPF model settings in \cref{subsec:priors} for $a$, $\alpha_{i k}$, $\varepsilon$, $c_0$, and $d_0$, and we set $K = 15$ for the upper bound on the number of signatures. Our goal is to compare the performance of different versions of the model and algorithm in order to assess the utility of various aspects.  For example, comparing \ref{simmod:add_MAP} to \ref{simmod:add_MAPTrue} quantifies any loss in performance due to including irrelevant covariates. Meanwhile, comparing \ref{simmod:add_MCMC200} to \ref{simmod:add_MCMC} evaluates whether there is a loss in performance due to approximating the integral in \cref{eq:logPosterior} using a coarser grid than the true data generating process.

\subsubsection{Computation details and evaluation metrics}
\textbf{Computation}.
For methods \ref{simmod:add_MAPTrue}, \ref{simmod:add_noCovs}, and \ref{simmod:add_MAP}, we run Algorithm \ref{algo:MAP_rules} until the relative difference in log-posterior is less than $10^{-7}$. For method \ref{simmod:add_compNMF}, we use the multiplicative rules in \cref{algo:compressive_nmf_map} with tolerance $10^{-7}$. For the MAP algorithms, we initialize by randomly drawing from the priors.
For methods \ref{simmod:add_MCMC}, \ref{simmod:add_MCMC200}, and \ref{simmod:add_MCMC500}, we initialize by using Algorithm \ref{algo:MAP_rules} to estimate the MAP, and then use it as a starting point for Algorithm \ref{algo:Gibbs_rules}, running one MCMC chain for $3{,}000$ iterations and discarding the first  $1{,}500$ as burn-in. We postprocess the output to eliminate redundant factors using the thresholding rule of \citetSupp{zito2024compressive} by discarding all signatures for which $\hat{\mu}_k \leq 5\varepsilon$. Thus, a latent process is excluded from the factorization when its average activity is close to $\varepsilon = 0.001$. The number of signatures remaining is an estimate of the rank, $\hat{K}$.
We let $\hat{R}$, $\hat{\Theta}$, $\hat{B}$, $\hat{\mu}$, and $\hat{\sigma}^2$ denote the parameter estimates, specifically, these are the posterior modes when using Algorithm \ref{algo:MAP_rules} and the posterior means when using Algorithm~\ref{algo:Gibbs_rules}. Finally, we fit model \ref{simmod:sig_auto} via the function \texttt{sig\_auto\_extract()} in the R package \texttt{sigminer} \citepSupp{sigminer_2020}.

\textbf{Evaluation metrics}.
For each dataset, we calculate $\textsc{rmse}(\hat{R}, R_0)$, $\textsc{rmse}(\hat{\Theta}, \Theta_0)$, and $\textsc{rmse}(\hat{B}, B_0)$, where $\textsc{rmse}(\hat{A}, A) = \big(\sum_{i j}(\hat{a}_{i j} - a_{i j})^2/ m n\big)^{1/2}$ denotes the root mean-squared error between two matrices of dimension $m\times n$.
Here, $\Theta_0\in\mathds{R}^{K\times J}$ is the matrix with entry $(k,j)$ equal to $\int_{0}^{T} \vartheta^{0}_{k j}(t) \mathrm{d} t$, where $\vartheta^{0}_{k j}(t)$ is defined as in \cref{eq:vartheta_t} but using the true parameter values;
likewise, $\hat{\Theta}$ is defined similarly but using the estimated parameter values.
These quantities represent the total activity of each signature across the genome.
We assess these integrated activities rather than the baseline activities because they have a consistent meaning across both PPF and NMF.
To calculate the RMSE, we first complete all matrices with zeros to avoid any potential misalignment between $\hat{K}$ and the relevant true value. Then, we match estimated signatures---along with the corresponding activities and coefficients---to the true signatures using the Hungarian algorithm with cosine similarity.

A second important evaluation metric is $\textsc{rmse}(\hat{\Lambda}, \Lambda_0) = \big(\sum_m (\hat{\Lambda}_{j m} - \Lambda_{j m})^2/J M\big)^{1/2}$ for each dataset, where $\Lambda_{j m}= \sum_{i}\int_{\tau_m}^{\tau_{m+1}}\lambda_{ij}(t)\dt$ is the total intensity in bin $m$, and $\hat{\Lambda}_{j m}$ is its estimated value. This measures how well the mutation counts are retrieved along the simulated genomes for each patient and mutation type.

Finally, we calculate the $F_1$ score for each replicate. The $F_1$ score is defined as $F_1 = 2 \times \mathrm{precision} \times \mathrm{sensitivity}/(\mathrm{precision} + \mathrm{sensitivity})$, where \emph{precision} is the proportion of estimated signatures that have a cosine similarity $\geq 0.9$ with at least one of the ground truth signatures, and \emph{sensitivity} is the proportion of ground truth signatures for which there is an estimated signature with cosine similarity $\geq 0.9$ \citepSupp{Islam_2022}. Hence, $F_1$ measures how similar $\hat{R}$ and $R_0$ are, with values close to 1 indicating high similarity.

\subsubsection{Simulation results}\label{subsec:sim_results}

\begin{figure}[t]
    \centering
    \includegraphics[width=\linewidth]{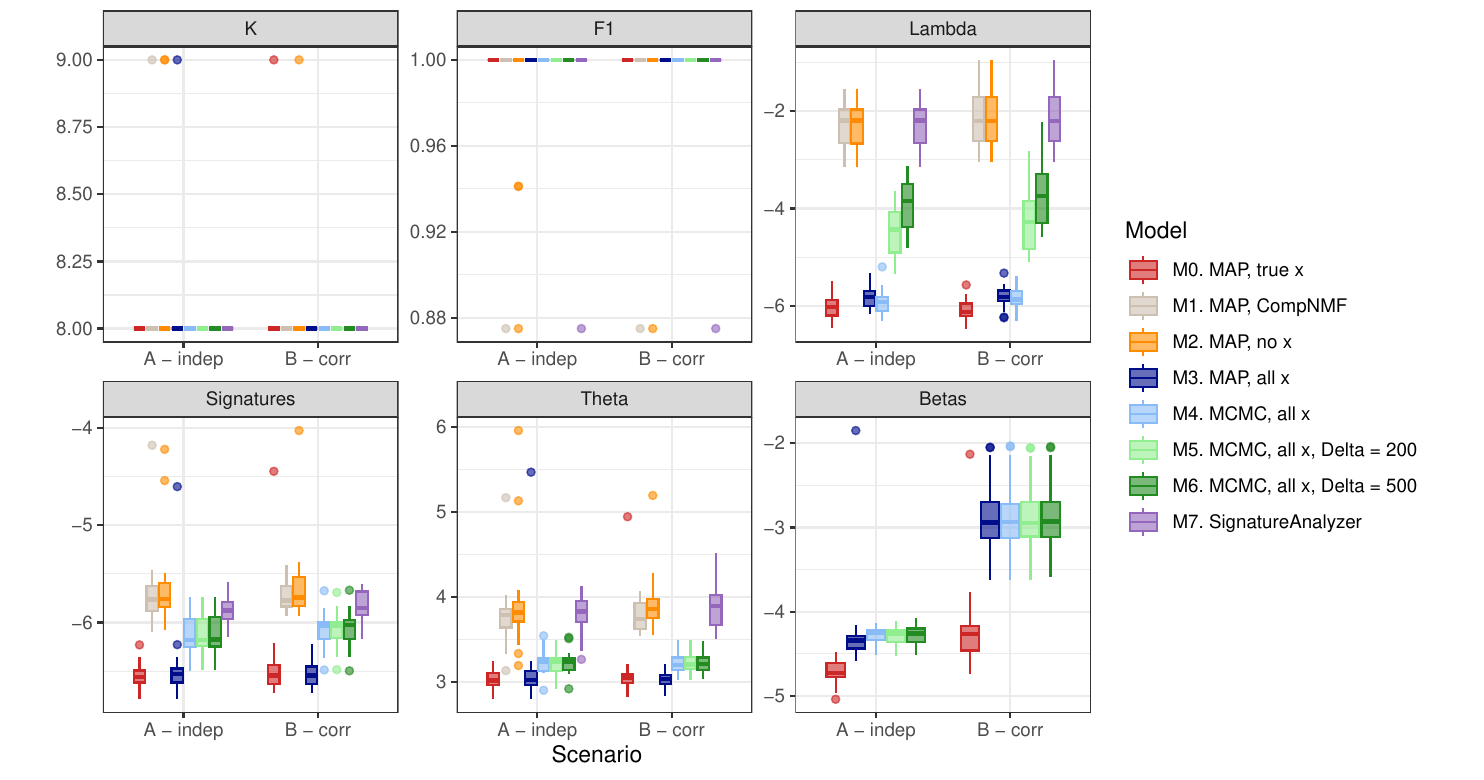}
    \caption{Results of the simulation. Boxplots display the values from 20 randomly generated datasets under both scenarios. The panels display the log(RMSE) of $(\hat{R}, R_0)$, $(\hat{\Theta}, \Theta_0)$, $(\hat{B}, B_0)$, and $(\hat{\Lambda}, \Lambda_0)$.  The panel titled ``K'' shows the estimated number of signatures across the 20 replicates. Panel ``F1'' shows the $F_1$ score.
    }
    \label{fig:Sim_results_add}
\end{figure}

\cref{fig:Sim_results_add} shows the estimation performance of each method in terms of recovering the data-generating parameters, plotted on the log scale to facilitate visualization. Several points are worth noting, which can be summarized as follows.

\begin{itemize}
    \item \textbf{Benefit of covariates}. PPF with covariates (e.g., \ref{simmod:add_MAPTrue} and \ref{simmod:add_MAP}) provides a major improvement in performance compared to the baseline NMF model (\ref{simmod:add_compNMF}), in terms of the estimated total intensity $\Lambda$, signatures $R$, and total activity $\Theta$. In contrast, using copy numbers without covariates (\ref{simmod:add_noCovs}) provides little to no benefit over the baseline NMF model (\ref{simmod:add_compNMF}). Hence, even though the baseline NMF retrieves the correct number of signatures in every case with high precision, accounting for covariates and copy numbers improves inference for $R_0$ and $\Theta_0$. Model \ref{simmod:sig_auto} behaves similarly to the CompNMF case.
    \item \textbf{Recovery of $K$}. On these data, all versions accurately recover the true number of factors ($K = 8$) and have a perfect $F_1$ score with only a few exceptions. Thus, any differences in performance are not due to incorrect recovery of $K$.
    \item \textbf{Impact of extra covariates}. As expected, including unneeded covariates (\ref{simmod:add_MAP}) can reduce performance  relative to including only the covariates used to generate the data (\ref{simmod:add_MAPTrue}), particularly in Scenario B (correlated covariates).  While this loss in performance is substantial for estimating the coefficients $B$, there is little difference in performance for estimating $\Lambda$, $R$, and particularly $\Phi$, for which there is essentially no difference.
    \item \textbf{MCMC versus MAP}.  Curiously, the MAP estimate (\ref{simmod:add_MAP}) tends to be more accurate than the posterior mean from MCMC (\ref{simmod:add_MCMC}) in terms of the signatures $R$ and total activity $\Theta$. This is a natural consequence of the optimization algorithm.
    \item \textbf{Resolution of covariate aggregation}. Using a coarser scale of aggregation for the covariates (\ref{simmod:add_MCMC200} and \ref{simmod:add_MCMC500}) has little impact on estimation of the signatures $R$, total activity $\Theta$, and coefficients $B$, compared to using a finer scale (\ref{simmod:add_MCMC}).  However, it does reduce the accuracy of the total intensity $\Lambda$.
    \item \textbf{Impact of correlated covariates}. For all methods, performance in Scenario B is comparable to Scenario A for estimating $K$, $\Lambda$, $R$, and $\Phi$. However, estimation of $B$ is noticeably worse in Scenario B, particularly when irrelevant covariates are included.
\end{itemize}
Overall, the best performance is obtained when using PPF wih the MAP estimate, including covariates and copy numbers.

\subsubsection{Computation time and effective sample sizes}
We further benchmark the performance in terms of computation time and number of iterations to run each method. \cref{tab:sim_time_results} shows the wall clock time for the MAP-based methods. The PPF models (\ref{simmod:add_MAPTrue}, \ref{simmod:add_noCovs}, and \ref{simmod:add_MAP}) do take considerably longer (up to 8 minutes) than the baseline NMF model (\ref{simmod:add_compNMF}).
As expected, including more covariates requires more time, as we go from $L = 0$ (\ref{simmod:add_noCovs}) to $L = 5$ (\ref{simmod:add_MAPTrue}) to $L = 10$ (\ref{simmod:add_MAP}) covariates.
For all MAP methods, the computational burden is comparable for Scenarios A (independent covariates) and B (correlated covariates).

\begin{table}[th]
\caption{Computation time and number of iterations for MAP-based methods. Shown are averages across 20 randomly generated datasets, with standard deviation in parentheses beneath each entry.}
\centering
\footnotesize
\begin{adjustbox}{max width=1\textwidth,center}
\begin{tabular}{|cl|cc|cc|}
\hline
 \multicolumn{2}{|c|}{\textsc{Method}} & \multicolumn{2}{c|}{\textsc{Scenario A}} & \multicolumn{2}{c|}{\textsc{Scenario B}} \\
& & \textsc{Time (min)} & \textsc{Iterations} & \textsc{Time (min)} & \textsc{Iterations} \\
\hline
\ref{simmod:add_MAPTrue}) & MAP, true $\bx$
  & 6.2  & 447.5
  & 8.3  & 487.0 \\
& & {\scriptsize (3.9)} & {\scriptsize (242.1)}
 & {\scriptsize (6.6)} & {\scriptsize (314.0)} \\
\hline
\ref{simmod:add_compNMF}) & MAP, CompNMF
  & 0.012  & 5495.1
  & 0.014  & 4775.6 \\
& & {\scriptsize (0.008)} & {\scriptsize (3361.0)}
 & {\scriptsize (0.009)} & {\scriptsize (2623.6)} \\
\hline
\ref{simmod:add_noCovs}) & MAP, no $\bx$
  & 2.8  & 750.0
  & 3.5  & 694.0 \\
& & {\scriptsize (1.6)} & {\scriptsize (400.4)}
 & {\scriptsize (2.1)} & {\scriptsize (331.9)} \\
\hline
\ref{simmod:add_MAP}) & MAP, all $\bx$
  & 6.6  & 442.5
  & 8.3  & 468.5 \\
& & {\scriptsize (3.8)} & {\scriptsize (251.2)}
 & {\scriptsize (6.4)} & {\scriptsize (317.4)} \\
\hline
\end{tabular}
\end{adjustbox}
\label{tab:sim_time_results}
\end{table}

For the MCMC algorithm, \cref{tab:mcmc_results} shows the total computation time and effective sample sizes for $R$, $\Phi$, $B$, $\mu$, $\sigma^2$, and the log-posterior density at the sampled parameters, for \ref{simmod:add_MCMC}, \ref{simmod:add_MCMC200}, and \ref{simmod:add_MCMC500}.
ESS is calculated using the samples from the last 1500 MCMC iterations, and computation time is based on all 3000 iterations. Scenarios A (independent covariates) and B (correlated covariates) are comparable in terms of MCMC computation time and ESS, except that Scenario B has lower ESS for the coefficients $B$, presumably since correlated covariates lead to posterior correlation between entries of $B$.

\begin{table}[th]
\caption{Computation time and effective sample sizes (ESS) for MCMC. Shown are averages across 20 datasets, with standard deviations in parentheses. ESS is calculated over 1500 samples.}
\centering
\begin{adjustbox}{max width=1\textwidth,center}
\small
\begin{tabular}{|l|cc|cc|cc|cc|cc|cc|cc|}
\hline
\multicolumn{1}{|c|}{\textsc{Method}}
& \multicolumn{2}{c|}{\textsc{Time (min)}}
& \multicolumn{2}{c|}{\textsc{ESS} $R$}
& \multicolumn{2}{c|}{\textsc{ESS} $\Theta$}
& \multicolumn{2}{c|}{\textsc{ESS} $B$}
& \multicolumn{2}{c|}{\textsc{ESS} $\mu$}
& \multicolumn{2}{c|}{\textsc{ESS} $\sigma^2$}
& \multicolumn{2}{c|}{\textsc{ESS} logPost}\\
& \textsc{a} & \textsc{b}
& \textsc{a} & \textsc{b}
& \textsc{a} & \textsc{b}
& \textsc{a} & \textsc{b}
& \textsc{a} & \textsc{b}
& \textsc{a} & \textsc{b}
& \textsc{a} & \textsc{b} \\
\hline
\ref{simmod:add_MCMC}) MCMC,
& 18 & 19 & 397 & 400 & 636 & 645 & 238 & 251 & 1403 & 1398 & 1488 & 1410 & 181 & 159 \\
all $\bx$, $\Delta_b=100$
& {\scriptsize (2)} & {\scriptsize (3)} & {\scriptsize (38)} & {\scriptsize (29)} & {\scriptsize (50)} & {\scriptsize (54)} & {\scriptsize (19)} & {\scriptsize (25)} & {\scriptsize (65)} & {\scriptsize (76)} & {\scriptsize (56)} & {\scriptsize (40)} & {\scriptsize (78)} & {\scriptsize (45)} \\
\hline
\ref{simmod:add_MCMC200}) MCMC,
& 15 & 12 & 396 & 399 & 636 & 642 & 236 & 250 & 1392 & 1418 & 1474 & 1408 & 155 & 152 \\
all $\bx$, $\Delta_b=200$
& {\scriptsize (2)} & {\scriptsize (3)} & {\scriptsize (37)} & {\scriptsize (30)} & {\scriptsize (54)} & {\scriptsize (56)} & {\scriptsize (23)} & {\scriptsize (25)} & {\scriptsize (63)} & {\scriptsize (57)} & {\scriptsize (51)} & {\scriptsize (53)} & {\scriptsize (49)} & {\scriptsize (20)} \\
\hline
\ref{simmod:add_MCMC500}) MCMC,
& 13 & 9 & 396 & 396 & 633 & 638 & 234 & 248 & 1419 & 1404 & 1459 & 1392 & 149 & 165 \\
all $\bx$, $\Delta_b=500$
& {\scriptsize (2)} & {\scriptsize (3)} & {\scriptsize (38)} & {\scriptsize (29)} & {\scriptsize (55)} & {\scriptsize (54)} & {\scriptsize (21)} & {\scriptsize (28)} & {\scriptsize (69)} & {\scriptsize (57)} & {\scriptsize (36)} & {\scriptsize (69)} & {\scriptsize (20)} & {\scriptsize (40)} \\
\hline

\end{tabular}
\end{adjustbox}
\label{tab:mcmc_results}
\end{table}

\subsection{Simulation II - Evaluation of PPF in misspecified settings}\label{subsec:Simulation_misspec}

\subsubsection{Description of misspecification scenarios}
We now benchmark PPF performance under misspecification. In particular, our model may face several sources of misspecification stemming from data complexity and underlying biology. These include i) patient-specific deviations from the reference epigenome; ii) \emph{kataegis} \citepSupp{Nik_zainal_2012}, which is a form of hypermutation; iii) noisy copy-number estimates; and iv) misspecified mutation opportunities for the trinucleotide channels.

PPF assumes that all genomic covariates come from a reference epigenome that is shared across patients. This is because most public data lack paired whole-genome sequencing and epigenetic tracks. Hence, we do not account for patient-specific covariates when inferring mutational signatures, activities, and regression coefficients. To see how deviations from the common epigenome affect the results, we simulate several scenarios with an increasing degree of misspecification. Each scenario adds one violation on top of the previous one, so that Scenario 6 is simultaneously affected by all of them.

\begin{enumerate}
    \item \textbf{Scenario 0} -- \underline{correctly specified model}. We simulate data from a PPF model with $J = 100$ patients, and  $L_{\mathrm{true}} = 5$, generating covariates, activities, regression coefficients, and mutation occurrences using the same approach as in \cref{subsec:sim_data} in the correlated-covariate setting. Instead of randomly simulating signatures $R^0$, we use $K_0 = 8$ true signatures from COSMIC: four sparse ones (SBS1, SBS2, SBS13, and SBS18), and four flat ones (SBS3, SBS5, SBS8, and SBS44). We chose this because, while sparse signatures usually carry distinctive signals, flat ones may be confused with one another because their mass is less concentrated over the simplex. Moreover, unlike \cref{subsec:sim_data}, we also simulate copy number deletions: for each segment $s = 1, \ldots, S_j$ we sample $V_{js}$ uniformly from $\{0.01, 0.5, 1\}$ with probability $0.2$, and $V_{js} = 2 + \mathrm{NegBin}(1, 10)$ otherwise, independently across segments and patients.
    \item \textbf{Scenarios 1, 2, and 3} -- \underline{noisy patient-specific covariates}. We simulate data using patient-specific epigenomes. Let $\bx_m$ denote the reference covariates in bin $[\tau_m, \tau_{m+1})$. For each patient $j$, we draw $\boldsymbol{\epsilon}_{jm} \sim \mathrm{N}(\mathbf{0}, v^2 I_L)$ independently for $m = 1, \ldots, M$, set $\tilde{\bx}_{jm} = \bx_m + \boldsymbol{\epsilon}_{jm}$, and let $\bx_j(t) = \{\tilde{\bx}_{jm} - \mathrm{mean}(\tilde{\bx}_j)\}/\mathrm{sd}(\tilde{\bx}_j)$ for $t \in [\tau_m, \tau_{m+1})$, where $\mathrm{mean}(\cdot)$ and $\mathrm{sd}(\cdot)$ are computed coordinate-wise over the $M$ bins $\tilde{\bx}_{j1}, \ldots, \tilde{\bx}_{jM}$, as in \cref{subsec:sim_data}. Hence, the patient-specific covariates are standardized noisy versions of the reference epigenome. Mutations for patient $j$ are then generated with $\bx_j(t)$ in place of $\bx(t)$, while PPF is given only the reference track $\bx(t)$. We consider $v^2 = 0.25$ (Scenario 1), $v^2 = 1$ (Scenario 2), and $v^2 = 4$ (Scenario 3).
    \item \textbf{Scenario 4} -- \underline{hypermutated regions}. We extend Scenario 3 by adding hypermutated regions for signatures SBS2 and SBS13 for 20\% of the patients in the data. This resembles the \emph{kataegis} phenomenon typical of APOBEC-induced dysregulations \citepSupp{Nik_zainal_2012}. Let $\mathcal{J}_{\mathrm{h}} \subset \{1, \ldots, J\}$ denote this subset, drawn uniformly at random without replacement, and let $\mathcal{K}_{\mathrm{h}} = \{\text{SBS2}, \text{SBS13}\}$. For each of these patients, we randomly select 10 bins (5 for SBS2, 5 for SBS13); that is, for every $j \in \mathcal{J}_{\mathrm{h}}$ and $k \in \mathcal{K}_{\mathrm{h}}$ we draw $H = 5$ bin indices $m_{jk1}, \ldots, m_{jkH}$ uniformly without replacement from $\{1, \ldots, M\}$, independently across patients and signatures and independently of the covariates and of the copy numbers. We then define the true intensity function for patient $j$ and signature $k$ as
    $$
      \tilde{\lambda}_{jk}(t) \;=\; \lambda_{jk}(t) \;+\; \sum_{h = 1}^{H} \varrho \, \mathds{1}\{t \in [\tau_{m_{jkh}}, \tau_{m_{jkh}+1})\},
      \qquad
      \lambda_{jk}(t) \;=\; \phi^0_{kj}\,\tfrac{1}{2} c_j(t) \, e^{{\bbeta^0_k}^{\top} \mathbf{x}_j(t)},
    $$
    where $\varrho = \varsigma / (\tau_{m+1} - \tau_m)$ with $\varsigma = 30$, so that each hotspot bin carries $30$ excess mutations in expectation, and where $\tilde{\lambda}_{jk}(t) = \lambda_{jk}(t)$ whenever $j \notin \mathcal{J}_{\mathrm{h}}$ or $k \notin \mathcal{K}_{\mathrm{h}}$. Then, each excess mutation is assigned to channel $i$ with probability $r^0_{ik}$, with a position drawn uniformly within its bin. Thus, every affected patient carries $300$ excess mutations in expectation, concentrated in $10$ of the $M = 20{,}000$ bins.
    \item \textbf{Scenario 5} -- \underline{noisy copy number}. We extend Scenario 4 by adding noise to the true patient-specific copy numbers. This mimics the output of an imperfect copy-number caller. Specifically, we set $\hat{c}_{jm} = \max\{ c_{jm} + \eta_{jm}, 0.01\},$ and $\eta_{jm} \sim \mathrm{N}(0, 0.25)$ independently across $j$ and $m$. We use $c_{jm}$ to generate mutations, and $\hat{c}_{jm}$ during inference.
    \item \textbf{Scenario 6} -- \underline{misspecified opportunity counts}. Trinucleotide motifs have unequal frequencies along the genome. To account for these differences, we extract the proportions $p_1, \ldots, p_I$ of the trinucleotide context associated with each of the $I = 96$ channels, counted on both strands over chromosomes 1--22, X and Y of genome \texttt{hg19}, and normalize them such that $\sum_{i = 1}^{I} p_i = 1$. Mutation opportunities are simulated as $\mathbf{w} \sim \mathrm{Dir}(\kappa p_1, \ldots, \kappa p_I)$ with $\kappa = 10^5$ and set $o_i = I w_i$, so that $\sum_{i = 1}^{I} o_i = I$. The intensity function in this scenario for channel $i$ is then equal to $o_i \sum_{k = 1}^{K} r^0_{ik} \tilde{\lambda}_{jk}(t)$, while every model is fitted under the implicit assumption that $o_i=1$. Thus, this scenario evaluates how incorrect opportunity counts bias the parameter estimates.
\end{enumerate}

 We again use $T = 2{,}000{,}000$, split into bins of size 100. To evaluate the model out-of-sample, we fit PPF and the alternative models on the first $16{,}000$ bins and hold out the last $4{,}000$ for evaluation. We simulate 20 replicate datasets per scenario. The next subsection describes which models are used in the comparison.

\subsubsection{Models and metrics of comparison}
For each scenario and each simulated dataset, we consider two different groups of models.
\begin{enumerate}
    \item \underline{\emph{De novo}} models. We compare PPF, fitted by both MAP and MCMC, against CompNMF and SignatureAnalyzer. We set $K=15$ in all cases. The goal is to determine how the misspecification affects retrieval of all model parameters, including the number of signatures and the signatures themselves, as well as estimates of the total mutation occurrence along the simulated genome, both in- and out-of-sample.
    \item \underline{Fixed-signatures models}. In this case, we keep the signature matrix $R$ fixed to a set of $K = 13$ signatures. These consist of the $K_0 = 8$ true ones and $5$ additional signatures not used to generate the data (SBS6, SBS20, SBS26, SBS30, and SBS40a), which act as distractors. We compare PPF (MAP and MCMC) against the baseline CompNMF. Our goal here is to assess how calibrated the assignment probabilities are under model misspecification.
\end{enumerate}
We estimate PPF using the reference $\mathbf{x}(t)$ and the observed copy-number track, while the patient-specific covariates $\mathbf{x}_j(t)$ or the true copy number are never observed. Since CompNMF and SignatureAnalyzer carry no covariates, we extrapolate predictions on the held-out bins using copy numbers only, with the per-copy baseline normalized over the training bins.

We use the following evaluation metrics. In the \emph{de novo} use case, we report the estimated number of signatures $\hat{K}$, the $F_1$ score for signature recovery, the RMSEs of $\hat{R}$, $\hat{\Theta}$, and $\hat{B}$ defined previously, the proportion of non-null regression coefficients whose sign is correctly recovered, and the root mean squared error between the estimated intensity $\hat{\Lambda}$ and the observed counts. All RMSEs are calculated by first matching estimated signatures to the true signatures using the Hungarian algorithm. In the fixed-signatures model, we evaluate calibration of attribution probabilities as follows.  Let $\hat{P}_{nk}$ be the estimated posterior probability that mutation $n$ was generated by signature $k$, and let $k_n$ be the signature that actually generated it. Letting $\hat{k}_n = \arg\max_{k} \hat{P}_{nk}$ and $\hat{p}_n = \hat{P}_{n \hat{k}_n}$ denote the predicted signature and associated probability, respectively, we define expected calibration error (ECE) score as
$$
  \mathrm{ECE} = \sum_{b = 1}^{B} \frac{|\mathcal{N}_b|}{N} \big| \mathrm{acc}(b) - \mathrm{conf}(b) \big|,
$$
which is obtained by partitioning the interval $[0, 1]$ into $10$ equally spaced bins and letting $\mathcal{N}_b$ collect the mutations whose confidence $\hat{p}_n$ falls in bin $b$, with
$$
  \mathrm{acc}(b) = \frac{1}{|\mathcal{N}_b|}\sum_{n \in \mathcal{N}_b} \mathds{1}\{\hat{k}_n = k_n\},
  \qquad
  \mathrm{conf}(b) = \frac{1}{|\mathcal{N}_b|}\sum_{n \in \mathcal{N}_b} \hat{p}_n.
$$
In a perfectly calibrated model, an attribution accuracy of $80\%$ should have an attribution probability around $0.8$, and we should expect $\mathrm{ECE} \approx 0$. Hence, the closer ECE is to zero, the better the calibration of the model probabilities.

\subsubsection{Results}

\begin{figure}
    \centering
    \includegraphics[width=\linewidth]{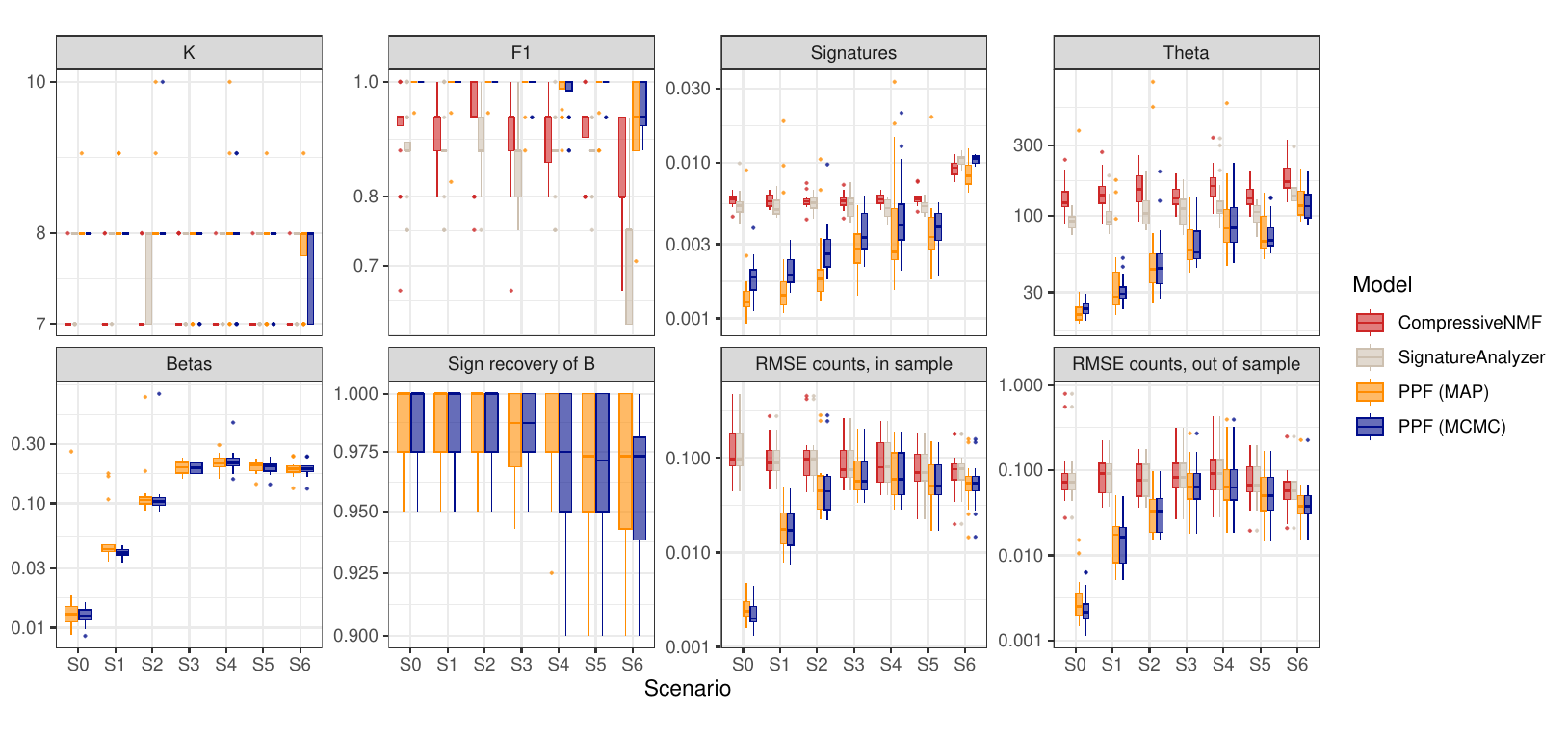}
    \caption{Results of the simulation with model misspecification. Boxplots show values from 20 randomly generated datasets across all scenarios. The panels display the log RMSE of $(\hat{R}, R_0)$, $(\hat{\Theta}, \Theta_0)$, $(\hat{B}, B_0)$, and $(\hat{\Lambda}, N)$ in- and out-of-sample, where $N$ indicates the mutation counts along the genome.  Panel ``K'' shows the estimated number of signatures across the 20 replicates. Panel ``F1'' shows the $F_1$ score. Panel ``Sign recovery of $B$'' calculates the fraction of coefficients where the sign agrees with the truth in each dataset.}
    \label{fig:Recovery_misspec}
\end{figure}
\begin{figure}[h]
    \centering
    \includegraphics[width=\linewidth]{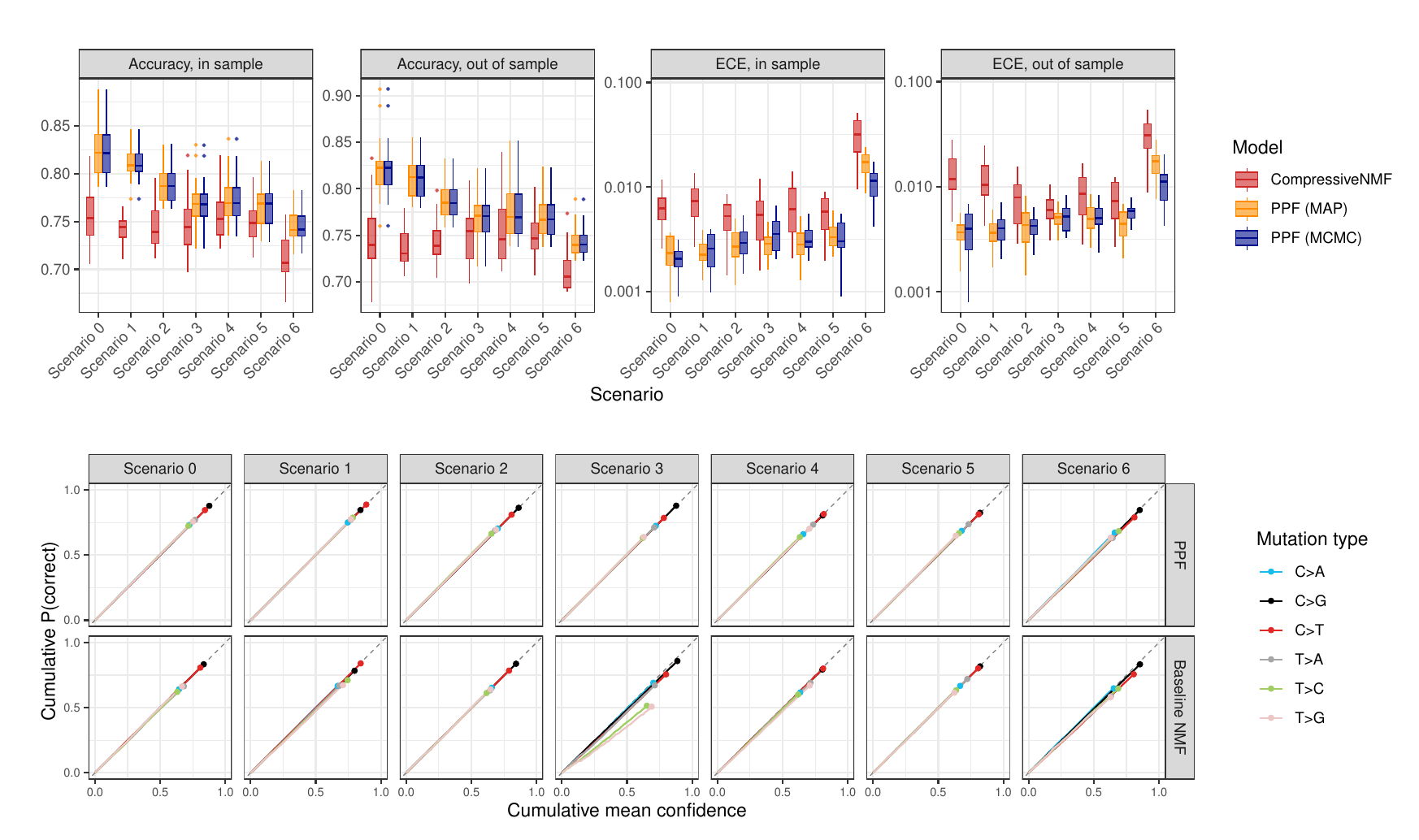}
    \caption{Top panels: accuracy and ECE score in- and out-of-sample for CompNMF and PPF when signatures are held fixed to $K = 13$ (the 8 true signatures and 5 additional redundant signatures). Bottom panels: calibration plots of PPF and CompNMF for one single dataset in each scenario, across the six aggregate mutation types. }
    \label{fig:calibration_misspec}
\end{figure}

Figures~\ref{fig:Recovery_misspec} and \ref{fig:calibration_misspec} illustrate the results across misspecification scenarios. Our findings can be summarized as follows.

\begin{itemize}
    \item \textbf{Performance of PPF vs NMF}. Figure~\ref{fig:Recovery_misspec} displays the results in the \emph{de novo} cases. Specifically, the number of estimated signatures $\hat{K}$ is around the true number $K_0 = 8$ in both the PPF solutions and SignatureAnalyzer, while CompNMF estimates a median of 7 signatures. In turn, SignatureAnalyzer has the worst $F_1$ score. While CompNMF's $F_1$ is slightly higher, it still remains lower than PPF, indicating that covariates help retrieve the correct number of signatures and profiles with higher cosine similarity than baseline NMF, even under misspecification. For the signature matrix $\hat{R}$ and the total activities $\hat{\Theta}$, CompNMF and SignatureAnalyzer each have relatively constant RMSEs across scenarios. This is expected because misspecification does not affect the total number of mutations by type, only their distribution along the genome. In contrast, while PPF performs better across all scenarios (with MAP performing slightly better than MCMC), its performance worsens as misspecification increases. As for the matrix of regression coefficients, $\hat{B}$, we see that the performance of PPF degrades with increasing noise for the patient-specific covariates (Scenarios 0 to 3), but the introduction of hypermutations (Scenario 4), noisy copy numbers (Scenario 5), and unequal mutation opportunities (Scenario 6) has a limited effect. This suggests robustness to all forms of misspecification, except patient-specific deviations, when estimating $\bbeta_k$. Moreover, the signs of the estimated coefficients generally agree with the truth with near-one probability, indicating that even under noisy covariates our model can infer the correct direction of the epigenetic effects in each signature. Finally, PPF outperforms CompNMF and SignatureAnalyzer (adjusted for copy numbers) when inferring mutational burden along the genome, indicating that covariates improve fit.
    \item \textbf{Calibration of PPF probabilities}. Figure~\ref{fig:calibration_misspec} displays the calibration and the accuracy for the mutation attribution in the fixed signature case. PPF is consistently more accurate and shows better calibration than CompNMF across all scenarios. In particular, the accuracy is highest in Scenario 0 (correct specification), and lowest in Scenario 6 (highest misspecification). Second, in- and out-of-sample calibration is relatively stable across Scenarios 0 to 5 and increases only in Scenario 6, when channels have different mutation opportunities. This is expected, since incorrect opportunities naturally translate into slightly worse assignment probabilities. Third, PPF and CompNMF are generally well calibrated across mutation types, while accuracy decreases as the degree of misspecification increases. As for SignatureAnalyzer, we do not include it in this comparison because \texttt{sig\_auto\_extract()} allows for \emph{de novo} analysis only.
\end{itemize}

We conclude by reporting computation times, convergence details, and effective sample sizes for all models in Tables~\ref{tab:sim_time_misspec} and \ref{tab:mcmc_misspec}. We note that MCMC chain mixing degrades slightly as misspecification increases.

\begin{table}[th]
\caption{Computation time and number of iterations for MAP-based methods under misspecification for the \emph{de novo} methods. Shown are averages across 20 datasets per scenario, with standard deviations in parentheses. SignatureAnalyzer reports no iteration count.}
\centering
\footnotesize
\begin{adjustbox}{max width=0.9\textwidth,center}
\begin{tabular}{|cl|cc|cc|cc|}
\hline
\multicolumn{2}{|c|}{\textsc{Scenario}} & \multicolumn{2}{c|}{\textsc{CompNMF}} & \multicolumn{2}{c|}{\textsc{SignatureAnalyzer}} & \multicolumn{2}{c|}{\textsc{PPF, MAP}} \\
& & \textsc{Time (min)} & \textsc{Iter.} & \textsc{Time (min)} & \textsc{Iter.} & \textsc{Time (min)} & \textsc{Iter.} \\
\hline
0. & correctly specified & 0.012 & 5676.8 & 1.4 & --- & 7.9 & 314.5 \\
 & & {\scriptsize (0.005)} & {\scriptsize (2693.0)} & {\scriptsize (0.4)} & --- & {\scriptsize (4.6)} & {\scriptsize (122.0)} \\
\hline
1. & $v^2 = 0.25$ & 0.022 & 5848.3 & 2.1 & --- & 7.0 & 293.5 \\
 & & {\scriptsize (0.016)} & {\scriptsize (3117.4)} & {\scriptsize (0.5)} & --- & {\scriptsize (3.1)} & {\scriptsize (104.6)} \\
\hline
2. & $v^2 = 1$ & 0.043 & 8306.8 & 1.9 & --- & 7.8 & 319.5 \\
 & & {\scriptsize (0.092)} & {\scriptsize (15620.8)} & {\scriptsize (0.8)} & --- & {\scriptsize (4.0)} & {\scriptsize (122.0)} \\
\hline
3. & $v^2 = 4$ & 0.028 & 7982.2 & 2.0 & --- & 8.4 & 362.5 \\
 & & {\scriptsize (0.017)} & {\scriptsize (6587.2)} & {\scriptsize (0.8)} & --- & {\scriptsize (3.4)} & {\scriptsize (100.0)} \\
\hline
4. & $+$ hotspots & 0.030 & 7453.4 & 2.1 & --- & 9.4 & 380.5 \\
 & & {\scriptsize (0.023)} & {\scriptsize (5399.7)} & {\scriptsize (0.5)} & --- & {\scriptsize (3.9)} & {\scriptsize (148.0)} \\
\hline
5. & $+$ CN noise & 0.019 & 6156.6 & 1.9 & --- & 7.4 & 332.0 \\
 & & {\scriptsize (0.014)} & {\scriptsize (4885.4)} & {\scriptsize (0.6)} & --- & {\scriptsize (2.6)} & {\scriptsize (101.5)} \\
\hline
6. & $+$ opportunity & 0.022 & 7335.7 & 2.4 & --- & 8.3 & 371.5 \\
 & & {\scriptsize (0.014)} & {\scriptsize (3298.4)} & {\scriptsize (0.8)} & --- & {\scriptsize (3.6)} & {\scriptsize (143.9)} \\
\hline
\end{tabular}
\end{adjustbox}
\label{tab:sim_time_misspec}
\end{table}

\begin{table}[th]
\caption{Computation time and effective sample sizes (ESS) of the PPF sampler under misspecification, for the $\emph{de novo}$ methods. Averages are computed across 20 datasets per scenario, with standard deviations in parentheses. ESS for $R$, $\Theta$, $B$, $\mu$, and $\sigma^2$ is calculated over the last 1500 post-burn-in; the log-posterior is recorded once every ten iterations, so its ESS is calculated over the 150 recorded values.}
\centering
\small
\begin{adjustbox}{max width=0.9\textwidth,center}
\begin{tabular}{|cl|c|cccccc|}
\hline
\multicolumn{2}{|c|}{\textsc{Scenario}} & \textsc{Time (min)} & \textsc{ESS} $R$ & \textsc{ESS} $\Theta$ & \textsc{ESS} $B$ & \textsc{ESS} $\mu$ & \textsc{ESS} $\sigma^2$ & \textsc{ESS} logPost \\
\hline
0. & correctly specified & 32 & 329 & 466 & 164 & 1049 & 1502 & 157 \\
 & & {\scriptsize (10)} & {\scriptsize (46)} & {\scriptsize (45)} & {\scriptsize (19)} & {\scriptsize (167)} & {\scriptsize (79)} & {\scriptsize (28)} \\
\hline
1. & $v^2 = 0.25$ & 31 & 300 & 431 & 154 & 1020 & 1479 & 165 \\
 & & {\scriptsize (5)} & {\scriptsize (33)} & {\scriptsize (36)} & {\scriptsize (10)} & {\scriptsize (157)} & {\scriptsize (45)} & {\scriptsize (32)} \\
\hline
2. & $v^2 = 1$ & 32 & 243 & 371 & 140 & 875 & 1458 & 150 \\
 & & {\scriptsize (6)} & {\scriptsize (23)} & {\scriptsize (31)} & {\scriptsize (11)} & {\scriptsize (121)} & {\scriptsize (71)} & {\scriptsize (38)} \\
\hline
3. & $v^2 = 4$ & 30 & 185 & 304 & 137 & 703 & 1503 & 175 \\
 & & {\scriptsize (5)} & {\scriptsize (18)} & {\scriptsize (35)} & {\scriptsize (19)} & {\scriptsize (168)} & {\scriptsize (28)} & {\scriptsize (154)} \\
\hline
4. & $+$ hotspots & 33 & 186 & 297 & 131 & 687 & 1464 & 133 \\
 & & {\scriptsize (5)} & {\scriptsize (28)} & {\scriptsize (34)} & {\scriptsize (15)} & {\scriptsize (103)} & {\scriptsize (56)} & {\scriptsize (56)} \\
\hline
5. & $+$ CN noise & 29 & 171 & 292 & 131 & 620 & 1506 & 142 \\
 & & {\scriptsize (4)} & {\scriptsize (16)} & {\scriptsize (34)} & {\scriptsize (14)} & {\scriptsize (141)} & {\scriptsize (41)} & {\scriptsize (42)} \\
\hline
6. & $+$ opportunity & 26 & 191 & 273 & 125 & 568 & 1500 & 146 \\
 & & {\scriptsize (6)} & {\scriptsize (15)} & {\scriptsize (36)} & {\scriptsize (12)} & {\scriptsize (126)} & {\scriptsize (25)} & {\scriptsize (48)} \\
\hline
\end{tabular}
\end{adjustbox}
\label{tab:mcmc_misspec}
\end{table}

\clearpage

\section{Extended analyses on breast cancer data}\label{sec:add_breast}

In this section, we present additional results and new analyses on the breast cancer application in \cref{sec:application}.
Section~\ref{subsec:additional_results} reports more details on the main application, including comparisons with SignatureAnalyzer, MCMC convergence diagnostics, and model checking. Section~\ref{sec:sensitivity_analysis} performs sensitivity analysis for the choice of the number of signatures $K$, and the values of $c_0$ and $d_0$ in the prior for the variance of the regression coefficient $\sigma^2_k$. Section~\ref{sec:breast80} presents an analysis of a novel cohort of breast cancer data, with the goal of assessing the replicability of our findings. Section~\ref{sec:tensorsignatures} compares the results of PPF against \texttt{TensorSignatures} \citepSupp{TensorSignatures_2021}, using categorical covariates representing chromatin states \citepSupp{Ernst2017_ChromHMM}.

\subsection{Additional results on the main application}\label{subsec:additional_results}

\subsubsection{Initialization details in the \emph{de novo} application}\label{subsec:details_denovo}
We run Algorithm \ref{algo:MAP_rules} from three different starting points with $K = 12$, generated at random
from the prior, stopping the updates when the relative difference in log-posterior
falls below $10^{-7}$. These runs took $1{,}200$, $1{,}890$, and $1{,}000$ iterations to
complete, over $303$, $336$, and $174$ minutes respectively (an average of 12 seconds per
iteration), run sequentially on an AMD Ryzen 9 3900 12-core dedicated server with 128GB of
memory, running Ubuntu 20.04, R version 4.3.1 linked to OpenBLAS. All three runs detect very
similar sets of signatures: the average cosine similarity between each pair of solutions is
$0.97$ after matching columns via the Hungarian algorithm. The estimated number of signatures
$\hat{K}$ (the number for which $\hat{\mu}^{\textsc{map}}_k > 5\varepsilon$) was 12, 10, and 11,
with log-posteriors $-12{,}346{,}824$, $-12{,}343{,}017$ and $-12{,}345{,}000$, respectively.
We selected the solution with the highest log-posterior density ($\hat{K} = 10$), which corresponded also to the run that took the most iterations.

\subsubsection{Comparison with \emph{de novo} Poisson NMF methods}\label{subsec:PPF_vs_SA}
\begin{figure}[t]
    \centering
    \includegraphics[width=\linewidth]{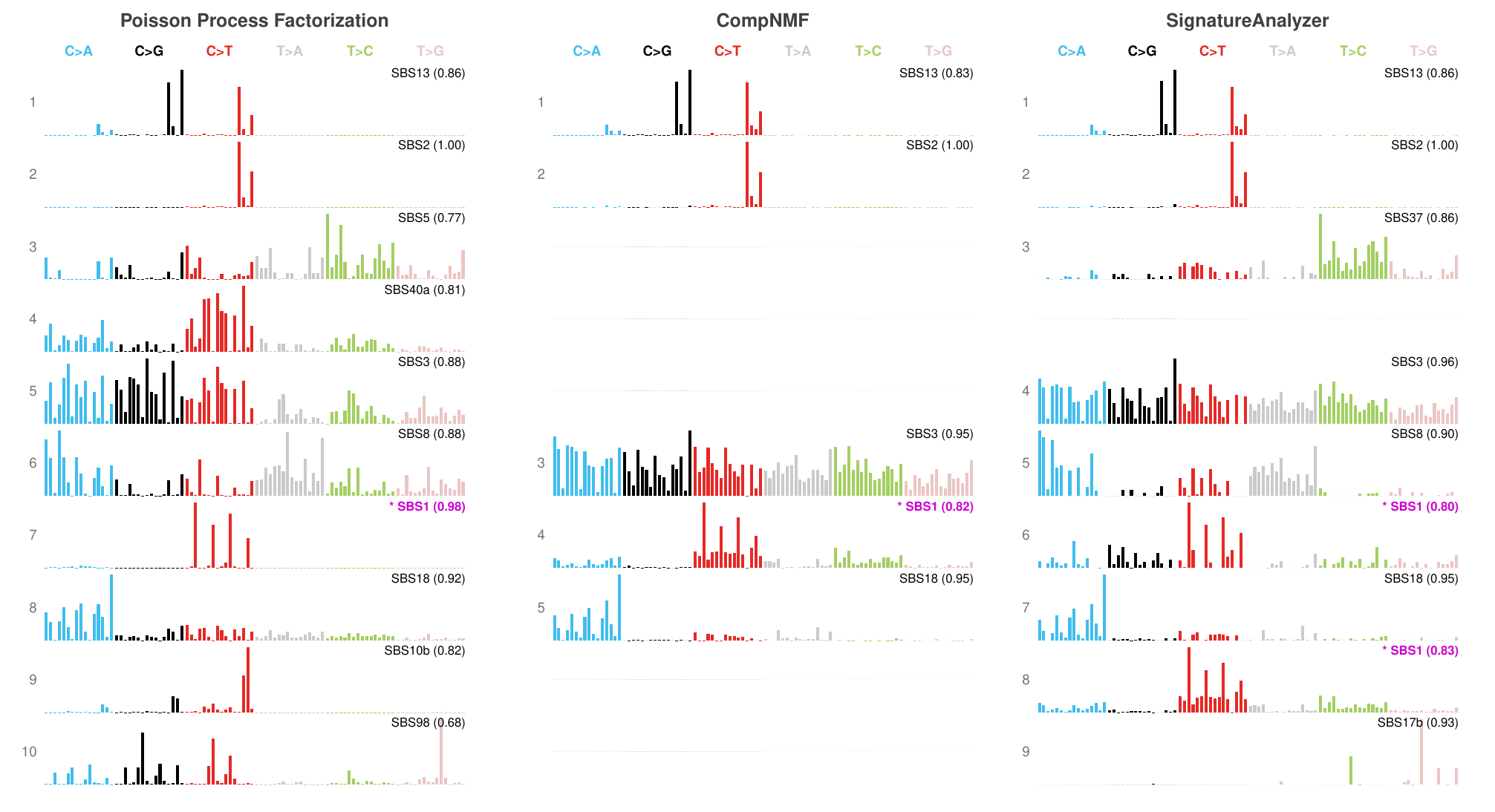}
    \caption{Estimated signatures in ICGC breast adenocarcinoma. Left: results from Poisson process factorization. Center: results from baseline Poisson NMF (CompNMF) using \cref{algo:compressive_nmf_map}. Right: results from \textsc{SignatureAnalyzer}.
    We highlight SBS1 with an asterisk in magenta to highlight the similarities and ensure a better comparison.
    }
    \label{fig:SigComparison}
\end{figure}

Since both PPF and NMF infer $R$ and $\Theta$, we can meaningfully compare these parameters to evaluate whether using covariates provides additional benefit when inferring signatures (Question Q1 from \cref{sec:breastCancer}).
We compare this to the solutions from CompNMF \citepSupp{zito2024compressive} and SignatureAnalyzer \citepSupp{kim2016somatic} (\texttt{sig\_auto\_extract} in \citetSupp{sigminer_2020}), both of which also automatically select the number of signatures.  To evaluate the fit in terms of aggregated counts, we compute the reconstruction error $\textsc{rmse}(\hat{\Lambda}, N)$ where $N \in \mathds{R}^{I\times J}$ is the matrix of total mutations from \cref{subsec:BaselinePoisson}
and $\hat{\Lambda}$ is the expected value of this matrix under the estimated model, computed using \cref{eq:expectation} with $A = [0, T)$. PPF, CompNMF, and SignatureAnalyzer have reconstruction errors $14.01$, $14.63$, and $13.37$, respectively, and estimated numbers of signatures 10, 5, and 9, respectively. Thus, PPF has comparable performance to methods designed to directly reconstruct the total counts.
The signatures under all three methods are displayed in \cref{fig:SigComparison}. PPF infers a more distinctive profile for SBS1, whereas SignatureAnalyzer splits the peaks at CpG sites across two signatures, indicating a propensity to mix SBS1 with relatively flat signatures; this behavior is known to occur under correlated exposures \citepSupp{Wu2022}. This is because methylation, the biological process generating SBS1, is included as a covariate in PPF but is unavailable to the other two methods; we defer the comparison against simulated data with known signatures to \cref{secSupp:simulations}.

\subsubsection{Diagnostics of MCMC algorithms}

\begin{figure}[t]
    \centering
    \includegraphics[width=\linewidth]{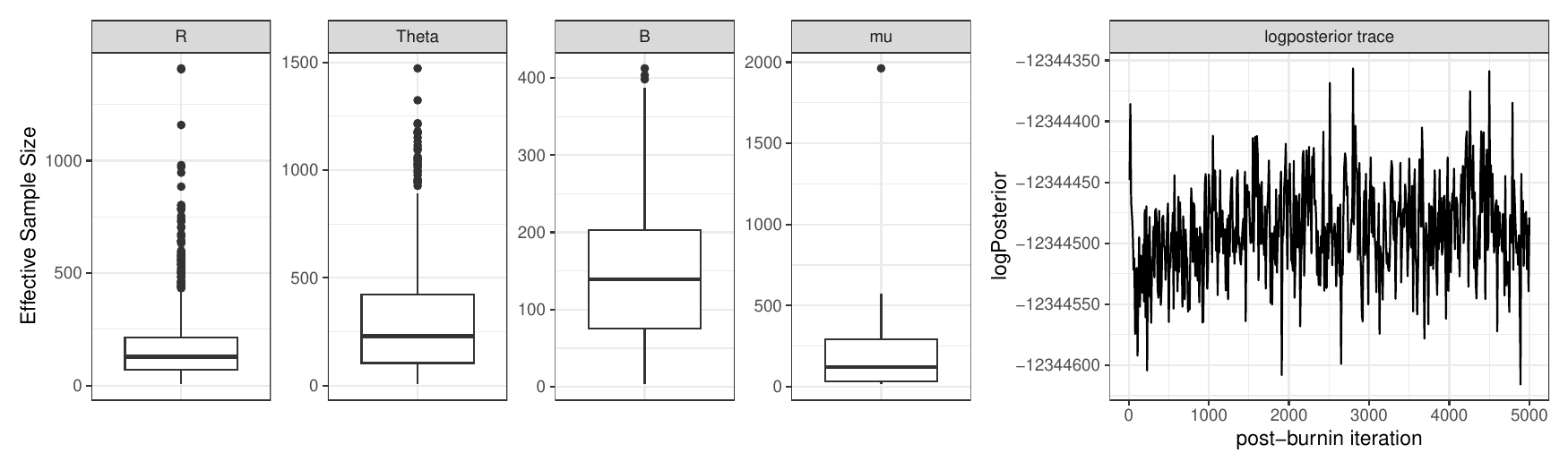}
    \caption{Left: effective sample sizes for model parameters in the \emph{de novo} analysis. Right: traceplot for the log posterior density. Values are computed over $5{,}000$ iterations after burn-in. The log posterior is evaluated every 10 iterations to speed up computation.}
    \label{fig:ESS_denovo}
\end{figure}

\begin{figure}[t]
    \centering
    \includegraphics[width=\linewidth]{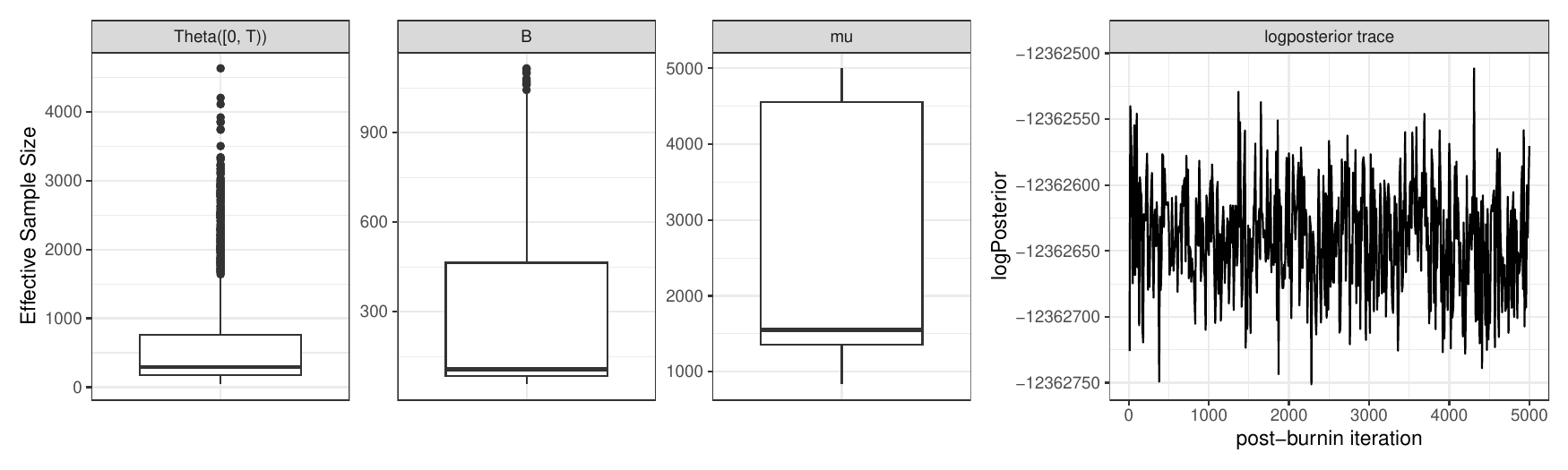}
    \caption{Left: effective sample sizes for model parameters in the fixed-signatures analysis. Right: traceplot for the log posterior. Values are computed over $5{,}000$ iterations after burn-in. The log posterior is evaluated every 10 iterations to speed up computation.}
    \label{fig:ESS_Fixed}
\end{figure}

\cref{fig:ESS_denovo} and \cref{fig:ESS_Fixed} display the effective sample sizes for the model parameters in both the \emph{de novo} and the refitting application. Convergence appears to be acceptable, and is improved by using the adaptive elliptical slice sampling algorithm proposed by \citetSupp{marco2026adaptivegeneralizedellipticalslice}.

\subsubsection{Additional model checking results}\label{subsec:checking}

\begin{figure}[t]
    \centering
    \includegraphics[width=\linewidth]{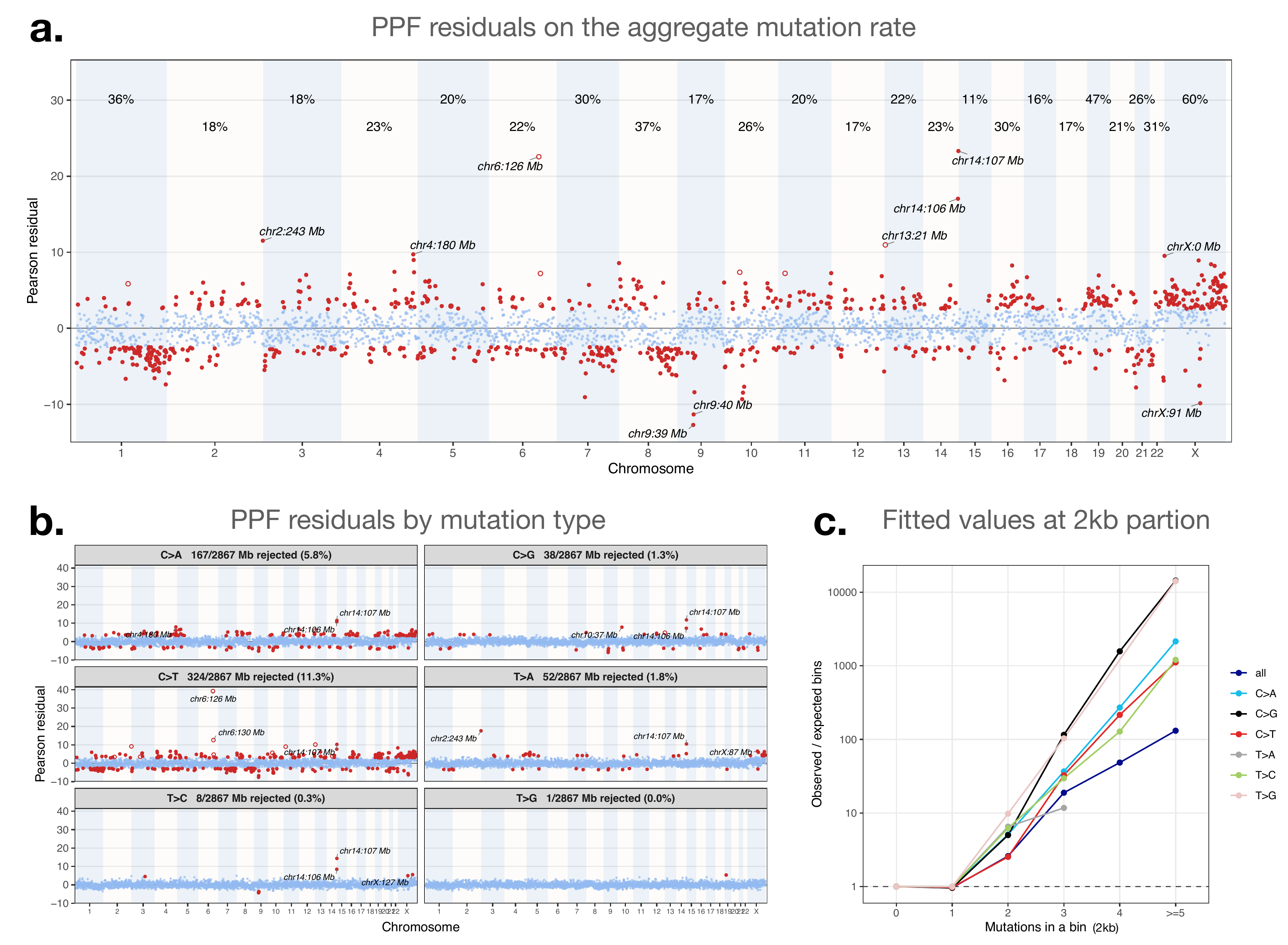}
    \caption{Model checking at the megabase and kilobase scales. \textbf{a.} Pearson residuals $(N_b - \hat{\Lambda}_b)/\sqrt{\hat{\Lambda}_b}$ for each of the $2{,}867$ megabase regions along the genome. Red points are rejected by an exact Poisson test after Benjamini--Hochberg correction at $5\%$; percentages above each chromosome indicate the fraction of megabases where the test was rejected; \textbf{b.} Pearson residuals computed separately for each macro mutation type. Panel titles display again the fraction of total megabases for which the Poisson test rejected the null. Alternating bands denote the different chromosomes; \textbf{c.} Ratio of the observed to the expected number of $2$kb bins containing a given number of mutations, by mutation type.}
    \label{fig:Resid_supplement}
\end{figure}

We now extend the model checking diagnostics presented in \cref{subsec:breast_denovo}. Our goal is to determine which regions show signs of misspecification in the aggregate mutation rate, and whether these deviations are also observed by mutation type.

All results are displayed in \cref{fig:Resid_supplement}. In particular, \cref{fig:Resid_supplement}a shows the Pearson residuals between the predicted intensity function and the observed mutations in each megabase region, calculated as $(N_b - \hat{\Lambda}_b)/\sqrt{\hat{\Lambda}_b}$ for region $b$. For each megabase region, we also calculate a two-tailed Poisson test to determine which regions show substantial deviation from the observed values. We see that $707$ of the $2{,}867$ regions ($24.7\%$) are rejected after Benjamini--Hochberg adjustment to control the false discovery rate at $5\%$, with $372$ showing underprediction and $335$ showing overprediction. The percentage of rejections per chromosome ranges from $11\%$ to $60\%$, with chromosome X having the highest rejection rate. \cref{fig:Resid_supplement}b performs the same analysis but stratified by mutation type. We see that the most misspecified channels are C$>$T ($324$ megabases, $11.3\%$) and C$>$A ($167$, $5.8\%$), while T$>$C and T$>$G show essentially no deviation. This is likely due to two main characteristics of our model: the absence of a specific component in the intensity function to capture hypermutation (often happening in C$>$T mutations and \emph{kataegis}) and the lack of explicit transcription and replication strand biases (which are often prominent in  C$>$A mutations). Finally, \cref{fig:Resid_supplement}c explores the fit by mutation type at the $2$kb resolution of the covariates. We see that the model reproduces the
number of empty and singleton bins almost exactly (ratios $1.00$ and $0.96$), but under-predicts bins carrying two or more mutations by factors of $2.6$ to $131$, confirming that the residual structure is local clustering rather than a bias in the aggregate rate.

\subsection{Sensitivity analyses}\label{sec:sensitivity_analysis}

\subsubsection{Sensitivity to model settings}\label{subsec:sens_model_setting}
We begin by elucidating the role of the hyperparameters in the priors of our model, and show how our results in the application are influenced by their choice. In particular, \cref{eq:R_prior,eq:baseline_Priors,eq:prior_beta,eq:compressive} depend on the following quantities: the upper bound on the number of signatures $K$, the prior concentration on the Dirichlet distribution $\alpha$, the quantities $c_0, d_0>0$ determining the concentration of the prior over the variance $\sigma^2_k$ of the regression coefficients $\bbeta_k$, and the values determining the compressive hyperprior over the relevance weights $\mu_k$, that is, $a$ and $\varepsilon$, which govern selection of the latent factorization rank.

\textbf{Sensitivity to the choice of $a$, $\alpha_{ik}$, and $\varepsilon$}. The behavior of the baseline compressive NMF model with respect to these quantities is extensively documented in \citetSupp{zito2024compressive}. Recall that $\mu_k$ controls the total contribution of signature $k$, that is, $\theta_{k j} = \int_{0}^T\vartheta_{k j}(t)\dt$, so that $\mathds{E}(\theta_{k j} \mid \mu_k, \bbeta_k) = \mu_k$ for every patient $j$ and every signature $k$. Hence, all the shrinkage properties discussed in \citetSupp{zito2024compressive} directly apply to PPF. In particular, $a$ determines the strength of the compressive hyperprior $\mu_k\sim \mathrm{InvGa}(a J + 1, \varepsilon a J)$, so that small values of $a$ lead to a higher number of active signatures, while large values compress the contribution of more factors to $\varepsilon$. We choose $a = 1.01$ as default. As for $\varepsilon$, we find that the model is relatively insensitive to its value as long as it remains small, typically 0.001 or 0.01. Hence, the choice is irrelevant, and all signatures for which $\hat{\mu}_k > C\varepsilon$ for any $C>1$ can be excluded from the factorization. As for $\alpha_{ik}$, we found the baseline NMF model is generally insensitive to it, and the same is expected for PPF. Setting $\alpha_{ik} > 1$ always ensures that the MAP exists. Refer to the Supplementary material of \citetSupp{zito2024compressive}.

\textbf{Sensitivity to the choice of $K$, $c_0$, and $d_0$}. We evaluate the sensitivity of the final 10 signature solutions with respect to a combination of choices of $K$ (i.e., the upper bound to the number of signatures), and the informativeness of the prior for the variance of the regression coefficients $c_0$ and $d_0$. Denoting $(K, c_0, d_0)$ as the triplet of hyperparameters, we specifically run the MAP algorithm from random starting points for the following seven settings: the reference in the main manuscript $(12, 100, 1)$, and then $(15, 100, 1)$, $(20, 100, 1)$, $(12, 10, 0.1)$, $(12, 2, 0.01)$, $(20, 10, 0.1)$ and $(20, 2, 0.01)$. Therefore, $\sigma_k^2$ has a prior mean approximately equal to 0.01, with varying degrees of concentration. We always keep the default choice of $a = 1.01$, $\alpha = 1.01$, and $\varepsilon = 0.001$, and exclude all signatures for which $\hat{\mu}_k < 5\varepsilon$.

\begin{table}[th]
\caption{Sensitivity of the \emph{de novo} fit to the maximum number of signatures $K$ and to the prior on $\sigma^2_k$. \textsc{Selected} is the number of active signatures, \textsc{Recovered} is the number of those paired with a signature of the reference solution in the main document, and \textsc{Mean cos.} is the average cosine similarity over the matched pairs; the reference row is matched against itself and so attains one by construction. $\min\mu$ is the smallest relevance weight among the retained signatures, \textsc{Iterations} is the number of MAP iterations to convergence, and \textsc{Sec./iter.} and \textsc{Hours} report the time per iteration and the total wall-clock time.}
\centering
\footnotesize
\begin{adjustbox}{max width=1\textwidth,center}
\begin{tabular}{|l|ccc|cccc|ccc|}
\hline
\multicolumn{1}{|c|}{\textsc{Scenario}} & \multicolumn{3}{c|}{\textsc{Setting}} & \multicolumn{4}{c|}{\textsc{Signatures}} & \multicolumn{3}{c|}{\textsc{Computation}} \\
& $K_{\max}$ & $c_0$ & $d_0$ & \textsc{Selected} & \textsc{Recovered} & \textsc{Mean cos.} & $\min\mu$ & \textsc{Iterations} & \textsc{Sec./iter.} & \textsc{Hours} \\
\hline
Reference & 12 & 100 & 1 & 10 & 10 & 1.000 & 34.4 & 1890 & 10.7 & 5.60 \\
\hline
\texttt{K15} & 15 & 100 & 1 & 13 & 8 & 0.944 & 13.6 & 2270 & 16.1 & 10.14 \\
\hline
\texttt{K20} & 20 & 100 & 1 & 11 & 9 & 0.976 & 29.8 & 2110 & 16.4 & 9.64 \\
\hline
\texttt{K12\_c0\_10} & 12 & 10 & 0.1 & 12 & 10 & 0.981 & 28.0 & 1140 & 9.7 & 3.08 \\
\hline
\texttt{K12\_c0\_1} & 12 & 2 & 0.01 & 12 & 10 & 0.981 & 27.9 & 1130 & 9.1 & 2.87 \\
\hline
\texttt{K20\_c0\_10} & 20 & 10 & 0.1 & 11 & 9 & 0.976 & 30.5 & 2380 & 15.9 & 10.53 \\
\hline
\texttt{K20\_c0\_1} & 20 & 2 & 0.01 & 11 & 9 & 0.967 & 24.1 & 2870 & 15.8 & 12.57 \\
\hline
\end{tabular}
\end{adjustbox}
\label{tab:sensitivity_denovo}
\end{table}

\begin{figure}[h]
    \centering
    \includegraphics[width=\linewidth]{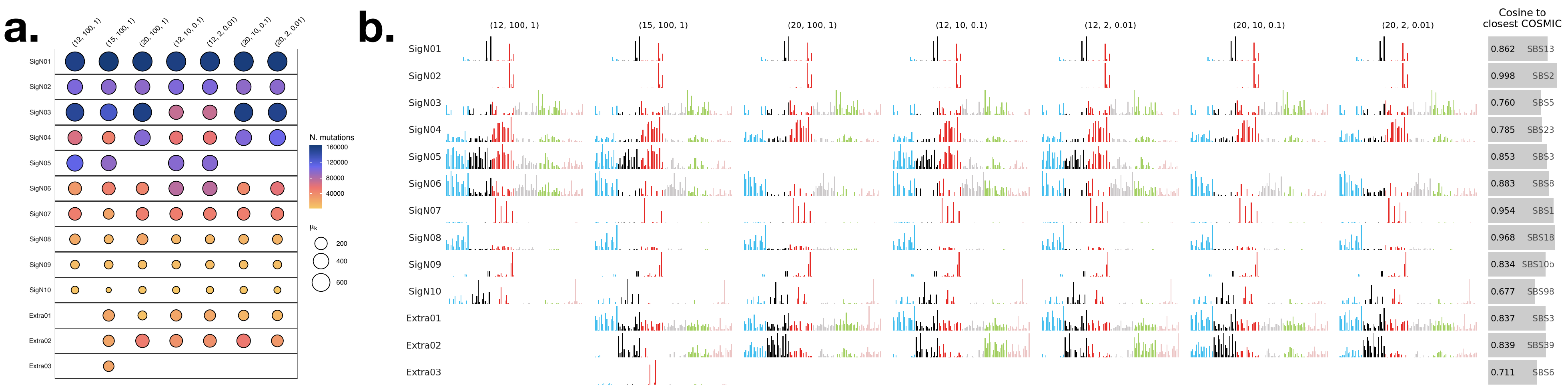}
    \caption{Solutions from the maximum-a-posteriori for varying combinations of $K$, $c_0$ and $d_0$. \textbf{a.} Relevance weights $\hat{\mu}_k$ estimated for each choice; \textbf{b.} Signatures in each solution, with the last column indicating the closest match to COSMIC and the associated cosine similarity. Signatures are matched via the Hungarian algorithm.}
    \label{fig:sensitivity_K_sigma}
\end{figure}

Table~\ref{tab:sensitivity_denovo} and Figure~\ref{fig:sensitivity_K_sigma} display the results. We see that the different settings do estimate varying numbers of signatures, ranging from 10 to 13. The solution is generally stable for the well-identified components, and most signatures have indistinguishable spectra: the average cosine similarity with the reference ranges from $0.944$ to $0.981$. Variations in the number of estimated signatures depend largely on the strong multimodality of the posterior and the random starting point. For example, with setting $(15, 100, 1)$ we have 13 signatures, but signatures SigN07 and Extra03 are an exact decoupling of SigN07 in the main manuscript. Moreover, raising $K$ to 20 instead removes the \sbs{3}-like signature SigN05, which is resolved in only four of the seven settings, while every setting other than the reference adds two flat components whose closest COSMIC matches are \sbs{3} and \sbs{39}, at cosine similarities of $0.84$. This indicates moderate sensitivity to the choices, which can be addressed by running the model from multiple random starting points and retaining the solution with the highest log-posterior. Finally, computation scales with the number of signatures: the cost per iteration rises from roughly nine to sixteen seconds as $K_{\max}$ goes from twelve to twenty, and the total computation time increases from under three hours to about twelve.

\subsubsection{Stability of estimates under covariate ablation}\label{subsec:sens_covariate_ablation}

We now turn to the pre-specified signatures application in \cref{subsec:breast_semi_sup} and illustrate how our estimates change as covariates are added sequentially to the model. Our goal is to assess how sensitive each $\bbeta_{k}$ is across different nested specifications, and how the model behaves on held-out regions of the genome when predicting the mutation rate and the signature attribution probabilities.

\begin{figure}%
    \centering
    \includegraphics[width=\linewidth]{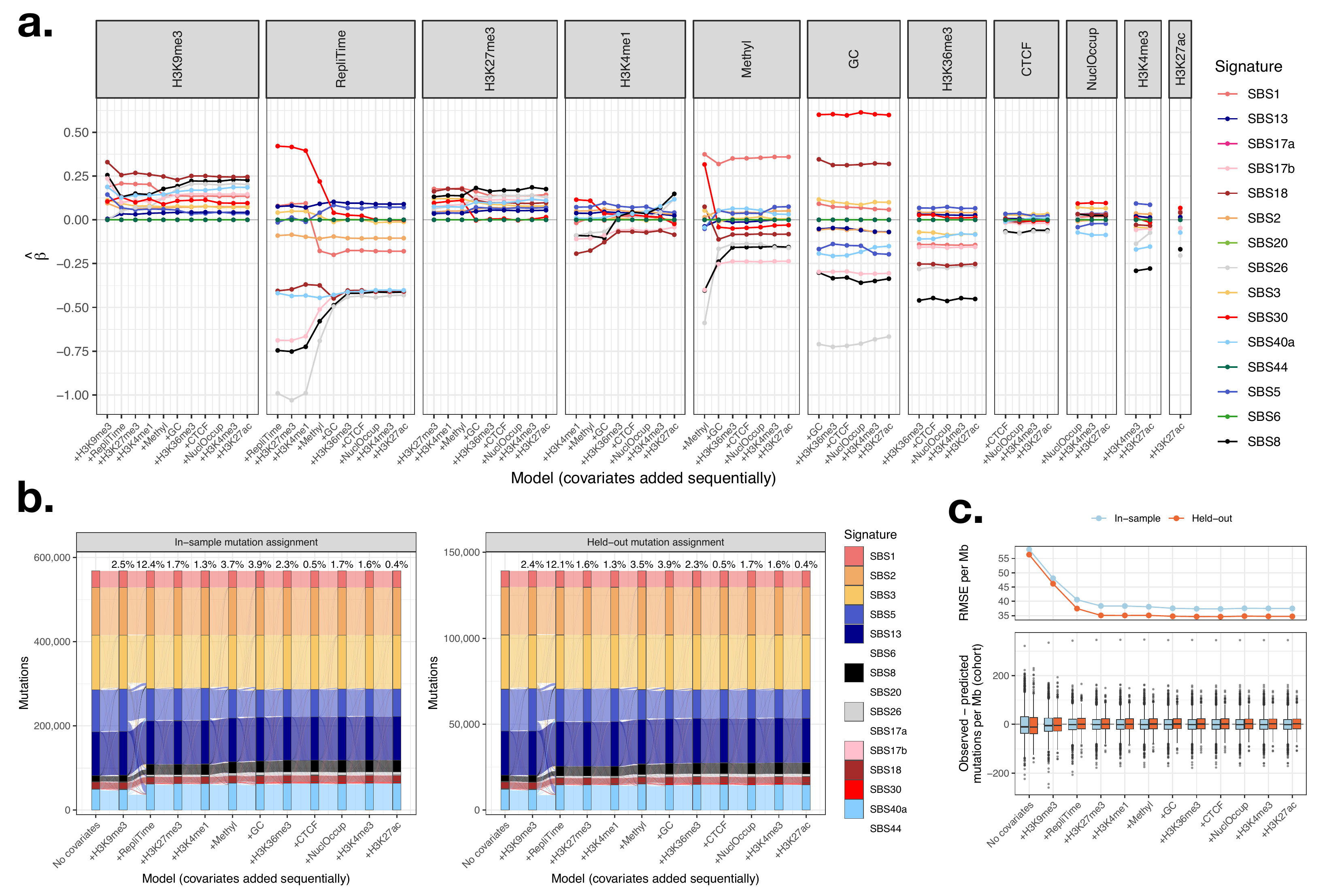}
    \caption{Model estimates under forward selection. \textbf{a.} Estimated regression coefficients $\hat{\beta}_{k\ell}$ in the sequence of models. Each panel shows the estimated coefficient for the covariate in the panel title, as covariates are sequentially added to the model. Colors denote the signature. The x-axis indicates the order in which covariates have been added. For example, the points under panel ``H3K9me3'' in correspondence to "+GC" indicate the value of $\beta_{k, \mathrm{H3K9me3}}$ when the covariate "GC content" is added to the model, after having added H3K9me3, RepliTime, H3K27me3, H3K4me1, and methylation.
    \textbf{b.} Mutation assignment to signatures in-sample (left) and on the held-out megabases (right) for the COSMIC signatures. Colors denote the mutation assignments to signatures, obtained as the argmax of \cref{eq:AssignentProbs}. Each line on the y-axis represents an individual mutation. Bands between the columns represent variations in assignment after the inclusion of a covariate, with numbers indicating the \% of total mutations changing attribution at each step. \textbf{c.} RMSE for the total counts at the Mb scale (top), and absolute difference between observed and predicted total mutations (bottom) as covariates are added, in- and out-of-sample.}
    \label{fig:Stability_betas}
\end{figure}

Our workflow is as follows. First, we bin the genome into megabase regions. We randomly sample 20\% of these regions and treat the mutations that fall into these bins as hold-out data; hence, the covariate values and the associated mutations for these held-out regions are not used when fitting the model. We estimate a baseline NMF model using Algorithm~\ref{algo:MAP_rules} without covariates, keeping the signatures fixed to those in \cref{subsec:breast_semi_sup} and using the remaining 80\% of the genome. Unlike the main analysis, we adopt a 10kb resolution for both copy numbers and genomic features, which allows quicker computations without substantially altering the results. We calculate the residuals between the estimated mutation rate from the baseline model and the total mutations in the 10kb bins, and compute their correlation with the covariates. We then estimate a new model by adding the single feature with the highest absolute correlation with the residuals. We proceed with this forward selection strategy until all $L= 11$ covariates are included.

The results of the forward selection are shown in Figure~\ref{fig:Stability_betas}. A few aspects are worth highlighting. First, we find that H3K9me3, replication timing, and H3K27me3 are the three most important predictors and have the largest impact on the predicted RMSE both in- and out-of-sample; see Figure~\ref{fig:Stability_betas}c, top panel. In particular, their inclusion nearly halves the RMSE at the Mb scale for the total mutation rate. Second, the effects of the covariates are relatively stable across models (Figure~\ref{fig:Stability_betas}a), especially for GC content, H3K36me3, and H3K9me3. The largest variability is in the replication timing estimates, which change significantly as more covariates are added. Specifically, including methylation has a very strong effect on the sign of replication timing (RepliTime) for SBS1, which switches from positive to negative. This is because methylation is the biological process generating SBS1, so including it as a covariate better isolates the effects of the other features. A similar pattern is observed for SBS30 methylation when GC content is included. Third, replication timing is the single most impactful covariate on the mutation attributions: its inclusion reassigns more than 12\% of mutations both in-sample and in the held-out megabases (Figure~\ref{fig:Stability_betas}c). In total, nearly 19\% of mutations change attribution from the baseline model with no covariates to the complete specification, with the largest migration being from SBS5 to SBS40a and SBS3, and from SBS3 to SBS8. Finally, adding replication timing allows the model to retain the MMRd signature SBS26, which is otherwise compressed out in the baseline. This suggests that adjusting for epigenetic covariates yields better extraction of signals from separate biological processes.

\subsection{Replication of pre-specified analysis in the Breast80 dataset}\label{sec:breast80}

\begin{figure}%
    \centering
    \includegraphics[width=\linewidth]{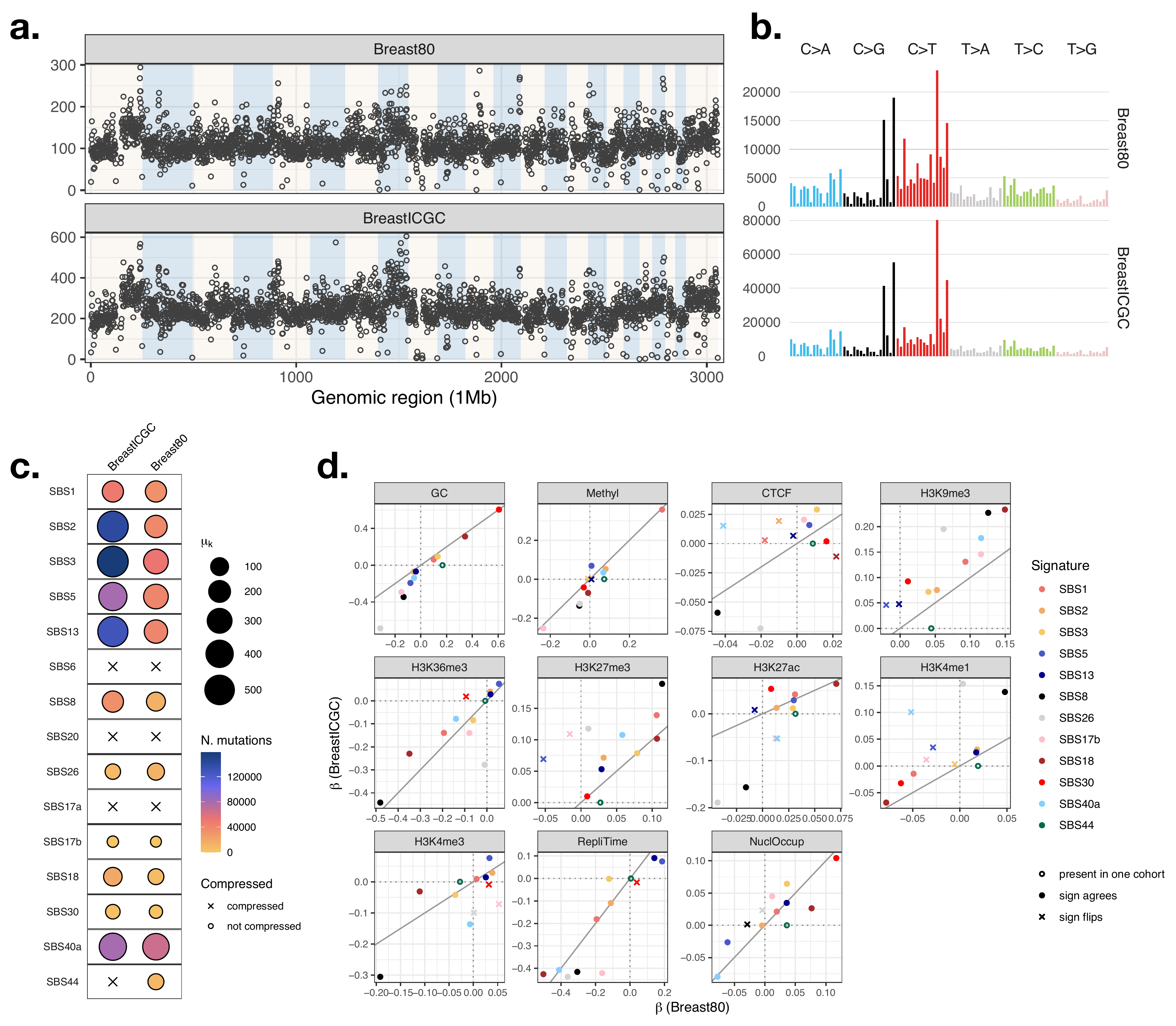}
    \caption{Replication of covariate effects in the Breast80 and BreastICGC cohorts. \textbf{a.} Total number of mutations at the Mb scale in Breast80 (top) and BreastICGC (bottom). \textbf{b.} Total number of mutations by trinucleotide type in both datasets. \textbf{c.} MAP for the relevance weights $\hat{\mu}_k$ from PPF. Colors denote the number of mutations assigned, and circle size indicates magnitude. They symbol ``$\times$'' denotes signatures excluded by the model, for which $\hat{\mu}_k\approx \varepsilon$. \textbf{d.} Scatterplot of $\hat{\beta}_{k\ell}$ in both datasets, colored by signatures. The gray line indicates the 45-degree line.}
    \label{fig:ReplicationBreast80}
\end{figure}

The pre-specified refitting analysis performed in \cref{subsec:breast_semi_sup} shows how the activity of known COSMIC signatures varies along the genome and the related effects of genomic covariates. In this section, we assess whether the estimated regression coefficients from PPF can be replicated in magnitude and direction on an independent collection of patients diagnosed with the same tumor. We specifically rely on data from 80 whole-genome-sequenced breast cancer samples from \citetSupp{Davies_2017}, which are further described in \citetSupp{Davies_MMRd}. We refer to the collection in the main manuscript as the ``BreastICGC'' cohort, and to the new one as the ``Breast80'' cohort. See Section~\ref{subsec:data_sources} for details on data access.

The Breast80 dataset features $J = 80$ patients and $323{,}436$ mutations in total, with a median of $2{,}328$, and with only 7 patients showing more than $9{,}000$ mutations (highest mutated patient is PD14455a, $39{,}410$). Figure~\ref{fig:ReplicationBreast80}a and \ref{fig:ReplicationBreast80}b display the variations of the total number of mutations at the Mb scale along the genome, and by mutation type in both collections, respectively. We observe a stark similarity in mutation distributions across the genome (Pearson correlation = 0.81), with jumps on chromosomes 1 and 8. However, only 504 mutations share the exact same position across both datasets (0.15\% of Breast80 and 0.07\% of BreastICGC), and there is no overlap with the donors identifiers. We also observe more pronounced peaks in C$>$T and C$>$G channels in BreastICGC than in Breast80, which likely point to higher APOBEC activity.

To assess differences in mutational signatures, we fit PPF with the same $L = 11$ covariates described in \cref{tab:Epi_covariates}, using a coarser 10 kb partition of the genome. We run Algorithm~\ref{algo:MAP_rules}, using a random starting point and a tolerance of $10^{-6}$, fix signatures to $K = 15$ known COSMIC ones: SBS1 and SBS5 (aging); SBS6, SBS20, SBS26 and SBS44 (mismatch repair deficiency, MMRd); SBS3 and SBS8 (homologous recombination deficiency, HRD); SBS2 and SBS13 (APOBEC); SBS17a, SBS17b and SBS18 (damage by ROS), SBS30 (base excision repair deficiency), and SBS40a (ubiquitous in cancer). All other model parameters are set to their default values.
Convergence to reach the MAP took 120 iterations (4:37 minutes) in BreastICGC, and 180 iterations (5:03 minutes) in Breast80.

Figures~\ref{fig:ReplicationBreast80}c displays the fitted relevance weights $\hat{\mu}_k$ for the fixed signatures in the two cohorts. Signatures where $\hat{\mu}_k < 5\varepsilon$ are those excluded by the posterior and are indicated as ``$\times$''. We see that  SBS6, SBS20, and SBS17a are not selected for either cohort, while SBS44 is selected only in Breast80. This is expected, as this collection features a high number of samples with MMRd-type mutations \citepSupp{Davies_MMRd}. Moreover, the activities of SBS1, SBS8, SBS17b, SBS18, SBS30, and SBS40a are comparable between the two datasets, while SBS2, SBS3, SBS5, and SBS13 are almost twice as large. This implies that aging, HRD, and APOBEC impose a greater burden on the ICGC cohort than on Breast80, explaining the earlier-described peaks in C$>$T and C$>$G motifs. Finally, \ref{fig:ReplicationBreast80}d displays the estimated regression coefficients $\beta_{k\ell}$ in both datasets. We see that the majority of estimates lie close to the 45-degree line, indicating strong agreement in effect. This is especially evident in GC content, methylation, H3K36me3, and replication timing. Moreover, coefficient signs agree in 79\% of the cases, and disagreements are primarily in values between -0.1 and 0.1. Overall, this indicates that the breast cancer results with PPF show good replicability.

\subsection{\emph{De novo} comparison between PPF and \texttt{TensorSignatures}}\label{sec:tensorsignatures}

In this section, we compare PPF against \texttt{TensorSignatures}, a tensor factorization method introduced by \citetSupp{TensorSignatures_2021}. Section~\ref{subsec:chromatins} presents a brief explanation of chromatin states and the model itself; Section~\ref{subsec:benchmarck_TS_PPF} discusses how the two models can be meaningfully compared; Section~\ref{subsec:TS_PPF_results} comments on the results.

\subsubsection{Overview of chromatin states and \texttt{TensorSignatures}}\label{subsec:chromatins}

Chromatin states are a discrete categorization of genomic segments in terms of specific genomic functions. The classification was introduced by \citetSupp{Ernst2017_ChromHMM} using a hidden Markov model to summarize data from several tissue-specific histone marks (H3K4me3, H3K4me1, H3K36me3, H3K27me3, H3K9me3), which produced a partition of the genome into non-overlapping contiguous regions. The simplest classification takes on 15 distinct categories and is reported in Table~\ref{tab:chromhmm_states} along with a brief description, the associated histone mark, and the macro-group of each state: active promoters, gene bodies and enhancers (TssA--Enh), heterochromatin (ZNF.Rpts, Het), poised or bivalent regulatory regions (TssBiv--EnhBiv), Polycomb-repressed chromatin (ReprPC, ReprPCWk), and quiescent regions carrying none of the five marks (Quies). This categorization allows us to assign each SBS mutation to a unique chromatin state, which we can then use as a categorical covariate in our PPF framework. In BreastICGC, we observe that Quies alone covers 58\% of the mutations, whereas TxFlnk, BivFlnk, and TssBiv together account for fewer than 500. Refer to \citetSupp{Ernst2017_ChromHMM} for a full overview.

\texttt{TensorSignatures} \citepSupp{TensorSignatures_2021} is a factorization method to infer mutational signatures and their activity by leveraging information about the chromatin states, and further distinguishes mutations by replication strand and transcription strand. Hence, mutations are arranged in a tensor   $\mathbf{C}^{\mathrm{SNV}}$ of dimension $96 \times 3 \times 3 \times 16 \times J$, where $J$ is the number of patients and 96 are the usual mutation types; the second dimension indicates replication strand (forward, backward, unknown); the third indicates transcription strand (transcribed, untranscribed, unknown);  the fourth dimension are chromatin states (15 states + unknown). The factorization model is obtained as

$$
\mathds{E}\big(\mathbf{C}^{\mathrm{SNV}}\big) = \mathbf{T}\times\mathbf{E},
\qquad
\mathbf{C}^{\mathrm{SNV}} \sim \mathrm{NegBin}\big(\mathbf{T}\times\mathbf{E},\ \tau\big),
$$
where $\mathbf{E}\in\mathds{R}_+^{K_{\mathrm{TS}}\times J}$ collects the exposures of the $K_{\mathrm{TS}}$ signatures in the $J$ patients, $\tau$ is a dispersion parameter, and $\mathbf{T}$ is a tensor of signatures built multiplicatively from low-dimensional factors as
$$
\mathbf{T} = \mathbf{T_1}\odot\mathbf{B}\odot\mathbf{A}\odot\mathbf{K}_{\mathrm{epi}},
$$
where $\odot$ denotes element-wise multiplication after broadcasting \citepSupp[see Equation 6 of ][]{TensorSignatures_2021}. Here, $\mathbf{T_1}$ is the strand-specific 96-dimensional spectra, $\mathbf{B}$ and $\mathbf{A}$ carry the transcriptional and replication strand biases and the activities in transcribed and early-replicating regions, and $\mathbf{K}_{\mathrm{epi}}$ is the reshaping of a matrix $\mathbf{k}_{\mathrm{epi}}\in\mathds{R}_+^{K_{\mathrm{TS}}\times L}$ whose entries $a_{k\ell}$ are referred to as \emph{amplitudes}. In particular, $a_{k\ell}$ multiplies the rate of signature $k$ in genomic state $\ell$, with the first row fixed to $a_{k1}=1$ for identifiability. Hence, $\log a_{k\ell}$ is a log-intensity ratio between state $\ell$ and the reference state. This is a quantity that is directly comparable with the regression coefficients $\beta_{k\ell}$ adopted by PPF, under categorical covariates. The model additionally accepts a normalization tensor $\mathbf{N}$ of the same shape as $\mathbf{C}^{\mathrm{SNV}}$, which multiplies the predicted counts and therefore acts as an offset. All parameters are estimated by maximum likelihood using \texttt{TensorFlow}, and the rank $K_{\mathrm{TS}}$ is chosen by an information criterion computed over a grid of candidate values.

\begin{spacing}{1}
\begin{table}[t!]
\caption{\footnotesize{The 15 chromatin states of the Roadmap Epigenomics core \textsc{ChromHMM} model \protect\citepSupp{Ernst2017_ChromHMM}. The \textsc{Marks} column indicates the histone modifications that primarily define each state. \textsc{Mutations} reports the number of single-base substitutions in the ICGC breast adenocarcinoma cohort falling in each state, with the proportion of the total in parentheses. The \textsc{Group} column divides the states into active, heterochromatic, bivalent, Polycomb-repressed, and quiescent regions.}}
\label{tab:chromhmm_states}
\centering
\footnotesize
\begin{adjustbox}{max width=\textwidth,center}
\begin{tabular}{p{0.16\textwidth}p{0.09\textwidth}p{0.17\textwidth}p{0.5\textwidth}r}
\toprule
\textsc{Group} & \textsc{State} & \textsc{Marks} & \textsc{Description and role} & \textsc{N. mutations} \\
\midrule
Quiescent & Quies & --- & Quiescent regions, with no detectable signal from any of the five marks. & 413,349  (58.46\%) \\
\midrule
\multirow{7}{*}{Active} & TssA & H3K4me3 & Active transcription start sites; promoters of expressed genes. & 5,381 \, (0.76\%) \\
\addlinespace[1pt]
 & TssAFlnk & H3K4me3, H3K4me1 & Regions immediately flanking active promoters. & 1,419 \, (0.20\%) \\
\addlinespace[1pt]
 & TxFlnk & H3K4me1, H3K36me3 & Transcribed regions at the $5'$ and $3'$ ends of genes. & 124 \, (0.02\%) \\
\addlinespace[1pt]
 & Tx & H3K36me3 (strong) & Gene bodies under strong transcriptional elongation. & 26,268 \, (3.72\%) \\
\addlinespace[1pt]
 & TxWk & H3K36me3 (weak) & Weakly transcribed gene bodies. & 117,901  (16.68\%) \\
\addlinespace[1pt]
 & EnhG & H3K4me1, H3K36me3 & Enhancers located within transcribed genes. & 2,527 \, (0.36\%) \\
\addlinespace[1pt]
 & Enh & H3K4me1 & Intergenic enhancers. & 28,620 \, (4.05\%) \\
\midrule
\multirow{2}{*}{\makecell[l]{Heterochromatin}} & ZNF.Rpts & H3K36me3, H3K9me3 & Zinc-finger gene clusters and associated repeats. & 1,012 \, (0.14\%) \\
\addlinespace[1pt]
 & Het & H3K9me3 & Constitutive heterochromatin. & 36,446 \, (5.15\%) \\
\midrule
\multirow{3}{*}{Bivalent} & TssBiv & H3K4me3, H3K27me3 & Bivalent (poised) promoters, carrying both activating and repressive marks. & 235 \, (0.03\%) \\
\addlinespace[1pt]
 & BivFlnk & mixed & Regions flanking bivalent promoters and enhancers. & 58 \, (0.01\%) \\
\addlinespace[1pt]
 & EnhBiv & H3K4me1, H3K27me3 & Bivalent (poised) enhancers. & 604 \, (0.09\%) \\
\midrule
\multirow{2}{*}{Polycomb} & ReprPC & H3K27me3 (strong) & Polycomb-repressed chromatin. & 6,777 \, (0.96\%) \\
\addlinespace[1pt]
 & ReprPCWk & H3K27me3 (weak) & Weakly Polycomb-repressed chromatin. & 66,322 \, (9.38\%) \\
\bottomrule
\end{tabular}
\end{adjustbox}
\end{table}
\end{spacing}

\subsubsection{Benchmarking PPF against \texttt{TensorSignatures}}\label{subsec:benchmarck_TS_PPF}

In order to ensure a meaningful comparison between PPF and \texttt{TensorSignatures}, we adopt the following model choices. First, we discard the transcriptional and replication strands by assigning all mutations to the ``unknown'' category (indexed by 3). This is because PPF does not account for either. This allows us to retain the 96-dimensional mutation-type classification, which we denote as $\mathbf{T}_0^{\mathrm{avg}}$. Second, we let $\mathbf{k}_{\mathrm{epi}}\in\mathds{R}_+^{15\times K_{\mathrm{TS}}}$, so that the matrix encoding the covariates has one amplitude per chromatin state, keeping ``Quies'' as baseline. Third, we use total copy numbers per patient as an offset. This compensates for the fact that \texttt{TensorSignatures} does not explicitly account for copy numbers, unlike PPF.

Let $B_1, \ldots, B_M$ be the segments from ChromHMM, $s(m) \in \{1, \ldots, 15\}$ the state of segment $m$, and take $\mathbf{x}(t) = \mathbf{e}_{s(m)} \in \{0,1\}^{14}$ for $t \in B_m$, where $\mathbf{e}_\ell$ indicates state $\ell$ and Quies is dropped, so that $\beta_{k,\mathrm{Quies}} = 0$ in PPF and $a_{k, \mathrm{Quies}} = 1$ is the quiescent amplitude in \texttt{TensorSignatures}. Writing $\mathrm{c}_{\ell j} = \sum_{m : s(m) = \ell} \frac{1}{2} c_{jm}\Delta_m$ the total copy numbers in state $\ell$ for patient $j$, the resulting expected number of mutations from the triplet $(i, j, \ell)$ is
\begin{equation}\label{eq:mean_TS}
\mathds{E}\big(C^{\mathrm{SNV}}_{i33\ell sj}\big)
= \mathrm{c}_{\ell j}\sum_{k=1}^{K_{\mathrm{TS}}}\big(\mathbf{T_0}^{\mathrm{avg}}\big)_{ik}\,E_{kj}a_{k\ell}\,
\end{equation}
in \texttt{TensorSignatures}, while Equation~\eqref{eq:expectation} in the main manuscript becomes
\begin{equation}\label{eq:mean_PPF_state}
\mathds{E}\big(Z_{ij}(\{t : s(t) = \ell\})\big) = \mathrm{c}_{\ell j}\sum_{k=1}^K r_{ik}\,\phi_{kj}\,e^{\beta_{k\ell}}.
\end{equation}
Since the factor $\frac{1}{2}c_j(t_{ijn})$ is common to all $k$, the PPF likelihood depends on the data only through the counts for the triplet (state, channel, patient) and the total copy numbers $\mathrm{c}_{\ell j}$. Moreover, it is straightforward to see that equations~\eqref{eq:mean_TS} and  \eqref{eq:mean_PPF_state} are equivalent under the correspondence $r_{ik}\leftrightarrow(\mathbf{T_0}^{\mathrm{avg}})_{ik}$, $\phi_{kj}\leftrightarrow E_{kj}$ and $e^{\beta_{k\ell}}\leftrightarrow a_{k\ell}$. Therefore, under these choices, the two models  estimate the same quantity, up to the reference state:
\begin{equation*}
\beta_{k\ell} - \beta_{k\ell_0} = \log a_{k\ell} - \log a_{k\ell_0}.
\end{equation*}
Hence, after augmenting \texttt{TensorSignatures} to account for copy numbers, the main differences are the likelihood (negative binomial versus Poisson process), the estimation algorithm, and the ability to use continuous versus discrete covariates.

We fit both models on the BreastICGC adenocarcinoma data using the ChromHMM segments found in the breast epithelium track (E028 in Roadmap) as bins, with ``Quies'' as the reference state and keeping the mutational channels to 96. We estimate the MAP of PPF using $K = 12$ and default parameters $a = 1.01$, $\alpha = 1.01$, $\varepsilon = 0.001$, $c_0 = 100$, and $d_0 = 1$. Convergence at a tolerance of $10^{-6}$ was reached after 559 iterations in about 42 minutes, starting from a random point drawn from the prior. As usual, we discard the signatures for which the posterior relevance weights are $\hat{\mu}_k < 5\varepsilon$ in PPF, since these carry a null contribution in the factorization. As for \texttt{TensorSignatures}, we rely on the Akaike information criterion (AIC) to determine the number of active signatures. To ensure a fair comparison, we pick the solution that minimizes the AIC among a grid of ranks from $K_{\mathrm{TS}} = 4$ to $K_{\mathrm{TS}} = 12$.

\subsubsection{Results}\label{subsec:TS_PPF_results}

\begin{figure}%
    \centering
    \includegraphics[width=\linewidth]{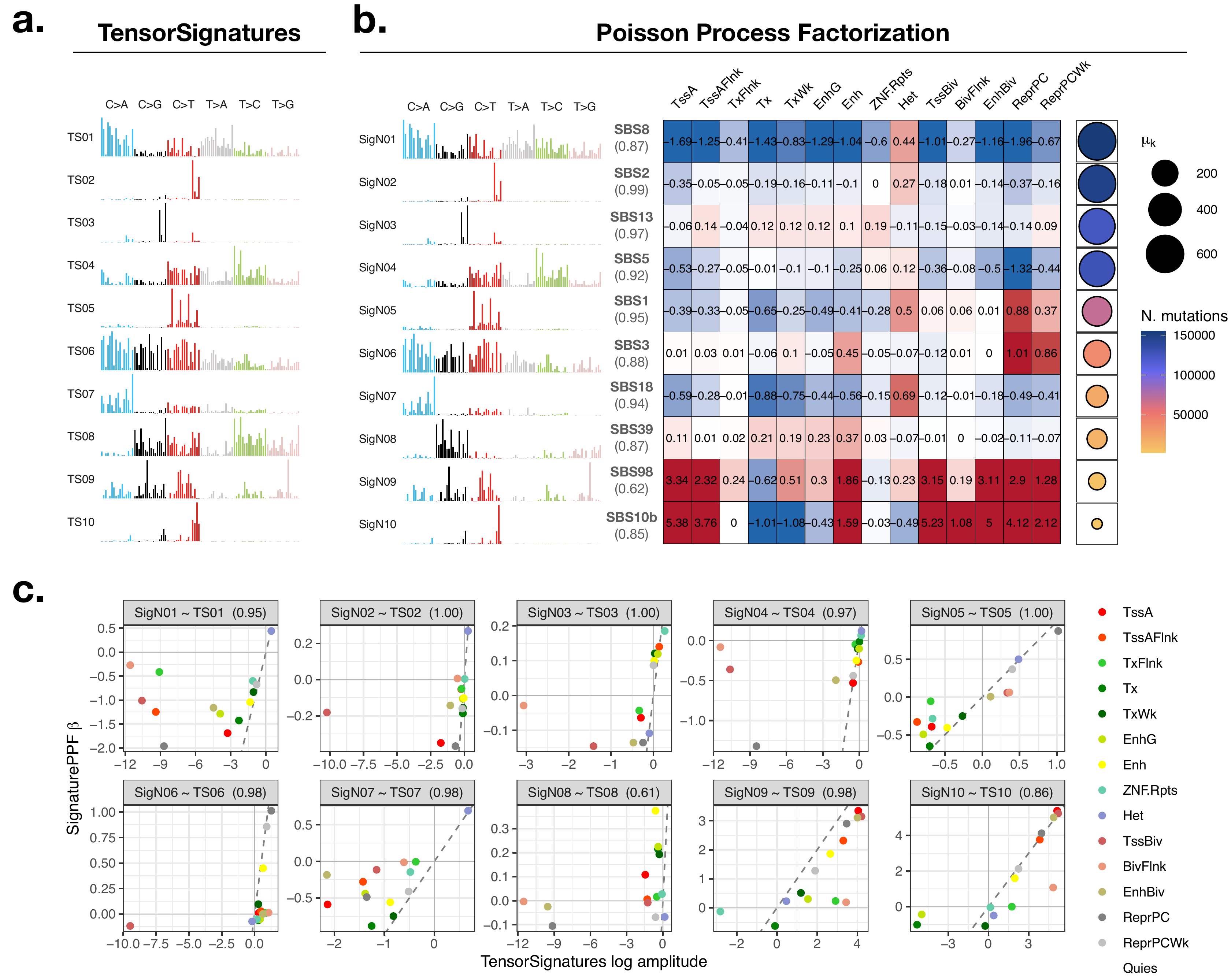}
    \caption{Results of the comparison between PPF and \texttt{TensorSignatures}, using chromatin states covariates on the ICGC data. \textbf{a.} Signatures inferred by \texttt{TensorSignatures}. \textbf{b.} Signatures, regression coefficients, and relevance weights estimated by PPF. Numbers to the right of the signature plot indicate the cosine similarity of the PPF solution with the closest signature in COSMIC. \textbf{c.} Estimated regression coefficients in PPF (y-axis) and log amplitudes from \texttt{TensorSignatures} (x-axis). Colors denote chromatin state, and each panel denotes the matched signatures. The panel titles indicate the two signatures plotted, with their cosine similarity in parentheses; e.g., ``SigN01 $\sim$ TS01 (0.95)'' indicates that the coefficient of SigN01 is plotted on the y-axis, and the log-amplitude of TS01 is on the x-axis, and the two signatures have cosine similarity equal to 0.95. The dashed gray line indicates the 45-degree line. }
    \label{fig:tensorcomparison}
\end{figure}

Figure~\ref{fig:tensorcomparison} displays the results of the comparison between the two approaches on the BreastICGC data. We begin by commenting on each solution, then interpret the regression coefficient for PPF.

\textbf{Model comparison}. We first see that both models infer exactly 10 signatures, with remarkably high cosine similarities. The estimated signatures are displayed in Figure~\ref{fig:tensorcomparison}a for \texttt{TensorSignatures} (labeled as TS01 up to TS10), and in the left panel of Figure~\ref{fig:tensorcomparison}b for PPF (using the nomenclature SigN01 up to SigN10).   We choose this order to match signatures based on closest cosine similarity. In particular, both models infer SBS8 and SBS3 (the HRD signatures) with a similarity of approximately 0.88, and infer the two APOBEC signatures SBS2 and SBS13 nearly perfectly. Both also detect SBS10b and SBS98. The starkest difference between the two profiles pertains to SBS39 (TS08 vs SigN09, cosine similarity of 0.61). This likely implies that this signature does not carry a strong signal in these data; for instance, we found evidence of SBS39 only in the sensitivity analysis of Section~\ref{sec:sensitivity_analysis}. Finally, the methods infer both SBS1 and SBS5, though SBS1 shows more contamination than in the main application because the methylation covariate is absent.

The most important distinction between the two models is highlighted by Figure~\ref{fig:tensorcomparison}c, which plots the estimated log-amplitudes $a_{k\ell}$ from \texttt{TensorSignatures} against the regression coefficients $\beta_{kl}$ estimated by PPF, for all signatures and chromatin states. We see that not only do all panels show a strong positive correlation among the coefficients, but many parameters also lie on the 45-degree line. This confirms the equivalence between equation~\eqref{eq:mean_TS} and equation~\eqref{eq:mean_PPF_state}, while differences are primarily attributed to estimation algorithms and likelihoods. In particular, the log-amplitudes from \texttt{TensorSignatures} are much more spread out than the PPF counterparts, and they reach very negative values. This suggests that PPF may produce more stable coefficient estimates.

\textbf{Effects of chromatin states on mutational signatures}. We now interpret the signatures discovered by PPF under chromatin states. Recall that $\hat{\beta}_{k\ell}$ is the log enrichment relative to the quiescent state. We see that the signatures split into three groups. First, the model detects two signatures depleted in every active state, SigN01 (SBS8) and SigN07 (SBS18), most strongly at promoters and in transcribed gene bodies ($\beta =-1.69$ at TssA and $\beta =-1.43$ at Tx for SigN01; $\beta =-0.88$ at Tx for SigN07). In these two, the heterochromatin state (Het) is the only one showing enrichment ($\beta = 0.44$ and $\beta = 0.69$). This reproduces, on categorical covariates, the association with late-replicating heterochromatin found in Section~\ref{subsec:breast_semi_sup}, and is consistent with transcription-coupled repair clearing lesions from gene bodies in both cases, despite the two signatures having distinct aetiologies. Second, SigN05 (SBS1) and SigN06 (SBS3) are enriched in Polycomb-repressed chromatin ($\beta = 0.88$ and $\beta =1.01$ at ReprPC) while SigN05 is also depleted in transcribed regions ($\beta =-0.65$ at Tx). This is consistent with deamination at methylated CpGs accumulating where repair is not transcription-coupled. Third, SigN09 (SBS98) and SigN10 (SBS10b) carry by far the largest coefficients, concentrated at active and bivalent promoters ($\beta = 5.38$ at TssA and $\beta = 5.23$ at TssBiv for SigN10) against depletion in gene bodies; both are supported by few, highly mutated patients, and the corresponding coefficients should be read with caution. Moreover, TxFlnk and BivFlnk are near zero for nearly every signature, which reflects their small mutation counts (Table~\ref{tab:chromhmm_states}) and the resulting shrinkage under the prior on $\bbeta_k$. Finally, the number of signatures and their profile are very similar to those found in the main application, indicating robustness to alternative epigenetic tracks.

\section{Breast cancer data sources and preprocessing}\label{sec:sources_prep}

\subsection{Mutation and copy number data sources}\label{subsec:data_sources}

\textbf{ICGC Breast-AdenoCA}.
Mutation data for the ICGC breast cancer cohort can be downloaded from \url{https://pcawg-hub.s3.us-east-1.amazonaws.com/download/October_2016_whitelist_2583.snv_mnv_indel.maf.xena.nonUS}. Copy number data for the same cohort of patients can be downloaded from \url{https://pcawg-hub.s3.us-east-1.amazonaws.com/download/20170119_final_consensus_copynumber_donor}.

\textbf{Breast80}. Mutations data and copy number data for the additional 80 whole-genome sequence breast samples from \citetSupp{Davies_2017} are accessible at
\url{https://ftp.sanger.ac.uk/pub/cancer/Nik-ZainalEtAl-560BreastGenomes/CopyNumber/Breast80/Caveman_80sample_Yclean_1Dec15.txt} and
\url{https://ftp.sanger.ac.uk/pub/cancer/Nik-ZainalEtAl-560BreastGenomes/CopyNumber/Breast80/}, respectively.

\begin{table}[ht!]
\centering
\small
\caption{Summary of sources used to assemble covariate data.}
\label{tab:details_data}
\begin{adjustbox}{max width=1\textwidth,center}
\begin{tabular}{l l l l l}
\toprule
\textbf{Repository} & \textbf{Covariate} & \textbf{Measurement} & \textbf{Type} & \textbf{File} \\
\midrule
ENCODE & CTCF     & Fold change over control   & Primary cell & ENCFF101RHP.bigWig \\
ENCODE & H3K27ac  & Fold change over control   & Primary cell & ENCFF981WTU.bigWig \\
ENCODE & H3K27me3 & Fold change over control   & Primary cell & ENCFF322JKF.bigWig \\
ENCODE & H3K36me3 & Fold change over control   & Primary cell & ENCFF175PNA.bigWig \\
ENCODE & H3K4me1  & Fold change over control   & Primary cell & ENCFF531OEA.bigWig \\
ENCODE & H3K4me3  & Fold change over control   & Primary cell & ENCFF094MQS.bigWig \\
ENCODE & H3K4me3  & Fold change over control   & Primary cell & ENCFF240TPI.bigWig \\
ENCODE & H3K4me3  & Fold change over control   & Primary cell & ENCFF925FUX.bigWig \\
ENCODE & H3K9me3  & Fold change over control   & Primary cell & ENCFF709GNN.bigWig \\
\midrule
ENCODE & CTCF     & Fold change over control   & Tissue       & ENCFF143AFG.bigWig \\
ENCODE & CTCF     & Fold change over control   & Tissue       & ENCFF663JVM.bigWig \\
ENCODE & H3K27ac  & Fold change over control   & Tissue       & ENCFF984VUL.bigWig \\
ENCODE & H3K27me3 & Fold change over control   & Tissue       & ENCFF353WUO.bigWig \\
ENCODE & H3K36me3 & Fold change over control   & Tissue       & ENCFF529OVO.bigWig \\
ENCODE & H3K4me1  & Fold change over control   & Tissue       & ENCFF467DBD.bigWig \\
ENCODE & H3K4me3  & Fold change over control   & Tissue       & ENCFF216DKX.bigWig \\
ENCODE & H3K9me3  & Fold change over control   & Tissue       & ENCFF764JHZ.bigWig \\
\midrule
ENCODE  & Nucleosome occupancy & Signal                  & K562  & GSM920557\_hg19\_wgEncodeSydhNsomeK562Sig.bigWig \\
ENCODE  & Replication timing   & Wavelet-smoothed signal & MCF7  & wgEncodeUwRepliSeqMcf7WaveSignalRep1.bigWig \\
BSgenome & GC content         & Percent                 & hg19  & BSgenome.Hsapiens.UCSC.hg19 \\
\midrule
GEO & Methylation & Beta score & & GSM5652347\_Breast-Luminal-Epithelial-Z000000V2.bigwig \\
GEO & Methylation & Beta score & & GSM5652348\_Breast-Luminal-Epithelial-Z000000VJ.bigwig \\
GEO & Methylation & Beta score & & GSM5652349\_Breast-Luminal-Epithelial-Z000000VN.bigwig \\
GEO & Methylation & Beta score & & GSM5652350\_Breast-Basal-Epithelial-Z000000V6.bigwig \\
GEO & Methylation & Beta score & & GSM5652351\_Breast-Basal-Epithelial-Z000000VG.bigwig \\
GEO & Methylation & Beta score & & GSM5652352\_Breast-Basal-Epithelial-Z000000VL.bigwig \\
GEO & Methylation & Beta score & & GSM5652353\_Breast-Basal-Epithelial-Z0000043E.bigwig \\
\midrule
Roadmap & Chromatin states & Categorical (15 levels) & &  E028\_15\_coreMarks\_dense.bed\\
\bottomrule
\end{tabular}
\end{adjustbox}
\end{table}

\subsection{Data preprocessing}\label{subsec:data_preprocessing}
\subsubsection{Continuous covariates for the main analysis}\label{subsec:cov_preprocessing}
We assembled the covariate data in \cref{tab:Epi_covariates} from various sources; see \cref{tab:details_data}. Specifically, we consider a set of data similar to the one previously analyzed by \citetSupp{Otlu_2023}, whose goal is to study the relationship between epigenetic markers and mutational signatures. We downloaded from \textsc{encode} all the \texttt{bigWig} files associated with the \texttt{bed} files indicated in the supplement of \citetSupp{Otlu_2023}. While \texttt{bed} files report the regions having statistically significant enrichments from a covariate, \texttt{bigWig} formats contain the signal for the whole genome; hence, they are better suited for our application. We use the files that measure the fold-change over control from the ChIP-seq assay for both histone modifications and CTCF binding.
This unit of measurement has a more transparent interpretation as opposed to $-\log_{10}$ p-values, which is instead used for visualization purposes and to identify significant peaks.
Replication timing is obtained from data on the MCF-7 breast cancer cell line, as in \citetSupp{Morganella2016}. As for nucleosome occupancy, we use data from an MNase-seq assay on the K562 leukemia cell line \citepSupp{Tolstorukov2011}. Although not breast-specific, it is one of the few datasets on nucleosome position derived from cancer cells \citepSupp{Otlu_2023}. Finally, methylation data are obtained from \citetSupp{Loyfer2023} and migrated from genome \texttt{hg38} to \texttt{hg19} using \texttt{liftOver}.

Our pre-processing steps are the following. We subdivide chromosomes 1 to 22 and X in the \texttt{hg19} reference genome into non-overlapping regions of $2,\!000$ nucleotides (2kb aggregation) via the \texttt{tileGenome} function in the \texttt{R} package \texttt{GenomicRanges} \citepSupp{GenomicRanges}. Chromosome Y is excluded because all patients are female. We assign to each bin a weight $\Delta_m$, calculated as the number of nucleotides in a region, excluding the portions that overlap with any of the problematic regions listed by \citetSupp{blacklist}. Hence, if a region is not problematic and does not have any missing values ``N'', we have $\Delta_m = 2{,}000$. Meanwhile, $\Delta_m = 0$ implies that no mutation can be detected in that bin. We subdivide the copy number data for each patient using the same partition, and call $c_{jm}$ the number of copies for patient $j$ in bin $m$. As for the covariates, we first import into memory each \texttt{bigWig} file in \cref{tab:details_data} using the \texttt{rtracklayer} package \citepSupp{rtracklayer} and then calculate the average value of the signal in every bin using the function \texttt{binnedAverage} in \texttt{GenomicRanges}. Signals from different files representing the same covariate are then averaged in each bin. As for GC content, we simply calculate the fraction of G or C nucleotides in the 2kb bins via the function \texttt{letterFrequency} in the package \texttt{Biostrings} \citepSupp{biostrings}. This results in 11 piecewise constant covariates aligned to the same coordinates, with a total of $M = 1{,}391{,}345$ bins having a non-zero weight $\Delta_m >0$ and with $T = \sum_{m=1}^M  \Delta_m = 2{,}780{,}909{,}076$. This is lower than the overall length of \texttt{hg19} due to the presence of blacklisted regions.
Finally, we cap the value of all covariates at their 99.9th percentile to avoid excessively long tails, and we standardize them to have a mean of zero and a variance of one across bins $m = 1,\ldots,M$.

\begin{figure}[t]
    \centering
    \includegraphics[width=\linewidth]{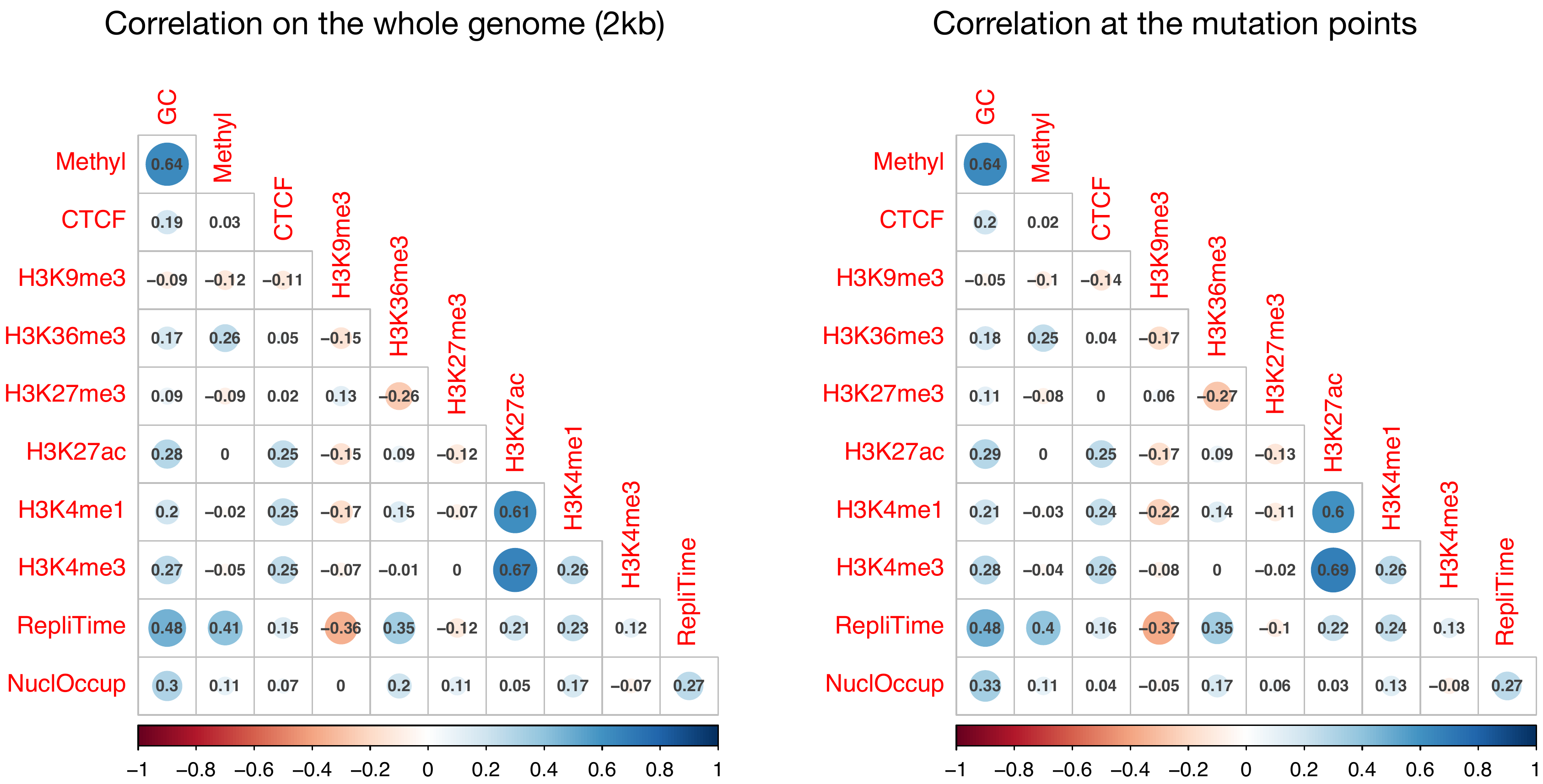}
    \caption{Correlation between genomic covariates at the 2kb resolution. Left: correlations over the whole genome. Right: correlation of the values at the mutation points. Regions that have $\Delta_m = 0$ have been excluded to compute the values.}
    \label{fig:correlation}
\end{figure}

The covariates in this analysis are not patient-specific, but instead correspond to a reference epigenome that is common to all patients. Moreover, except for nucleosome occupancy, all assays are run on tissues and cell lines from breast epithelium obtained from a separate group of healthy donors. Hence, our analysis detects associations that reflect mutational biases in the cancer genomes that are not driven by patient-specific alterations in the chromatin. The 2kb scale is motivated by the need to balance between information at the original \texttt{bigWig}s resolution and computational feasibility when running estimation algorithms. In particular, when $A = [0, T)$, the integral in \cref{eq:expectation} is equal to
$$
\int_0^T \frac{1}{2} \,c_{j}(t) \,e^{\bbeta_k^\top\bx(t)}\mathrm{d}t = \sum_{m = 1}^M \Delta_m \frac{1}{2} \,c_{j m} \,e^{\bbeta_k^\top\bx_m},
$$
where $\bx_m$ and $c_{j m}$ denote the constant values of the covariates and the copy numbers in each bin. Using finer partitions leads to heavier computational and memory requirements, while coarser partitions may obscure the effect of covariates operating at smaller scales, such as CTCF. Furthermore, a 2kb window size centered at each mutation ($\pm$1000 bases) is the proposed size in the analyses of \citetSupp{Otlu_2023} when analyzing mutational signature contributions near histone modifications and transcription factors. The correlation structure of the covariates along the genome is displayed in Figure~\ref{fig:correlation}.

\subsubsection{Covariates on chromatin state }
We download chromatin states from
\url{https://egg2.wustl.edu/roadmap/data/byFileType/chromhmmSegmentations/ChmmModels/coreMarks/jointModel/final/}. We use the track called \texttt{E028\_15\_coreMarks\_dense.bed}, which consists of non-overlapping genomic regions classified into 15 core marks, obtained via hidden Markov models using information of histone modification, methylation, and transcription factors, from breast variant human mammary epithelial cells \citepSupp{Ernst2017_ChromHMM}. Since the file covers the whole hg19 genome, we rely on the pre-defined genomic binning reported in the file, so that $\Delta_m$ varies depending on $m$. In each bin, we convert the categorical covariate into 14 binary variables representing each single chromatin state, using the quiescent state (``{Quies}'') as baseline. Similarly to \cref{subsec:cov_preprocessing}, we exclude blacklisted regions from \citetSupp{blacklist}. A brief description of the chromatin states and their interpretation is presented in  \cref{tab:chromhmm_states}.

\bibliographystyleSupp{chicago}
\bibliographySupp{references}

\end{document}